\documentclass[usenatbib]{mn2e}
\usepackage{graphicx}
\usepackage{pslatex}
\def\deg{{\rm o}}
\def\idm#1{{\mbox{\scriptsize #1}}}
\newcommand\Chi{{(\chi^2_\nu)^{1/2}}}
\newcommand\huaqr{{HU~Aqr}}

\def\v#1{{\bf{#1}}}
\newcommand\Pb{P_{\idm{bin}}}
\usepackage{color}
\definecolor{myred}{rgb}{0.7,0.1,0.1}
\definecolor{myblue}{rgb}{0.2,0.0,0.7}
\definecolor{mybrown}{rgb}{0.5,0.2,0.0}

\newcommand\corr[1]{{\color{black} #1}}

\newcommand\hide[1]{}
\newcommand\dcl{{day cycle$^{-2}$}}
\author[K. Go\'zdziewski et al.]{
Krzysztof~Go\'zdziewski$^1$, 
Ilham~Nasiroglu$^{2,3}$, 
Aga S\l{}owikowska$^4$, 
Klaus~Beuermann$^5$,
\newauthor
\hspace*{2mm}Gottfried~Kanbach$^2$, 
Bartosz~Gauza$^{6,7}$, 
Andrzej~J.~Maciejewski$^4$, 
Robert~Schwarz$^8$, 
\newauthor
\hspace*{2mm}Axel~D.~Schwope$^8$, 
Tobias~C.~Hinse$^9$,
Nader~Haghighipour$^{10}$,
Vadim Burwitz$^{2,11}$,
\newauthor
\hspace*{2mm}Mariusz S\l{}onina$^1$ and
Arne Rau$^2$\\
\\
$^1$Toru\'n Centre for Astronomy, Nicolaus Copernicus University, 
Gagarin Str. 11, 87-100 Toru\'n, Poland \\
$^2$Max Planck Institute for Extraterrestrial Physics, Giessenbachstrasse, 85748 Garching,
Germany\\
$^3$University of Cukurova, Department of Physics, 01330 Adana, Turkey\\
$^4$Kepler Institute of Astronomy, University of Zielona G\'ora, Lubuska 2,
65-265 Zielona G\'ora, Poland\\
$^5$Universit\"at Goettingen, Institut f\"ur Astrophysik,
Friedrich-Hund-Platz~1, DE 37077 G\"ottingen, Germany \\ 
$^6$Instituto de Astrof$\acute{i}$sica de Canarias (IAC), E-38200 La Laguna,
Tenerife, Spain\\
$^7$Dept. Astrof$\acute{i}$sica, Universidad de La Laguna (ULL), E-38206 La
Laguna, Tenerife, Spain\\
$^8$Leibniz-Institute f\"ur Astrophysik (AIP), An der Sternwarte 16, 14482
Potsdam, Germany\\
$^9$Korea Astronomy and Space Science Institute (KASI), Optical Astronomy
Research Center, 776 Daedukdae-ro, Daejeon, 305-348, 
Republic of Korea\\
$^{10}$Institute for Astronomy and NASA Astrobiology Institute,
University of Hawaii-Manoa, 2680 Woodlawn Drive, Honolulu, HI 96822, USA\\
$^{11}$Observatorio Astronomico de Mallorca, 07144 Costitx, Mallorca, Spain 
}
\begin{document}
%
%
\title{On the HU~Aquarii planetary system hypothesis}
%
\maketitle

\begin{abstract}
In this work, we investigate the eclipse timing of the polar binary 
HU~Aquarii that has been observed for almost two decades.  Recently, Qian 
et al. attributed large (O-C) deviations between the eclipse ephemeris and 
observations to a compact system of two massive jovian companions.  We 
improve the Keplerian, kinematic model of the Light Travel Time (LTT) 
effect and re-analyse the whole currently available data set.  We add 
almost 60 new, yet unpublished, mostly precision light curves obtained 
using the time high-resolution photo-polarimeter OPTIMA, as well as 
photometric observations performed at the MONET/N, PIRATE and TCS 
telescopes.  We determine new mid--egress times with a mean uncertainty at 
the level of 1~second or better.  We claim that because the observations 
that currently exist in the literature are non-homogeneous with respect to 
spectral windows (ultraviolet, X-ray, visual, polarimetric mode) and the 
reported mid--egress measurements errors, they may introduce systematics 
that affect orbital fits.  Indeed, we find that the published data, when 
taken literally, cannot be explained by any unique solution.  Many 
qualitatively different and best-fit 2-planet configurations, including 
self-consistent, Newtonian $N$-body solutions may be able to explain the 
data.  However, using high resolution, precision OPTIMA light curves, we 
find that the (O-C) deviations are best explained by the presence of a 
single circumbinary companion orbiting at a distance of $\sim 4.5$~AU with 
a small eccentricity and having $\sim 7$ ~{Jupiter-masses}.  This object 
could be the next circumbinary planet detected from the ground, similar to 
the announced companions around close binaries HW~Vir, NN~Ser, UZ~For, DP 
Leo or SZ~Her, and planets of this type around Kepler-16, Kepler-34 and 
Kepler-35.
\end{abstract}

%
%
\section{Introduction}
%
%
Magnetic cataclysmic variables (CVs, polars, a.k.a. AM Her stars) are 
interacting close binary systems. They  consist of {a} main sequence red 
dwarf secondary filling {its} Roche lobe, and {a} strongly magnetized white 
dwarf (WD) primary, with typical magnetic field values of 10--80~MG \citep
{Schwope2001}. The strong magnetic field {of the primary} interacts with 
{the} weaker magnetic field of the secondary and locks the two stars 
together. Hence, the synchronously rotating WD spins at the same rate as 
the orbital mean motion of the binary. Under the gravitational field of the 
primary, material flows from {the} donor {star} initially along the binary 
orbital plane, and finally is accreted quasi-radially {onto the magnetic 
poles of} the WD. {The  variable HU Aquarii system ({hereafter} \huaqr{})} 
belongs to this class {of} CV {binaries} hosting a strongly magnetic WD 
accompanied by {a red dwarf (spectral type M4V)} with {an} orbital period 
of about 125 minutes. This system is one of the brightest polars in {the} 
optical domain with visual magnitudes ranging from 14.6 to 18 \citep
{Warner1995, Hellier2001}, as well as in the X-ray energy range. Therefore, 
it {has also been} one of the most studied systems so far.

Accreted matter leaving from the red dwarf is initially not affected by the 
magnetic field of the WD. The matter follows a ballistic trajectory up to 
the moment when the WD magnetic field begins to dominate. Because the WD 
magnetosphere extends beyond the $L_1$ radius, the plasma stream  {cannot} 
orbit freely, and thus does not form an accretion disk, unlike {in} other 
non-magnetic cataclysmic variables. The accreted matter follows {the} 
magnetic field lines and forms an accretion spot at the magnetic poles of 
the WD. In many systems, the WD magnetic field is tilted in a such way that 
one magnetic pole {is oriented toward} the direction of flowing matter. 
Eclipses observed in highly inclined polars provide information about the 
stream geometry.

According to the most recent work of \cite{Schwope2011} the inclination of
the binary is $\sim 87^{\deg} \pm 0.8^{\deg}$. This special geometry is
important for the planetary hypothesis investigated in this work. Assuming
that a planetary companion (or companions) have formed in the circumbinary
disc, the inclination constraint removes the mass indeterminacy inherent to
the eclipse timing method.

Recently, the \huaqr{} system {has} received much attention in the 
literature. \cite{Schwarz2009} carried out an analysis of the light curves 
of the system and derived mid--egress times of the polar. They proposed a 
planetary companion as \corr{one possible} explanation of the detected (O-C) 
variability. Shortly after this work, \cite {Qian2011} presented and 
discussed 10~new light curves in the optical domain. These authors 
confirmed the deviations of the observed mid--egress times from a linear or 
quadratic ephemeris, concluding that the large (O-C) residuals may be 
explained by the Light Travel Time [LTT aka Roemer effect, \citet 
{Irwin1952}] due to two jovian-mass planetary companions in orbits with 
semi-major axes of a few AU and a moderate eccentricity of $\sim 0.5$ for 
the outer planet. The orbit of the inner planet was fixed to be circular. 
The ratio of the orbital periods of these massive putative planets would be 
presumably in a low--order 2c:1b mean motion resonance (MMR). The latter 
points to significant mutual interactions between these objects which 
\corr{strongly affects the orbital stability} of the system. Indeed, shortly after 
that work was published, \cite{Horner2011} performed dynamical analysis of 
the putative \huaqr{} 2-planet system, exploring the parameter space within 
$3\sigma$ uncertainty levels of the derived Keplerian elements. They found 
that none of the best-fit configurations {presented by \citet {Qian2011}} 
were dynamically stable implying that the planetary hypothesis proposed by 
these authors is hard to maintain. After a few months, in a new paper, \cite
{Wittenmyer2011} also re-analysed data in \cite{Qian2011} confirming that 
the 2-planet configuration is mathematically consistent with the 
observations, but inferred  orbits are catastrophically unstable over a 
$\sim 10^3$-- $10^{4}$~year time--scale. Furthermore, in a very recent 
paper, \cite{Hinse2011} improved the Keplerian fit models of this system by 
imposing orbital stability constraints on the objective function $\Chi$. 
Although these authors were able to find a {\em stable} 2-planet 
configuration consistent with the linear ephemeris model, orbital 
parameters of these planets were relatively distant from the formal 
best-fit solution by more than 3$\sigma$. Because the results of extensive 
dynamical analysis contradict the 2-planet hypothesis, an alternative 
explanation of the (O-C) \corr{diagrams needs to be} considered.

Long term monitoring of \huaqr{} shows large variations of the accretion 
rate that could be correlated with a migration of the accretion spot. 
Taking into account the observed changes of the accretion geometry during 
different accretion states, high and intermediate ones, \cite{Schwope2001} 
estimated the possible time--shift of eclipses to be on the level of 
2~seconds, which is still much smaller than \corr{the} deviations
\corr{between the theoretical ephemeris} and observed mid--egress moments.
{These results suggest that} the migration of the accretion spot cannot be
responsible for the (O-C) deviations, and we therefore ruled it out. 

The (O-C) variability of \huaqr{} could be \corr{also attributed to} other complex 
astrophysical phenomena in the binary, like the Applegate mechanism and/or 
magnetic braking discussed by \cite{Schwarz2009} and \cite{Wittenmyer2011}. 
The timing signal might also be affected by non-Gaussian red-noise, which 
is a~well known effect present in the precision photometry of transiting 
planets and timing of millisecond pulsars \citep
[e.g.,][]{Pont2006,Coles2011}. Hence, it should be stressed that we focus 
here on the planetary hypothesis, as one of \corr{the} possible, simple and somehow 
attractive \corr{explanations} of the (O-C) variability. We try to solve ``the 
puzzle'' of unstable 2-planet models through a new and independent analysis 
of available data, conducted along three basic directions.  

The first one relies on the re-analysis of published data, because we found 
a~few inconsistencies in the literature.  Surprisingly, while in the recent 
paper, \cite{Wittenmyer2011} take into account 82 mid--egress points from 
\cite{Schwarz2009} and \cite{Qian2011}, this is not the full data set 
available in the literature at that time. In fact, 72 egress times 
published by \cite{Schwarz2009} extend the data set in \cite{Schwope2001} 
that included 31~measurements.  Although the early data of \cite 
{Schwope2001} spanning cycles 0--22478 overlap with measurements 
in \cite{Schwarz2009} in the time window covering cycles 1319--60097, they 
may be helpful to constrain the best-fit models.  Up to now, the full list 
of published observations {consists of} 113 points, including data in \cite
{Qian2011}. Yet it is not quite obvious whether \cite{Qian2011} included 
measurements in \cite{Schwope2001} in their analysis. \cite {Hinse2011} 
considered the full data set available at that time, but in terms of the 
linear ephemeris LTT model only. In this  context, a direct comparison of 
the results in the published papers is difficult.

The second aspect of our study is a new kinematic model of the ephemeris 
that properly approximates orbits of putative companions in \corr{multi-body 
systems} (to the lowest possible order in the masses), as compared to  the 
full $N$-body model. The kinematic model used in all cited papers refers 
back to Keplerian parametrisation by \cite{Irwin1952} {\em for the ``one 
companion'' case}. That model, though commonly used in the literature \citep
[e.g.,][]{Lee2011}, seems nowadays redundant, as it was introduced to 
quantify a similarity between the LTT and the radial velocity curves, 
in a particular reference frame with the origin at the center of the 
two--body LTT orbit (instead of the dynamical barycenter). Indeed, the 
recent, although short history of modeling precision \corr{radial velocities}
teaches us that 
multiple planetary systems should be modeled either using kinematic 
formulation in a proper coordinate frame \citep 
[e.g.][]{Lee2003,Gozdziewski2003}, or using the most general and accurate 
full $N$-body model \citep{Laughlin2001}. The dynamical stability can be 
further incorporated as an additional, {\em implicit} observable to the 
objective function \citep[e.g.,][]{Gozdziewski2001,Gozdziewski2008}. In 
this work we are focused on the kinematic modeling though the 
self-consistent $N$-body approach was also used to analyse the \huaqr{} 
mid--egress times (see Appendix). Our results indicate that the Newtonian 
model may be required for other systems presumably exhibiting the LTT 
effect, indeed.

The third and, actually, critical direction of our work, is a careful 
independent analysis of the significantly extended data set including 
already published egress times, and new high-precision timing of the 
egresses {obtained} with the ultra-fast photometer OPTIMA 
\citep {Kanbach2003,Kanbach2008}, as well as {the} MONET/N, PIRATE and TCS 
telescopes. We collected almost 60 new egress times with superior accuracy 
at the sub-second level.  Moreover, we found that the literature data are 
non-homogeneous, as they come from different instruments with different 
time resolutions, as well {as} working in different spectral windows (from 
the visual range, through the UV, to the X-ray domain) and non/polarimetric 
modes. {Taking into account the above mentioned inhomogeneities factors and 
new data, we present the results from a quasi-global optimization of two 
basic LTT models, leading us to the conclusion that the measured (O-C) data 
of \huaqr{} may be best explained {by} a 1-planet configuration}. 
Simultaneously, it would resolve the 2-companions instability paradox in 
the simplest way.

This paper is structured as follows. In Sect.~2 we derive 2-planet LTT 
models on the basis of Jacobi coordinates which describes kinematic orbits 
\corr{in multiple systems} properly, as well as a hybrid optimisation algorithm and 
numerical setup that makes it possible to explore the $\Chi$ parameter 
space in a quasi-global manner. We also briefly describe the $N$-body 
formulation of the LTT effect. In Sect.~3, we re-analyse the data set 
published in the literature, following the 2-planet hypothesis by \cite 
{Qian2011} and further investigated by \cite{Wittenmyer2011}. Two examples 
of highly degenerate best-fit solutions are found. In Sect.~4, possible 
effects of different spectral windows for the light curves and 
determination of egress times are studied. Furthermore, we describe the new 
data set derived with the OPTIMA and other instruments. In  Sect.~5, we 
propose the 1-planet model that best explains the (O-C) variability. We 
briefly discuss the effect of red-noise  in Sect.~6 and present a summary 
of our work in the Conclusions, Sect. 7. The Appendix contains extensive 
supplementary material to Sect.~5, including the results of kinematic and 
$N$-body modeling of 2-planet systems, accompanied by the long-term 
stability tests.
%
%
\section{LTT model for \corr{a} 2-planet system}
%
We briefly develop the Keplerian model of the LTT signal in the three-body 
configuration, assuming that a compact binary (like HU~Aqr) has {\em two} 
planetary companions.  More technical details and a generalization of that 
model will be published elsewhere (Gozdziewski et al., in preparation). We 
consider the compact binary as a {\em single} object having the mass of 
$m_*$, which is reasonable in {accordance} with the extremely short orbital 
period {($\sim 125$~min)} of the polar.  A single companion, as well {as} 
multiple-planet models are particular cases of this problem.  The key point 
is that the Keplerian (or {\em kinematic}) model requires special 
coordinates in order to preserve the sense of Keplerian elements as an 
approximation of the {\em exact} $N$-body initial condition. That can be 
accomplished by expressing the dynamics through particular canonical 
coordinates in which the mutual planetary interactions are possibly small 
with respect to the main, ``pure'' Keplerian part.  The barycentric 
formulation \citep{Irwin1952} in fact ignores the interactions which could 
be  adequate for low-mass circumbinary objects, but it might fail when they 
have stellar masses as in the SZ~Her system \citep{Lee2011} where 
companions are as massive as 20\% of $M_{\sun}$, and can shift the system 
barycenter significantly. The reason for introducing this improved model is 
in fact the same as in the precision \corr{radial velocities} analyses \citep 
[e.g.,][]{Lee2003,Gozdziewski2003}.
%
%
\subsection{Kinematic parametrization of the LTT effect}
\begin{figure}
\centerline{
   \vbox{
     \hbox{
         \hbox{\includegraphics[width=3.3in]{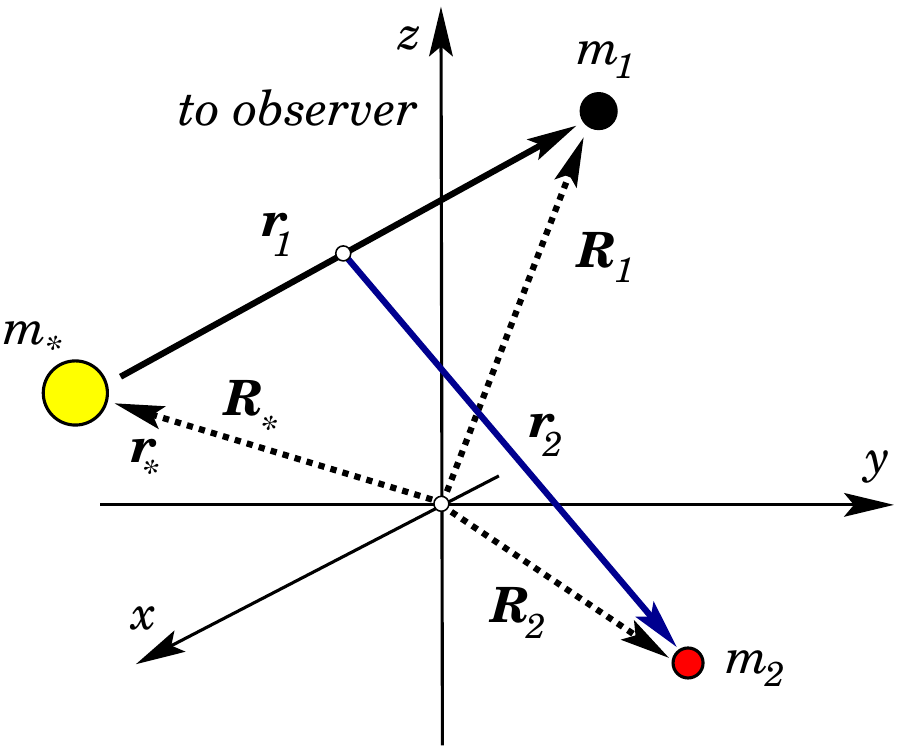}}
     }
   }
}
\caption{
The geometry of the system. The binary has a total mass $m_*$ and {because 
of its} short orbital period {it can be considered as a} point-like object 
accompanied by {planets as} point-masses. The {origin of the} coordinate 
system is fixed {at} the barycentre of the three-body {system}. The 
line-of-sight is along the $z$-axis. See the text for more details.
}
\label{fig:fig1}
\end{figure}

%
One of the well known frames that provides a proper description of 
kinematic orbits \corr{in multiple systems} is Jacobi coordinates. Let us assume 
that {$m_*$, $m_1$ and $m_2$ represent}  the masses 
of {the} compact binary $m_*$ and two planets, respectively.  
{Let us also assume that the} Cartesian coordinates {of these objects} 
with respect to the three-body barycentre are 
$
\v{R}_*, \v{R}_1, \v{R}_2,
$
{and their} Jacobi coordinates are denoted by
$
\v{r}_* \equiv \v{R}_*, \v{r}_1, \v{r}_2
$
(see Fig.~\ref{fig:fig1}). {Here $\v{R}_*$ is the position of the centre
of mass of the binary (CMB) in the barycentric frame, and $\v{r}_1$, $\v{r}_2$ 
are position vectors of the planetary companions in the Jacobi frame.}
{In this formalism,} the barycentric position of the binary is:
\begin{equation}
\label{eq:R}
 \v{R}_* = -\kappa_1 \v{r}_1 - \kappa_2 \v{r}_2,
\end{equation}
where the mass factor coefficients $\kappa_1\geq0$, $\kappa_2\geq 0$ are given by:
\begin{equation}
\label{eq:kappa}
\kappa_1 = \frac{m_1}{m_1+m_*}, \quad \kappa_2 = \frac{m_2}{m_1+m_2+m_*}.
\end{equation}
The coordinate transformation $\v{R} \rightarrow \v{r}$ is 
{taken from} \citet{Malhotra1993}:
\begin{eqnarray}
\label{eq:jtor}
\v{r}_* & \equiv & \v{R}_*,  \nonumber \\
\v{r}_1 & = & \v{R}_1 - \v{R}_*, \\
\v{r}_2 & = & \v{R}_2 - \frac{m_* \v{R}_* + m_1 \v{R}_1}{m_1+m_*}, \nonumber
\end{eqnarray}
and the inverse transformation is derived from the integral of the
barycentre:
\begin{eqnarray}
\label{eq:rtoj}
\v{R}_* & = & -\kappa_1 \v{r}_1 - \kappa_2 \v{r}_2, \nonumber \\
\v{R}_1 & = & (1-\kappa_1) \v{r}_1  - \kappa_2 \v{r}_2, \\
\v{R}_2 & = & (1-\kappa_2) \v{r}_2. \nonumber
\end{eqnarray}
To the first order in the {mass-ratio} ($\sim m_{1,2}/m_*$), the true $N$-body
orbit of body $i$, $(i=1,2)$ is described through geometric Keplerian
elements as follows:
\[
 \v{r}_i(t) = \v{P}_i \left[ \cos E_i(t) - e_i \right] + 
              \v{Q}_i \sqrt{1-e_i^2} \sin E_i(t),
\]
where 
\[
\v{P}_i = a_i \left( \v{l}_i \cos\omega_i + \v{k}_i \sin\omega_i \right), \quad
\v{Q}_i = a_i \left( -\v{l}_i \sin\omega_i + \v{k}_i \cos\omega_i \right),
\]
and geometric elements are defined through:
\[
 \v{l}_i = \left[ \begin{array}{c} 
                      +\sin\Omega_i \\ +\cos\Omega_i \\ 0 
                 \end{array} 
           \right], \qquad
 \v{k}_i = \left[ \begin{array}{c} 
                    +\cos i_i \cos\Omega_i \\ -\cos i_i \sin\Omega_i \\ \sin i_i 
                 \end{array} 
           \right].           
\]
{Here, $E_i(t)$ is the eccentric anomaly derived from the Kepler equation
\[
 n_i (t - T_i) = E_i(t) - e_i \sin E_i(t), 
\]
where $n_i=2\pi/P_i$ is the mean motion, in accordance with  Kepler 
\corr{3rd} law, $n_i^2 a_i^3 = \mu_i$, where $P_i$ is the orbital period of a
given object.}
Two tuples $(a_i,e_i,i_i,\Omega_i,\omega_i,T_i)$, $i=1,2$, that consist of 
the semi-major axis, eccentricity, inclination, nodal angle, argument of 
pericentre, and the time of pericentre passage, respectively, are for the 
geometric Keplerian elements.  These are related to the  Cartesian 
coordinates in the Jacobi frame through the usual two-body formulae \citep 
[see, e.g.][]{Morbidelli2002}, with an appropriate mass parameter $\mu_i$ 
(see below).

From Eq.~\ref{eq:R}, the $Z_*$ component of the CMB with respect to {the 
system barycentre} is:
\begin{equation}
   Z_*(t) \equiv \v{R}_* \cdot \v{e}_z = -\kappa_1 z_1(t) - \kappa_2 z_2(t), 
\label{eq:zstar}  
\end{equation}
where {$\v{e}_z$ is the unit vector along the $z$--axis of the reference 
frame, directed toward the observer. The signal contribution due to a given 
companion is}:
\begin{equation}
 z_i(t) = a_i \sin i_i \left[ \sin \omega_i \left( \cos E_i(t) - e_i \right) +
      \cos \omega_i \sqrt{1-e_i^2} \sin E_i(t)
     \right]
\label{eq:signal}
\end{equation}
(for planets $i=1,2$).  The $z_i(t)$ are then combined to obtain the
$Z_*(t)$ component of the CMB position vector w.r.t. 
the system barycentre. 
{The} LTT signal {is then} expressed {as}:
\[
\tau(t) = {-}\frac{1}{c} Z_* \equiv {+}\frac{1}{c}\left( \frac{m_1}{m_1+m_*} z_1 + 
           \frac{m_2}{m_1+m_2+m_*} z_2 \right),
\]
where $c$ is the speed of light. Note that we used the planetary version of 
the three-body system, with {\em one} dominant mass ($m_*$), hence the 
gravitational Keplerian parameters are:
\[
\mu_1 = k^2 (m_1+m_*), \quad \mu_2 = k^2 \frac{m_* (m_1+m_2+m_*)}{m_1+m_*},
\]
consistent with the expansion of \corr{the} Hamiltonian {perturbation} for the 
planetary version of the problem \cite[see, e.g.,][]{Malhotra1993}, {and 
the quantity} $k$ denotes the Gauss constant.

We introduce the signal semi-amplitude factors, $K_1$ and $K_2$ {as}:
\begin{eqnarray}
\label{eq:k}
K_1 &=& \left(\frac{1}{c} \right) \frac{m_1}{m_1+m_*} a_1 \sin i_1, \\  
K_2 &=& \left(\frac{1}{c} \right) \frac{m_2}{m_1+m_2+m_*} a_2 \sin i_2.
\label{eq:k1}
\end{eqnarray}
Using Eq.~\ref{eq:signal}, the single-planet signal contributions $\zeta_i$ 
are then given {by}:
\begin{equation}
 \zeta_i(t) = K_i \left[ \sin \omega_i \left( \cos E_i(t) - e_i \right) +
      \cos \omega_i \sqrt{1-e_i^2} \sin E_i(t)
     \right].
\label{eq:ltt}
\end{equation}
In {this equation}, the set of free orbital parameters is
($K_i,P_i,e_i,\omega_i,T_i$), $i=1,2$, {similar} to the common
kinematic \corr{radial velocity} model. The orbital period $P_i$
and the time of pericenter passage are introduced indirectly
through \corr{the} time dependence expressed by $E_i(t)$. 

We would like to note here that the contribution of the planet as expressed 
in Irwin's model has an extra term $e_i \sin\omega_{i}$ that appears due to 
{the} particular choice of the coordinate {system} with the origin at {the} 
center of the binary orbit around {the} common {center of mass} of the 
system. It should  also {be} stressed, that {no} simple superposition of 
kinematic orbits does account for the mutual gravitational interactions 
directly, but in our formulation, the Keplerian elements are the {closest} 
to the osculating $N$-body initial condition within the kinematic model.
%
%
\subsection{The (O-C) formulation}
%
From Eq.~\ref{eq:ltt}, the fit model of the planetary-induced LTT signal is
\[
 \tau(t,K_1,P_1,e_1,\omega_1,T_1,K_2,P_2,e_2,\omega_2,T_2) = 
 \zeta_1(t) + \zeta_2(t).
\]
Now let us assume that the observational data are given through eclipse 
cycle number $l$ ($l=0,\ldots,N$), the date of the eclipse time-mark $t_l$, 
and its uncertainty $\sigma_l$.  Then the $l$-cycle eclipse ephemeris with 
respect to the reference epoch $t_0$ ($l=0$), at time $t\equiv t_l$ may be 
written as follows:
\[
T_{\idm{ep}}(l) = t_0 + 
l P_{\mbox{\scriptsize bin}} 
+ \tau(t_l,K_{1,2},P_{1,2},e_{1,2},\omega_{1,2},T_{1,2}) 
+ \mbox{``physics''} ,
\]
where $P_{\mbox{\scriptsize bin}}$ is the orbital period of the binary. It 
should not be assumed as known in advance, hence must be fitted, as well as 
the initial epoch $t_0$ corresponding to cycle number $l=0$, simultaneously
with other parameters of the model. The term coded as ``physics'' 
contains {\em non-Keplerian} effects, such as the period damping or other 
phenomena that may/should be included in the fit model.  Here, we introduce 
two instances of such a model. Following \cite{Hilditch2001}, the {\em 
linear ephemeris} model, as above,
\begin{equation}
\mbox{(O-C)}  = T_{\idm{ep}}(l) - t_0
- l P_{\mbox{\scriptsize bin}}
 = \tau(t_l,K_{1,2},P_{1,2},e_{1,2},\omega_{1,2},T_{1,2}), 
\label{eq:linear}
\end{equation}
and the {\em quadratic ephemeris} model, the {simplest}, yet non-trivial
generalization of the polynomial ephemeris \citep{Hilditch2001}
\begin{equation}
\mbox{(O-C)} 
= T_{\idm{ep}}(l) - t_0 - l P_{\mbox{\scriptsize bin}} -\beta l^2 =
\tau(t_l,K_{1,2},P_{1,2},e_{1,2},\omega_{1,2},T_{1,2}).
\label{eq:parabolic}
\end{equation}
The quantity $\beta$ {in Eq.~\ref{eq:parabolic}} is a factor that describes 
the binary period damping (change) due to the {mass-transfer}, magnetic 
braking, gravitational radiation, and/or influence of a very distant 
companion
\[
\beta =\frac{1}{2}{P_{\mbox{\scriptsize bin}}}{\dot{P}_{\mbox{\scriptsize bin}}}.
\]
Let us note that also $\beta$ should be fitted simultaneously
with other free parameters of the model.
In the rest of this paper, we use a common notation in the extrasolar 
planets literature, that enumerates the planets by subsequent letters, 
i.e., ``b'' $\equiv$ ``1'', ``c'' $\equiv$ ``2'', etc., to avoid any 
confusion.
%
%
\subsection{Newtonian model of the LTT effect}
%
%
A derivation of the $N$--body model of the LTT is basically very simple. It
requires \corr{the computation of} the planetary contribution $\tau$ to the
(O-C) signal through the numerical integration of the equations of motion, a
computation of the star barycentric vector $\v{R}_*$ and its $Z_*$
-component, in accord with Eq.~ \ref {eq:zstar}. This formulation accounts
for the mutual interactions between all bodies in the system. A serious
computational drawback of this model is a significant CPU overhead,
nevertheless, as we will show in the Appendix, its application for systems
with massive companions presumably involved in low order mean motion
resonances can be indispensable, To solve the equations of motion
efficiently, we used the ODEX2 integrator \citep {Hairer2009} designed for
conservative, second order \corr{ordinary differential equations (ODEs)}.
The imposed variable time step accuracy preserved the total energy and the
angular momentum better than $10^{-11}$. In terms of the Newtonian model,
the planetary bodies are parametrised through the mass $m_i$, semi-major
axis $a_i$, eccentricity $e_i$ and three Keplerian angles \corr{describing
the orientation} of the orbit, for each companion in the system. We also
assume that the binary is \corr{a point mass with the} prescribed total mass
of the binary. Assuming a coplanar configuration (\corr{in the $N$--body
model the same inclinations are ``absorbed'' in the planetary masses}), we
have 5 free orbital parameters for each planet, \corr{similar} to the
kinematic model. Here, they are then represented as ``usual'' osculating,
astrocentric Keplerian elements at a given initial epoch, but other types of
the osculating elements may be  used as well.
%
%
\subsection{The optimization method and numerical setup}
%
Having the egress times measured with a great precision (at the 1~second 
level, or even better), the next step is to determine the set of primary 
parameters of the kinematic model, usually with the least squares 
{approach}, by constructing the \corr{reduced $\Chi$--squared} function
\[
 \Chi \equiv 
 \Chi(K_{1,2},P_{1,2},e_{1,2},\omega_{1,2},T_{1,2},t_0,
 P_{\mbox{\scriptsize bin}},\beta),
\]
and searching for its minimum in the space of the model parameters.  It is 
well known, however, that the $\Chi$ function may possess many local 
minima, {particularly if the model is not well constrained, as it might be 
in {our} case. To seek {a} global solution, we apply a hybrid algorithm 
that consists of two steps: a quasi-global method, the Genetic Algorithm 
[GA, \cite{Charbonneau1995}] that is relatively slow and inaccurate, but 
makes it possible to find good approximations to the second step, a fast 
local method.} Here, we use the Levenberg-Marquardt (L-M) algorithm with 
analytically computed derivatives. The idea of the hybrid code comes from 
our earlier works on modelling radial velocity observations \citep 
[e.g.,][]{Gozdziewski2004}.  We used freely available Fortran codes of the 
Genetic Algorithm (PIKAIA\footnote 
{http://www.hao.ucar.edu/modeling/pikaia/pikaia.php}, by Phil Charbonneau \&
Barry Knapp) and of the L-M method from the well known MINPACK\footnote 
{http://www.netlib.org/minpack/} package.

\begin{figure*}
\centerline{
   \vbox{
     \hbox{
         \hbox{\includegraphics[width=3.5in]{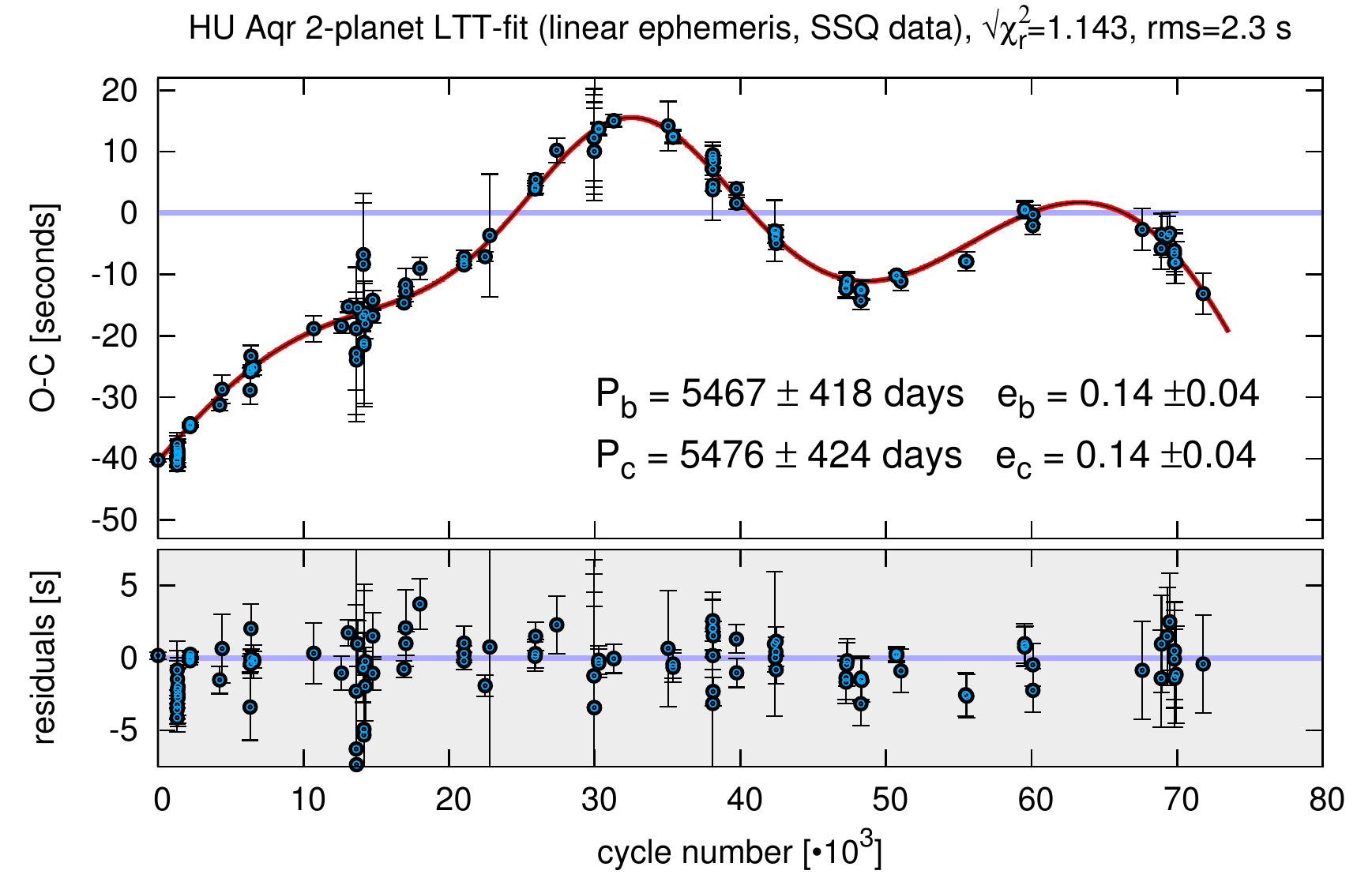}}
         \vspace*{0.2cm}
         \hbox{\includegraphics[width=3.5in]{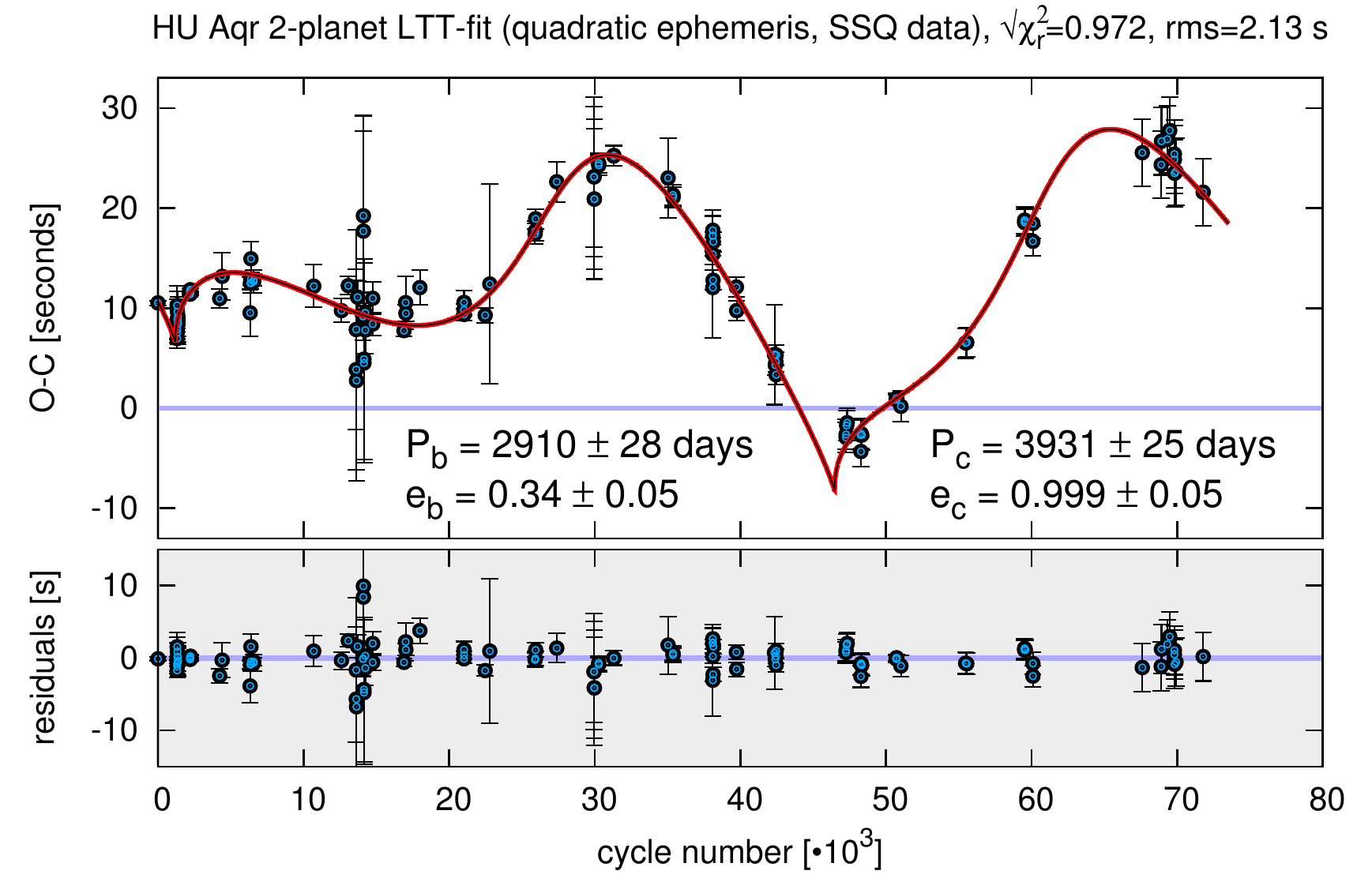}}
     }
   }
}
\caption{
Synthetic curves of the best-fit, 2-planet solutions to the mid--egress BJD
times of the SSQ data set.  The left panel corresponds to a linear ephemeris
model (Eq.~\ref{eq:linear}) and the right panel corresponds to a quadratic
(parabolic) ephemeris model (Eq.~\ref{eq:parabolic}, the parabolic trend has
been removed).  Panels are labeled with the orbital periods and
eccentricities of the putative companions.  Bottom plots with shaded
background show the residuals after subtracting planetary and astrophysical
contributions from the LTT signal. Discontinuous-like features of the
parabolic ephemeris model around cycles $0$ and $46,000$ appear due to the
extreme eccentricity of the outer body. See Table~\ref {tab:tab1} for the
orbital and inferred elements (Fit A and B, respectively).
}
\label{fig:fig2}
\end{figure*}

\begin{figure*}
\centerline{
   \vbox{
     \hbox{
         \hbox{\includegraphics[width=3.3in]{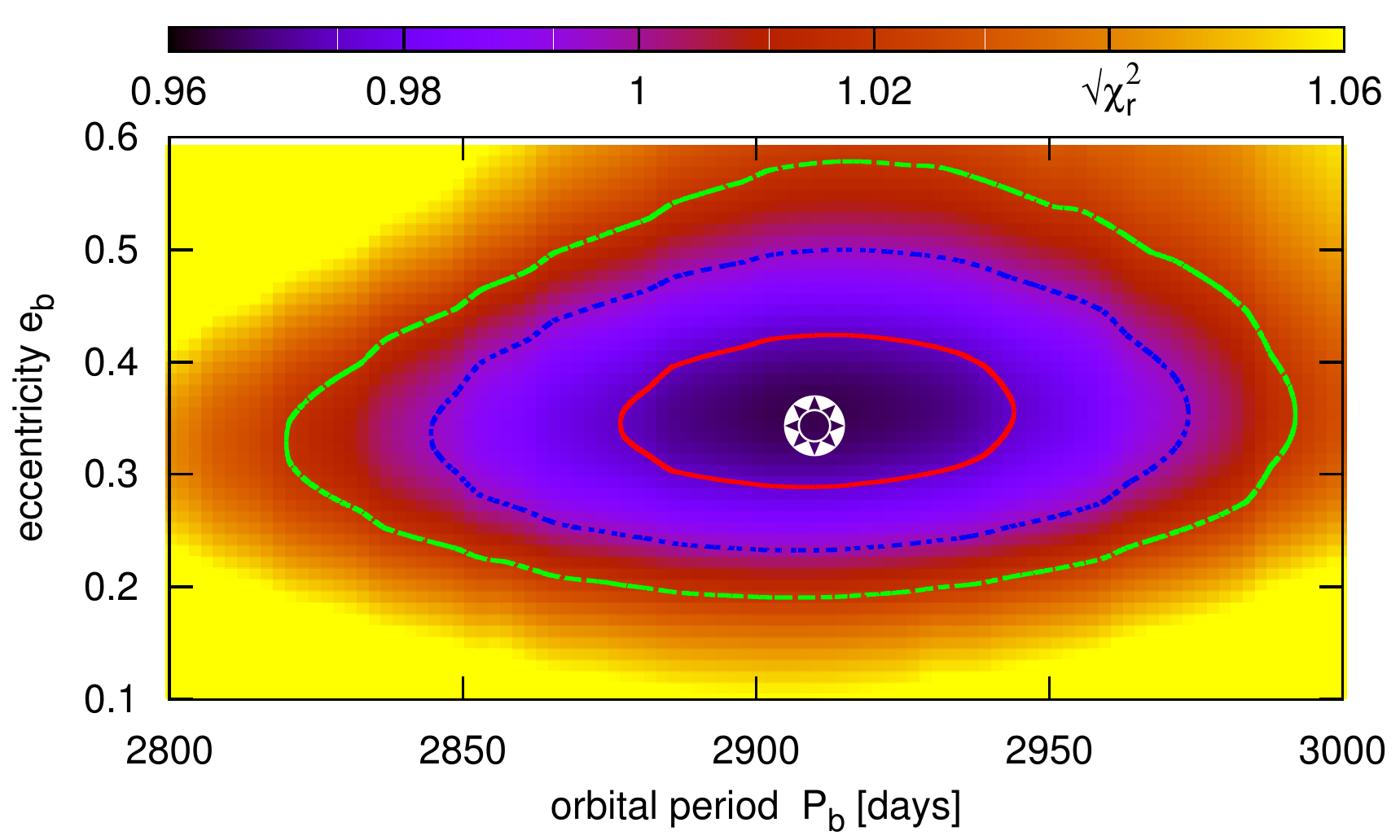}}
         \vspace*{0.2cm}
         \hbox{\includegraphics[width=3.3in]{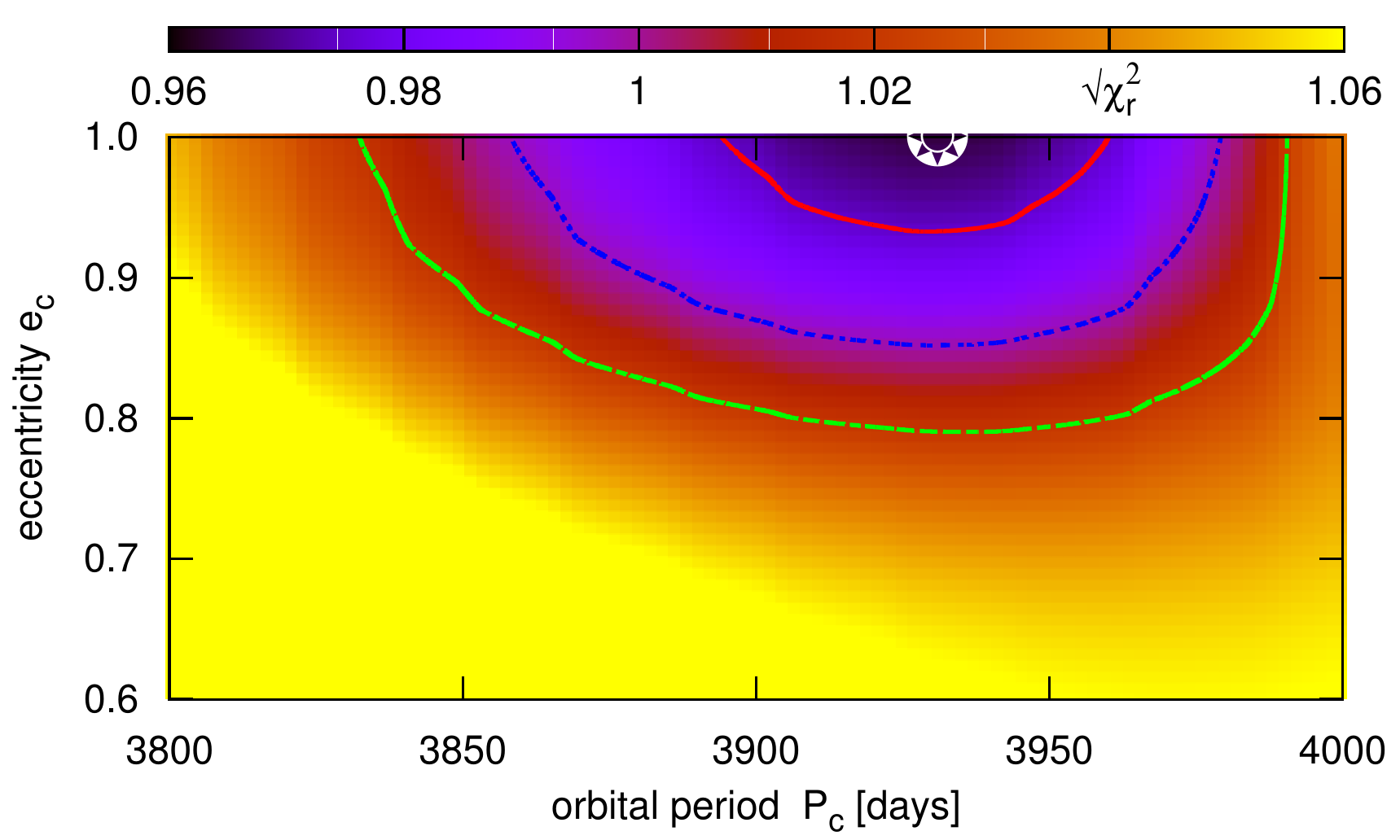}}
     }
   }
}
\caption{
Parameter scans of the $\Chi$ function computed around the best fit solution 
to the SSQ data for the quadratic ephemeris model {(}see the right-hand 
panel in Fig.~\ref{fig:fig2} and Fit~B in Table~\ref{tab:tab1}{)}.  Colour 
curves are for the formal 1,2,3$\sigma$--levels of the best-fit solution 
{whose} parameters are marked with an asterisk. 
}
\label{fig:fig3}
\end{figure*}

Once the primary set of the orbital model parameters are determined in the 
form of two five-tuples ($K_i,P_i,e_i,\omega_i,T_i$), $i=1,2$, we may also 
derive inferred Keplerian elements, {such as} {\em minimal} planetary 
masses and semi-major axes, by solving nonlinear equations expressing $K_i$ 
(Eqs.~\ref{eq:k},~\ref{eq:k1}) and the \corr{3rd} Kepler law in terms of the 
primary model parameters. The inclination has to be held fixed. Hence 
usually one assumes {$i_i=90^{\circ}$}. {Let us underline that while the 
LTT-model (Eq.~\ref{eq:ltt}) formulated in the barycenter frame has the 
same mathematical form as in the Jacobi frame, the orbital, geometrical 
(Keplerian) elements in multiple systems should be {related to Jacobian, 
canonical coordinates. {If} in the $N$-body numerical integrations and 
stability studies, {initial conditions have to be in \corr{the} form of osculating 
elements,} {one} should transform {these Jacobian elements into} the 
Cartesian coordinates w.r.t. the Jacobi frame \citep 
[e.g.,][]{Morbidelli2002}, and then, if necessary, to the astrocentric or 
barycentric coordinates.} In {this} sense, ``barycentric'' and ``Jacobian'' 
two-body elements may closely coincide for small, Jovian-mass planets. 
{But} for {more} massive companions when the LTT signal is easier to 
detect, or for very compact (resonant) systems, the semi-major axes, 
masses, Keplerian angles and inferred $N$-body initial conditions may be 
significantly different in both frames. We {will} discuss {this issue} in 
{more} detail in {a} forthcoming {article} (Gozdziewski et al., in 
preparation).

Each run of the hybrid code has been initiated by random selection of the GA 
population {(between 512 and 4096 individuals), considering possibly wide 
parameter ranges}.  For instance, {the range of} orbital periods was set 
blindly to [800, 63600]~days, {and} angles and eccentricities were set to 
their whole possible ranges. The original ``population'' was then 
transformed by GA operators over 512--1024 generations.  Each member of the 
final set {was then} used as an initial condition {for} the L-M algorithm, 
and the resulting solutions {were sorted and} stored.  The hybrid procedure 
was repeated hundreds of times for each combination of model--data set. We 
examined whether the obtained solutions converged to the same minima. Due 
to the semi-deterministic nature of the GAs, one should interpret the 
results in a statistical sense.

The same procedure may be applied to the Newtonian model, as the
planetary contribution $\tau$ can be computed independently of the 
optimisation method. (In this case the derivatives to the LM algorithm were
approximated numerically).

Finally, uncertainties of the best-fit parameters {were} determined {using} 
the bootstrap algorithm \citep{Press2002}, as the variances of parameters 
in a tested solution that has been re-fitted to 4096 synthetic data sets 
drawn randomly with replacement from the original sample. We found that
\corr{due to the particular distribution} of OPTIMA observations that
are grouped in small ``clumps'' of a few data points, the bootstrap
algorithm tends to underestimate the uncertainties when compared to
the formal error determination through the diagonal elements of
$\Chi$ curvature (covariance) matrix.
%
%
\section{Kinematic modeling the literature data}
%
To verify the literature models of the \huaqr{} system, we gathered 
Barycentric Julian Dated (BJD) egress times published by \cite 
{Schwope2001}, \cite{Schwarz2009} and \cite{Qian2011}.  That data set 
consists of 113 points, and will be called the SSQ set hereafter.  Our 
first attempt was to reproduce the results of \cite {Qian2011} with our 
formulation of the LTT model. We did not expect this to be straightforward, 
since their model assumes the inner planet to be on a circular orbit.
We conducted calculations for two ephemeris models, 
linear and quadratic (Eqs.~\ref{eq:linear} and~\ref{eq:parabolic}), 
respectively.
%
\subsection{The linear kinematic ephemeris 2-planet model}
%
In the linear ephemeris case, we found many, almost equally good 2-planet
solutions {with} $\Chi \sim 1.15$ and an rms $\sim 2.3$~sec. {In} these best
fits the inner planet {has a} period of $\sim 5500-6000$~days. However, the
period of the outer planet varies between 7000--20000~days. The resulting
systems imply (O-C) residuals \corr{caused individually by the planets} in
wide ranges, up to $\sim 6000$~seconds, and companions in basically any
mass, eccentricity and period range {while} still preserving excellent rms
$\sim 2.3$~sec and similar ``flat'' behaviour of the residuals. The left
panel of Fig.~\ref {fig:fig2} shows the most exotic and actually the
best--fit solution found in {our} experiment. {The} Keplerian fit parameters
{of this solution}, as well as {its} inferred elements are given in Table~
\ref{tab:tab1} (Fit A). This fit is very different from those found by \cite
{Qian2011}, \cite {Wittenmyer2011}, \corr{and even in} the last paper by
\cite{Hinse2011}. This configuration has $\Chi \sim 1.143$ and an rms $\sim
2.3$~sec, and is characterised by {\em almost equal} orbital periods {of}
$\sim 5470$~days. {The} pericenter arguments of the planets {in this fit}
{differ} by nominal value of $180.2^{\circ}$ {and as a result,} the
Keplerian barycentric orbits are almost exactly anti-aligned, with planets
placed close to their periastrons at the initial epoch. This configuration
could be understood as a pair of {\em Trojan--planets} in 1c:1b mean motion
resonance (MMR). Although the resulting LTT signal has apparently small
amplitude $\sim 60$ ~seconds as shown in the (O-C) diagram (see the
left-hand panel in Fig.~\ref {fig:fig2}), the LTT semi-amplitudes {
$K_{\idm{b,{c}}}$} are excessively large (up to $\sim 6000$~seconds),
implying just absurdly massive companions of $\sim 10,000$~Jupiter mass each
($\sim$10~M$_{\odot}$ !). \corr{This solution reveals that an inherent
degeneracy of the LTT signal (and its model) may appear because the signal
is the result of the {\em differential} gravitational tugs of the companions
on the binary.  Indeed}, in this particular Trojan configuration, even small
deviations from the anti-alignment of orbits leads to large changes in the
planetary masses (over 3 orders of magnitude) and semi-major axes (within a
range of a few AU), indicating that as they are not supported by the
currently available observations, these dynamical parameters are poorly
constrained. The mathematical fit permits putative companions as massive as
stars but in reality, such objects should influence dynamical and spectral
properties of the binary system. Such solutions are therefore excluded.

The 1c:1b~MMR solution is a vivid example demonstrating that due to the 
possibility of configurations involved in extremely strong mutual 
interactions, modeling the LTT signal globally (without any \corr{{\em a priori}} 
assumptions on the system configuration) cannot be studied in terms of 
the kinematic model.  In general, {an} exact, self-consistent $N$-body 
model should be used to determine the initial conditions. 
%
\subsection{The quadratic kinematic ephemeris 2-planet model}
%
In the case of {a} quadratic ephemeris, we found {a} well defined 
minimum of {$\Chi \sim 0.972$}, which is {an} apparently statistically 
perfect solution. Its synthetic curve with measurements over-plotted is 
}{shown} in the right--hand panel of Fig.~\ref{fig:fig2}, and orbital 
parameters are given in  Table~\ref{tab:tab1} as Fit~B. That solution has 
been frequently found in different runs of the hybrid code, which 
reinforces its global character. To show the latter, we computed parameter 
scans of $\Chi$ in the ($P_{\idm{b,c}}, e_{\idm{b,c}}$)--planes (Fig.~\ref 
{fig:fig3}), by fixing points of a grid in a given plane and minimizing 
$\Chi$ over all remaining free parameters of the model. This {made} it 
possible to obtain standard confidence levels as marked with coloured 
curves. The best-fit solution is again very different from solutions found 
in the literature. While the elements of the inner planet are well 
constrained, the orbit of the outermost companion reveals extremely large 
eccentricity ($\sim 1$). That points again to highly degenerate 
(unrealistic) best-fit {solutions}, with  near-parabolic or even 
hyperbolic, open orbit of one ``planet'' --- as the fit implies --- being a 
\corr{low-mass stellar object of $\sim 160$ Jupiter-masses}. 
Other solutions with slightly worse $\Chi \sim 1.1$ and still very 
similar rms $\sim 2.2$~sec may be found too, which means that the quadratic 
ephemeris model is unconstrained by the SSQ data. 

In the quadratic ephemeris model, the orbital periods are close to the \corr
{4:3~ratio, which is equivalent to the low-order 4c:3b mean motion
resonance. In addition, the eccentricity of the outer planet is extreme,
close to~1. Hence again}, the kinematic formulation seems inadequate to
derive the proper initial condition of the multiple-planet configuration.
We  conclude here that when we only have the SSQ data at our disposal, there
seems to be no \corr{unique and physically meaningful solution} explaining
the LTT variability. Or, the planetary fit model and its assumptions are
incorrect.

\begin{table}
\caption{
Jacobian geometric parameters {with the} inferred masses and semi-major axes
of 2-planet LTT fit models {\em in the barycenter frame} with the linear and
parabolic ephemeris to the SSQ data set. Synthetic curves with data sets are
{shown} in two panels of Fig.~\ref{fig:fig2} and $\Chi$ scans in Fig.~\ref
{fig:fig3}. Numbers in parentheses {represent} the {uncertainties} at the
last significant digit. Total mass of the binary is 1.08~$M_{\sun}$ \citep
{Schwarz2009}. Indices ``b'' and ``c'' refer here to planets ``1'' and ``2''
in the mathematical model Eqs.~10 and~11. See the text for more details.
}
\centering
\begin{tabular}{|c|c|c|}
\hline
Model &  Fit~A & Fit~B  \\
parameter   &  linear ephemeris  &  parabolic ephemeris  \\
\hline
$K_b$~[seconds]       &  5928 $\pm$ 15  &    10.2 $\pm$ 0.5      \\
$P_b$~[days]       &  5467 $\pm$ 418    &    2910 $\pm$ 28      \\
$e_b$             &  0.138 $\pm$ 0.034  &    0.34 $\pm$ 0.07         \\
$\omega_b$~[degrees]  &  207 $\pm$  21 &         22  $\pm$ 8    \\
$T_b$~[BJD 2,440,000+]   
                & 11694 $\pm$ 175   &         5652 $\pm$ 97 \\
\hline
$K_c$~[seconds]       &  5942 $\pm$ 17  &     322 $\pm$ 30     \\
$P_c$~[days]       &  5476 $\pm$ 424 &  3931 $\pm$ 50     \\
$e_c$             &  0.141 $\pm$ 0.035 &  0.99 $\pm$ 0.05      \\
$\omega_c$~[degrees]  &  27 $\pm$  20 &       358.3 $\pm$ 0.2 \\
$T_c$~[BJD 2,440,000+]   
                &  6214  $\pm$ 451  &    9207 $\pm$ 42   \\
\hline
$\Pb$~[days]    &  0.086820400(4)   &    0.0868204250(8)    \\
$T_0$~[BJD 2,440,000+]    
                &  9102.92004(2)    &   9102.91988(2)     \\ 
$\beta$ [$\times 10^{-13}$~\dcl] 
                &    ---          &    -3.06(6)     \\
\hline
$a_b$~[au]        &   1.375                 &   4.08    \\
$m_b\sin i$~[M$_{\idm{Jup}}$]        
                &    9780                &    5.69    \\
$a_c$~[au]        &   1.374                 &  4.58   \\
$m_c\sin i$~[M$_{\idm{Jup}}$]        
                &    9811                &    159    \\
\hline
$N$             &     113            &     113   \\
$\Chi$          &     1.143          &     0.972   \\ 
rms [seconds]       &    2.31            &     2.13   \\
\hline
\end{tabular}
\label{tab:tab1}
\end{table}
%
%
\section{New observations and data reduction}
%
\subsection{Observations with OPTIMA and other instruments}
%
\begin{figure}
\centerline{
   \vbox{
     \hbox{
         \hbox{\includegraphics[width=3.0in]{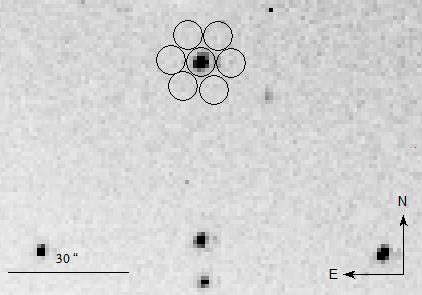}}
     }
   }
}
\caption{
OPTIMA hexagonal fibre bundle centred on \huaqr{}. The ring fibres (1--6) 
are used to monitor the background sky simultaneously.
}
\label{fig:fig4}
\end{figure}
To resolve the model degeneracies as described above, we gathered new, yet 
unpublished observations of the \huaqr{} binary. The new collected 
mid--egress BJD times are given in Table~\ref{tab:tab2}. These data extend 
the work of \cite{Schwope2001}, \cite {Schwarz2009} and \cite{Qian2011}. 
The \corr{currently} available data set of \huaqr{} egress times consists of 171 
measurements in total, including 10~points presented in \cite{Qian2011}. 
Among {these} measurements, {68} were obtained with the Optical Timing 
Analyzer (OPTIMA) instrument that operates mostly at the 1.3~m telescope at 
Skinakas Observatory, Crete, Greece.

\begin{figure*}
\centerline{
\hbox{
   \vbox{
      \hbox{\includegraphics[width=0.48\textwidth]{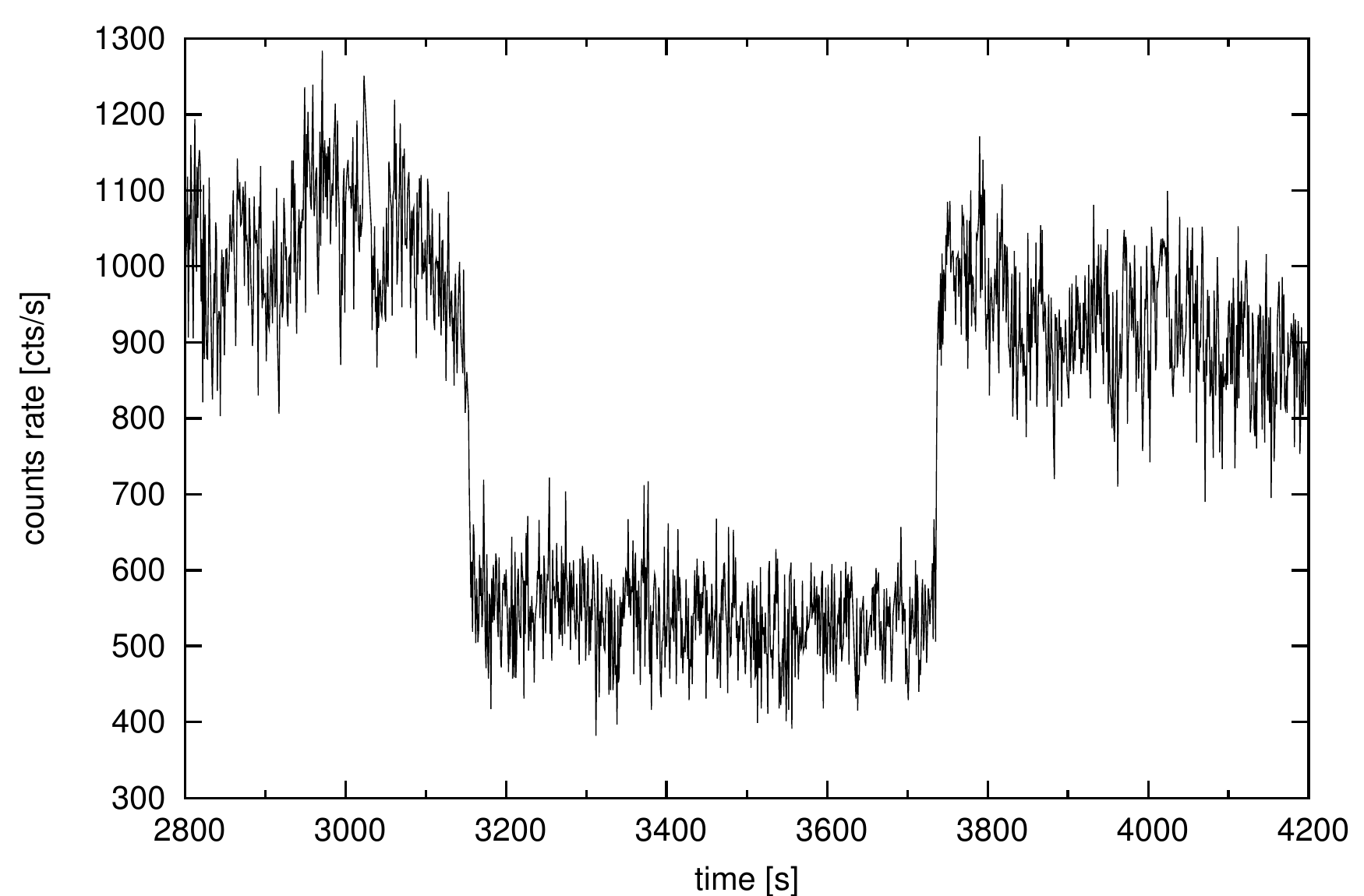}}   
      \hbox{\includegraphics[width=0.48\textwidth]{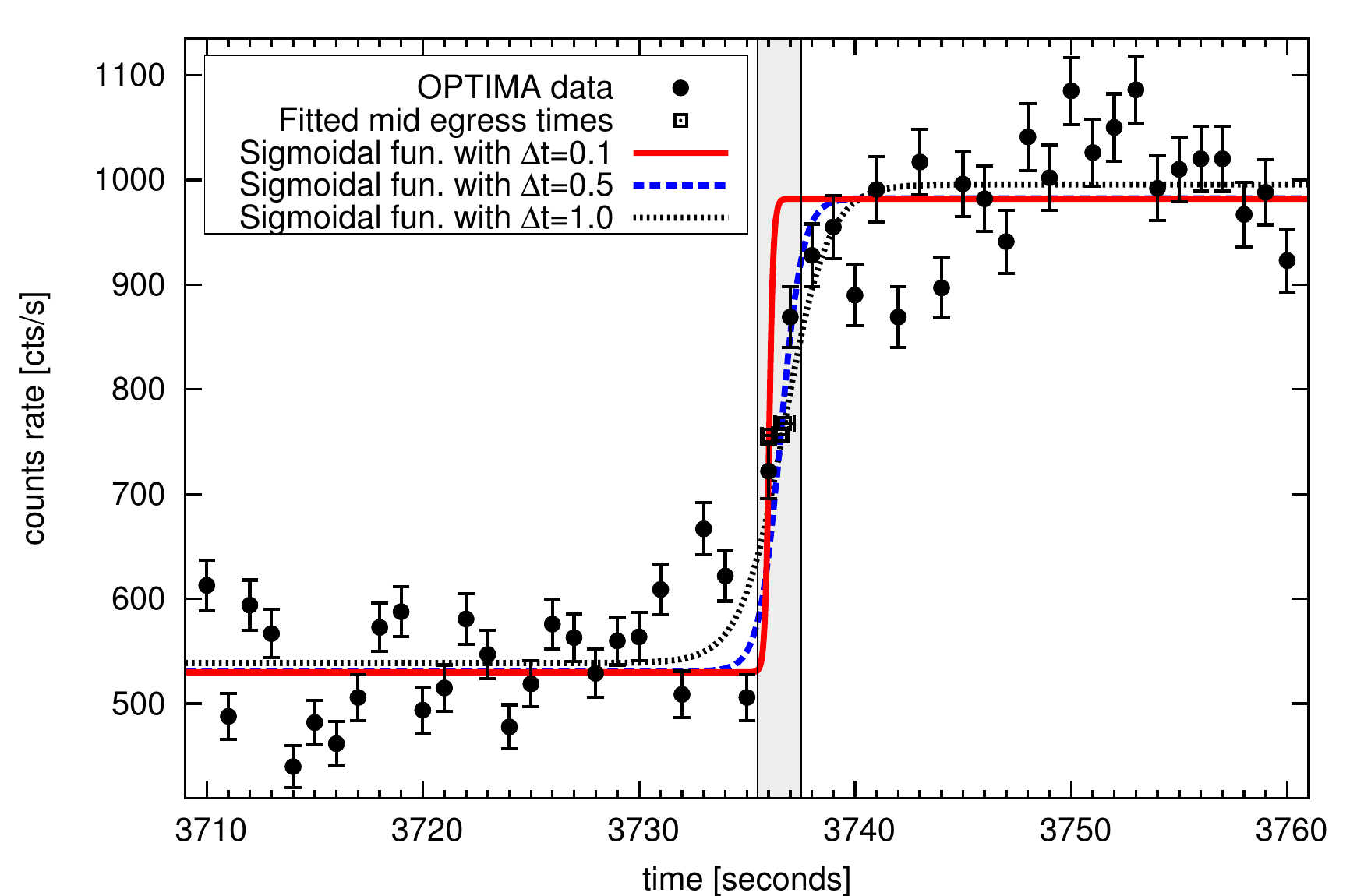}}   
   }  
   \vbox{
      \hbox{\includegraphics[width=0.48\textwidth]{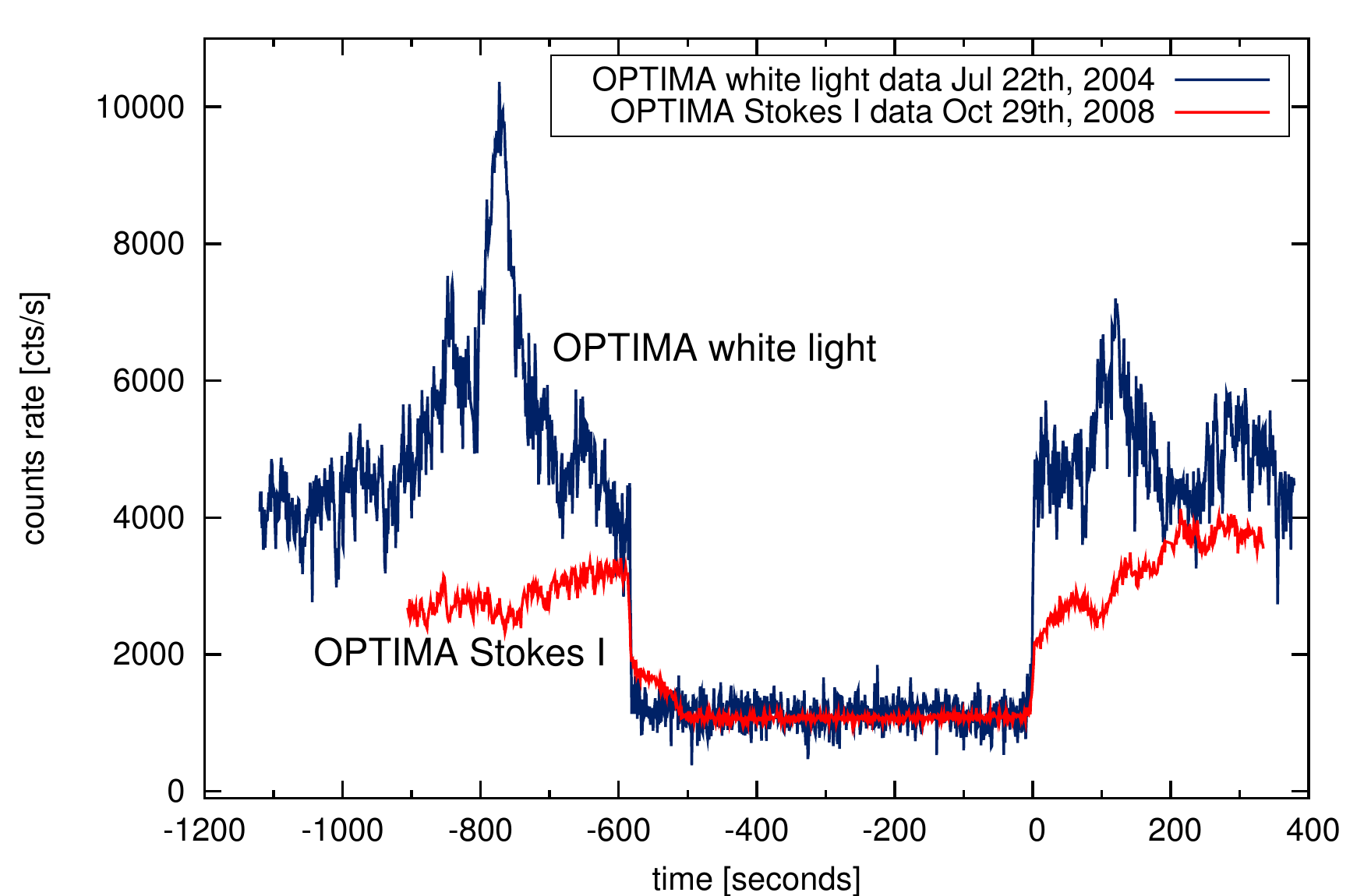}}
      \hbox{\includegraphics[width=0.48\textwidth]{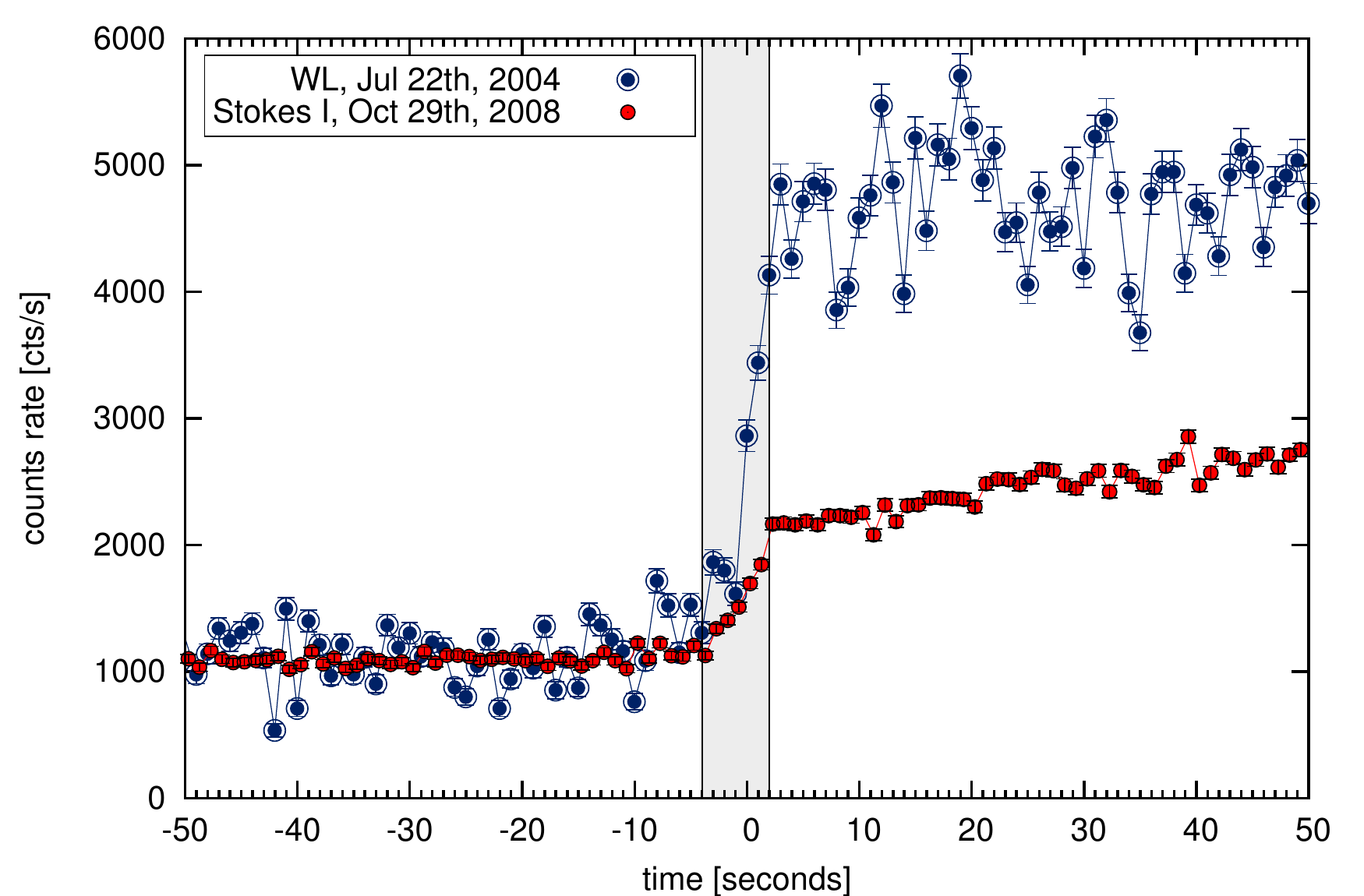}}  
   }
 }
}
\caption{
High time resolved OPTIMA light curves of \huaqr{}. \emph{The upper left 
panel:} photometric eclipse in the white light. \emph{The bottom left 
panel:} a close-up around the mid--egress overplotted with three fitted 
sigmoid functions with different values set {for the} $\Delta t$ parameter 
(0.1, 0.5, 1.0, respectively). The fitted mid--egress times are {denoted 
by} open squares. The shaded region corresponds to {a} time span of 
2~seconds. \emph{The upper right panel:} a comparison of photometric and 
polarimetric OPTIMA light curves of \huaqr{} obtained in 2004 and 2008, 
respectively. \emph{The bottom right panel:} a close-up of photometric and 
polarimetric \huaqr{} light curves around the mid--egress time showing the 
difference between the egress shapes. The shaded region covers 6~seconds. 
}
\label{fig:fig5}
\end{figure*}

The high--speed photometer OPTIMA is a sensitive, portable detector to 
observe extremely faint optical pulsars and other highly variable 
astrophysical sources.  The detector contains eight fibre--fed single 
photon counters --- avalanche photodiodes (APDs), and a GPS for the time 
control. There are seven fibres in the bundle (Fig.~\ref {fig:fig4}) and 
one separate fibre located at a distance of $\sim 1 \arcmin$. Single 
photons are recorded simultaneously and separately in all channels with 
absolute UTC time--scale tagging accuracy of $\sim 4~\rm{\mu s}$.  The 
quantum efficiency of the APDs reaches a maximum of 60$\%$ at 750 nm and 
lies above 20$\%$ in the range 450--950~nm \citep{Kanbach2003,Kanbach2008}. 
During the \huaqr{} observations, OPTIMA was pointed at RA(J2000) = 
$21^{\rm h} 07^{\rm m} 58 \fs 19$, Dec(J2000) = $-05\degr 17\arcmin 40 
\farcs 5$, corresponding to the central aperture of the fibres bundle (Fig.~
\ref {fig:fig4}). For sky background monitoring, we usually choose one out 
of the six hexagonally located fibres. We look for the fibre that is not by 
chance pointed to any source, therefore records sky background, and its 
response is the most similar to the central fibre response when the 
instrument is targeted at the dark sky. An example of \corr{a} sky background 
subtracted light curve is shown at the top left-hand panel in 
Fig.~\ref{fig:fig5}. 

We derived new fits to the \huaqr ~eclipse egress times, as well as 
reanalysed many of the already published OPTIMA data.  There are 26 
eclipses obtained by OPTIMA published by \cite{Schwarz2009}.  We were able 
to reanalyse only \corr{21 out of the 26 light curves}, because only those were 
available in the OPTIMA archive.  Our completely new data set includes 42 
precision photometric observations, starting from cycle $l \sim 29,900$, 
overlapping in time window with the literature data.  We derived 23 new 
eclipse profiles from the OPTIMA data archive spanning 1999--2007 and 
obtained 19 new OPTIMA optical \huaqr{} light curves in 2008--2010. Note 
that some of the OPTIMA observations have been already published in the 
very recent literature \citep{Nasiroglu2010}.

We also gathered and reduced 11 observations performed at the MONET 
(MOnitoring NEtwork of Telescopes) project which is \corr{a} network of two 1.2 m 
telescopes operated by the Georg-August-Universit\"at G\"ottingen, the 
McDonald Observatory, and the South African Astronomical Observatory. These 
precision data in white light (500--800~nm) were binned in 5~s intervals, 
with $10^{-6}$~days (0.1~s) accuracy, separated by 3~s readout. 
%
%
The most recent observations were performed at November 18th, 2011.

An additional three egress times were obtained from the eclipse observations 
carried out with the PIRATE telescope equipped with the SBIG STL1001E CCD 
camera \citep{Holmes2011}. PIRATE, funded by the Open University Department 
of Physics and Astronomy, is a remote-controlled telescope located at the 
Astronomical Observatory of Mallorca (OAM), Spain. 

\begin{figure*}
\centerline{
   \vbox{
     \hbox{
         \hbox{\includegraphics[width=1\textwidth]{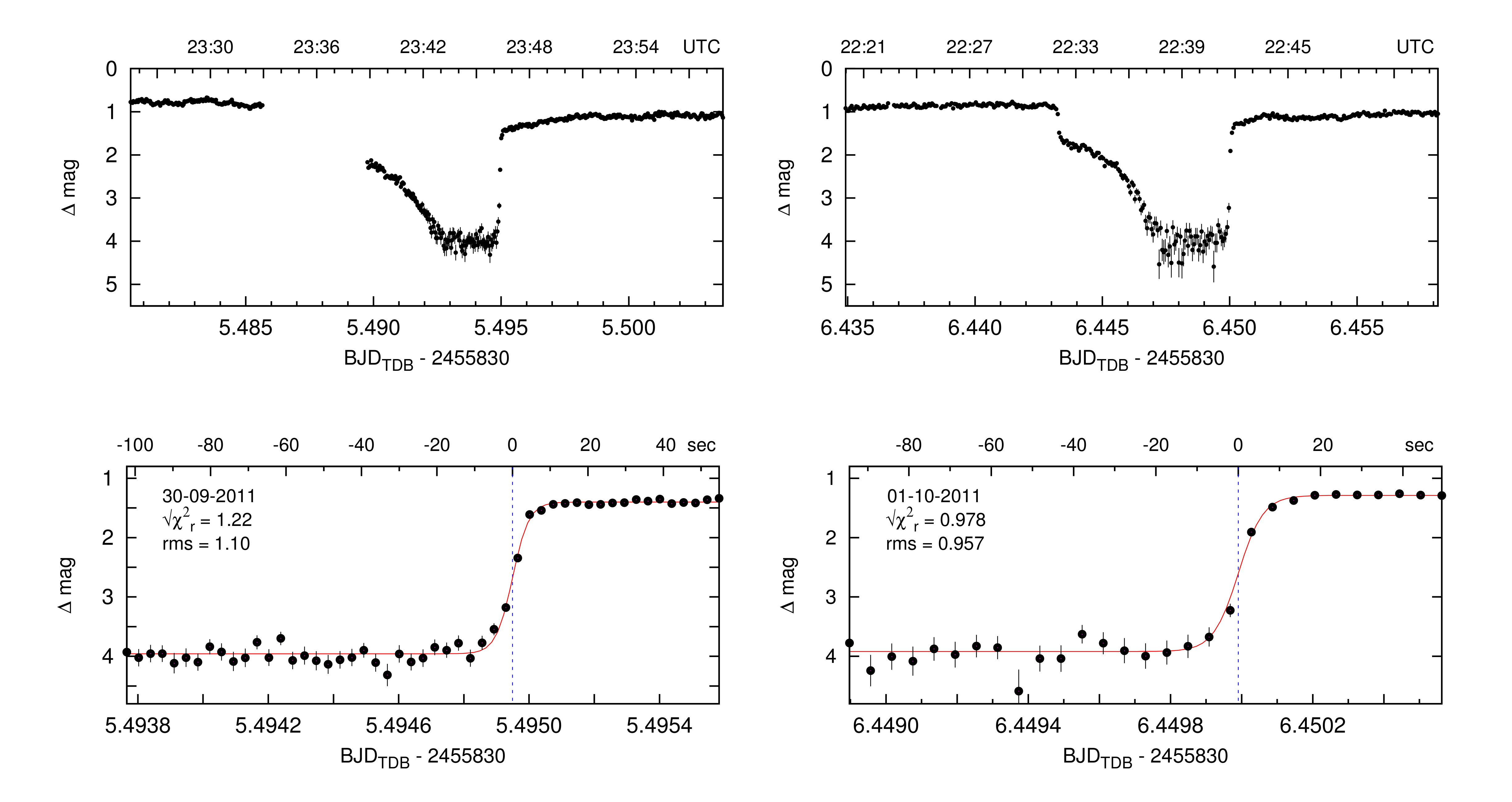}}         
     }
   }
}
\caption{
\huaqr{} eclipse light curves obtained with the WFC mounted on the TCS. 
Observations were performed on 30/09/2011 and 01/10/2011 with integration 
times of 3 and 5 seconds, respectively.  Bottom panels are for the close-up 
of eclipse egress, with fitted sigmoid function (solid line, see
Eq.~\ref{eq:sigmoid}). Blue vertical lines mark the determined mid--egress
times.
}
\label{fig:fig6}
\end{figure*}

We also performed optical observations of \huaqr{} in white light with the 
1.5-m Carlos S\'anchez Telescope (TCS) equipped with Wide FastCam (WFC, Fig.
\ref{fig:fig6}). The WFC is a 1k $\times$ 1k -- pixel camera with optics 
offering a FOV of 12 arcmin with a scale of 0.6 arcsec/pix. \huaqr{} 
eclipses were observed on September 30,  and October 01, 2011 with 
integration times of 3 and 5 seconds, respectively.  WFC works in frame 
transfer mode, therefore the readout time is effectively null or in other 
words is equivalent to the exposure time. UTC mid-exposure times of the 
photometric measurements were converted to the Barycentric Julian Dates in 
Barycentric Dynamical Time using the procedure developed by \cite
{Eastman2010}.

Some technical details of the observations performed with the MONET/N,
TCS, and PIRATE telescopes are given in Table~\ref{tab:tab3}.

\begin{table}
\caption{
New HU~Aqr BJD mid--egress times on the basis of light curves collected 
with: OPT-ESO22 --- OPTIMA photometer installed at ESO (Chile), OPT-SKO --- 
OPTIMA operated at the Skinakas Observatory (Crete), PIRATE --- a telescope 
at the Astronomical Obs. of Mallorca, MONET/N --- the network of telescopes 
at the McDonald Obs. and the SAO (South Africa), and WFC --- the 1.5-m  TCS 
(Canary Islands).
}
\centering
\begin{tabular}{|c c c c|}
\hline
Cycle & BJD & Error [days] & Instrument  \\
\hline
29946 & 2451702.8443352 & 0.0000037 & OPT-ESO22\\
29957 & 2451703.7993545 & 0.0000038 & OPT-ESO22\\
29958 & 2451703.8861705 & 0.0000034 & OPT-ESO22\\
30265 & 2451730.5400324 & 0.0000041 & OPT-SKO\\
30287 & 2451732.4500902 & 0.0000023 & OPT-SKO\\
30299 & 2451733.4919357 & 0.0000033 & OPT-SKO\\
30300 & 2451733.5787554 & 0.0000054 & OPT-SKO\\
30310 & 2451734.4469740 & 0.0000031 & OPT-SKO\\
30311 & 2451734.5337856 & 0.0000018 & OPT-SKO\\
35469 & 2452182.3533852 & 0.0000030 & OPT-SKO\\
38098 & 2452410.6041626 & 0.0000084 & OPT-SKO\\
42486 & 2452791.5719484 & 0.0000015 & OPT-SAO \\
42487 & 2452791.6587715 & 0.0000024 & OPT-SAO \\
44534 & 2452969.3800760 & 0.0000033 & OPT-NOT\\
44557 & 2452971.3769377 & 0.0000085 & OPT-NOT\\
51020 & 2453532.4971595 & 0.0000100 & OPT-SKO\\
51066 & 2453536.4909030 & 0.0000064 & OPT-SKO\\
51067 & 2453536.5777278 & 0.0000033 & OPT-SKO\\
55535 & 2453924.4913426 & 0.0000102 & OPT-SKO\\
55627 & 2453932.4788164 & 0.0000061 & OPT-SKO\\
55661 & 2453935.4307071 & 0.0000064 & OPT-SKO\\
55719 & 2453940.4662754 & 0.0000162 & OPT-SKO\\
60085 & 2454319.5242409 & 0.0000074 & OPT-SKO\\
64657 & 2454716.4671496 & 0.0000053 & OPT-SKO\\
64885 & 2454736.2622085 & 0.0000038 & OPT-SKO\\
64886 & 2454736.3490181 & 0.0000016 & OPT-SKO\\
65265 & 2454769.2539926 & 0.0000023 & OPT-SKO\\
67791 & 2454988.5622710 & 0.0000029 & OPT-SKO\\
67917 & 2454999.5016391 & 0.0000017 & OPT-SKO\\
67918 & 2454999.5884526 & 0.0000054 & OPT-SKO\\
68009 & 2455007.4891162 & 0.0000018 & OPT-SKO\\
72099 & 2455362.5844371 & 0.0000032 & OPT-SKO\\
72110 & 2455363.5394546 & 0.0000019 & OPT-SKO\\
72121 & 2455364.4944885 & 0.0000029 & OPT-SKO\\
72133 & 2455365.5363444 & 0.0000015 & OPT-SKO\\
72225 & 2455373.5238044 & 0.0000048 & OPT-SKO\\
72237 & 2455374.5656456 & 0.0000040 & OPT-SKO\\
72248 & 2455375.5206715 & 0.0000040 & OPT-SKO\\
72305 & 2455380.4694292 & 0.0000030 & OPT-SKO\\
72351 & 2455384.4631748 & 0.0000024 & OPT-SKO\\
72352 & 2455384.5499944 & 0.0000022 & OPT-SKO\\
72421 & 2455390.5406108 & 0.0000013 & OPT-SKO\\
73409 & 2455476.3190971 & 0.0000578 & PIRATE\\
73559 & 2455489.3421698 & 0.0000578 & PIRATE\\
73560 & 2455489.4290151 & 0.0001156 & PIRATE\\
75467 & 2455654.9954277 & 0.0000040 & MONET/N\\
75812 & 2455684.9484608 & 0.0000023 & MONET/N\\
76721 & 2455763.8681410 & 0.0000035 & MONET/N\\
77031 & 2455790.7824571 & 0.0000039 & MONET/N\\
77066 & 2455793.8211556 & 0.0000077 & MONET/N\\
77067 & 2455793.9079841 & 0.0000055 & MONET/N\\
77078 & 2455794.8630179 & 0.0000065 & MONET/N\\
77546 & 2455835.4949490 & 0.0000179 & WFC\\
77557 & 2455836.4499905 & 0.0000295 & WFC\\
77789 & 2455856.5922852 & 0.0000038 & MONET/N\\
77802 & 2455857.7209399 & 0.0000090 & MONET/N\\
77823 & 2455859.5441786 & 0.0000066 & MONET/N\\
78100 & 2455883.5934038 & 0.0000022 & MONET/N\\
\hline
\end{tabular}
\label{tab:tab2}
\end{table}

\begin{table}
\centering
\caption{Technical data of the MONET/N, PIRATE and TCS observations of \huaqr{}.
{T$_{\idm{obs}}$ {represents} the observation time span, {and}
$\Delta$T is {the mean} exposure time of a single frame.}
}
\begin{tabular}{|l c c c c|}
\hline
Date & Instrument & Filter & T$_{\idm{obs}}$ [hours] & $\Delta$T [s] \\
\hline
2010 Oct 06 &  PIRATE   &  WL  & 1 & 10\\
2010 Oct 19 &  PIRATE   &  WL  & 3.5 & 10, 20\\
2011 Sep 30 & {TCS/WFC} &  WL  & 1 & 3 \\
2011 Oct 01 & {TCS/WFC} &  WL  & 1 & 5 \\
2011 Apr 03 & {MONET/N} &  WL  & 0.60 &  8 \\ 
2011 May 03 & {MONET/N} &  WL  & 0.50 &  8 \\
2011 Jul 21 & {MONET/N} &  WL  & 0.32 &  8 \\
2011 Aug 17 & {MONET/N} &  WL  & 0.60 &  8 \\
2011 Aug 20 & {MONET/N} &  WL  & 0.46 &  8  \\
2011 Aug 20 & {MONET/N} &  WL  & 0.10 &  8  \\
2011 Aug 21 & {MONET/N} &  WL  & 0.58 &  8  \\
2011 Oct 22 & {MONET/N} &  WL  & 0.25 &  8  \\
2011 Oct 23 & {MONET/N} &  WL  & 0.50 &  8  \\
2011 Oct 25 & {MONET/N} &  WL  & 0.16 &  8  \\
2011 Nov 18 & {MONET/N} &  WL  & 0.45 &  8  \\
\hline
\end{tabular}   
\label{tab:tab3}
\end{table}


%
%
\subsection{Determining time markers of the eclipses}
%
In Fig.~\ref{fig:fig5} we show an example of \huaqr{} high time resolution 
OPTIMA photometric (see the right top panel, and blue curve in the 
left-hand top panel) and polarimetric (Stokes I, red curve in the 
right-hand top panel) light curves. These {graphs} are to be compared with 
the light curves from TCS, obtained with 3 and 5 seconds exposures 
illustrated in Fig.~\ref{fig:fig6}. Obviously, the OPTIMA resolution makes 
it possible to track the egress phase closely, {which enabled us} to 
determine the mid--egress moments very precisely.

Measuring the time of mid--egress properly is critical to obtain the (O-C) 
diagrams, since {it is the time marker of the eclipse} \citep
{Schwope2001,Schwarz2009}. To derive the mid--egress moment $t_0$, the 
sigmoid function
\begin{equation}
I(t) = a_1 + \frac{(a_2-a_1)}{(1.0 + \exp([t_0-t]/\Delta t)},
\label{eq:sigmoid}
\end{equation}
parametrised by $a_1, a_2$ and $t_0$ was fitted to the light curve points in 
the egress phase of the eclipse, spanning preselected {exponential scaling 
parameters} $\Delta t$ . We found {that there is no strong dependence of 
{the} derived $t_0$ \corr{on the adopted} $\Delta t$}. {This 
can be seen} in the bottom left-hand panel of Fig.~\ref{fig:fig5} {where} 
three mid--egress times $t_0$ are marked with black open squares. These 
moments are derived for three different choices of $\Delta t${:} 0.1, 0.5 
and 1.0, respectively. While these times depend on $\Delta t$, they fall 
within {a} 2~second range, as marked by a shaded strip at the bottom 
left-hand panel of Fig.~\ref{fig:fig5}. A half of that range ($\sim 1$ 
~second) may be typically estimated as the maximum possible error of $t_0$ 
in the OPTIMA data set.  The formal $1\sigma$ uncertainty of the sigmoid 
fit {in this case is still \corr{smaller} and} at the level of $\sim 0.1$~second. 
Moreover, the shape of the eclipse may significantly depend on the spectral 
window. Panels in the right column of Fig.~\ref{fig:fig5} illustrate the 
light curves of \huaqr{} derived in the optical, white band domain (the 
blue curve) and in the polarimetric domain (Stokes~I, the red curve). In 
the latter case, the egress looks quite different and spans {over a} longer 
time. {Given that} these two data sets were taken four years apart, the 
observed difference might {have been} caused by different emission states 
of the source. To derive the mid-egress moments gathered in \corr{Table}~\ref
{tab:tab2}, the sigmoid function was fitted to all single light curves.
%
%
\subsection{On the light curves in different spectral windows}
%
\cite{Vogel2008a} and \cite{Vogel2008b} obtained high time resolved and 
accurate light curves of \huaqr{} during its low state {using the} ULTRACAM 
\citep {Beard2002,Dhillon2007} at the Very Large Telescope (VLT, May 13, 
2005). {These authors decomposed} the light curve {into} three emission 
components emerging from the accretion spot, the photosphere surrounding 
it, and from the white dwarf itself.  As {a} result, they {were able to 
derive} the temperature of the WD $\sim13500(200)$~K and the temperature of 
the accretion spot $\sim25500(1500)$~K. They also estimated the ratio of 
the spot area to the WD surface to be on the level of $5\%$. The black body 
{spectra} of the WD and of the spot {have their maxima} at 215~nm and 
113~nm, respectively. The accretion spot and the accretion stream are time 
variable in brightness, as well as in the geometric position in the 
system.  Therefore, the orbital phase at which they occur is not constant. 
The ULTRACAM delivers \corr{simultaneous} light curves in three colours: $u$, $g$
and $r$. An example of {such a} three colour \huaqr{} light curve can be 
{found} in Fig.~6 of \cite{Schwarz2009}, where the shape---energy 
dependence can be easily seen. Thus, a comparison of egress times in 
different wavelength domains was possible. In {the two} cases {with} 
filters $u$ and $r$, the WD constitutes the main contribution to the egress 
intensity, because it can be seen unperturbed when it comes out of 
eclipse.  However, during high and intermediate accretion states, the WD 
might be out-shined by the accretion stream. In the $u$--band, the spot 
contributes 25\% of the emission, while in the $r$--band {its contribution 
is} only 12\% \citep{Vogel2008a,Vogel2008b}. This suggests that as the time 
marker of the eclipse, it is better to use more ``reddish'' than 
``blueish'' data, particularly in our case as we have broadband 
observations gathered in X-rays, UV and optical domains at our disposal. 

There exists evidence that the EUVE light curves differ from 
quasi-simultaneous ROSAT/HRI light curves, as it can be seen, for example, 
for eclipses recorded on October 1996 and May 1997, as shown in Fig.~2 of 
\cite{Schwope2001}.  The eclipse ingress is often not measured because of 
strong suppression of soft X-rays by absorbing matter along the 
accretion stream.  When the eclipse duration can be determined, the eclipse 
duration seems shorter in the case of EUVE data.
Thus the derived mid--egress moments can be 
shifted by a~few seconds.  According to \cite{Schwope2001}, an expected 
variation of the eclipse span should be not more than 0.001 of the orbital 
phase which corresponds to not more than $\sim 8$ seconds. \corr{Also \cite 
{Schwope2004} show in their Fig.~3 evidence of a different eclipse 
length as well as phase folded egress shapes at soft X-rays, HST UV, and} 
\corr{high-speed optical photometry with \corr{a} multichannel multicolor
photometer (the MCCP 2.2~m telescope at Calar Alto)}
during the 1993 high state and the 1996 low accretion state. 
The scatter of the egress times resulting from changes of the accretion 
geometry during high and intermediate accretion states is estimated on the 
level of 2 seconds \citep{Schwarz2009}.
\corr{Large differences between light-curves
due to the eclipse of the accretion stream are also visible in 
the optical photometric measurements performed in parallel with the ROSAT 
observations \citep[see Fig.~3 in][]{Schwope2001}.}
Some of those light 
curves were obtained with rather poor time resolution, \corr{e.g. 53
and 12 seconds.}

It is worth mentioning that time stamps calculated by \cite {Schwope2001} 
and \cite{Schwarz2009} for the photon counting UV and X--ray detectors were 
computed from the mean of the arrival times of the first three photons 
after the eclipse, while for the optical observations, they used the 
moments of the egress half intensity, which is common in the literature. 
Examples of the \huaqr{} light curves obtained by \corr{the XMM-Newton
EPIC--PN and Optical Monitor detectors} 
are presented in Figures 2, 3 and~4 of \cite{Schwarz2009}. XMM observations, 
contrary to the bright state ROSAT observations, were not resolved at time 
scales shorter than 2 seconds due to \corr{the} low count \corr{rates}.  

We first used all available egress times, archival as well as new ones,
to model the (O-C) diagram.  However, given the above-mentioned 
arguments, we decided to select only those measurements that were obtained 
in the white light or photometric V band, in order to keep the data more 
uniform and homogeneous.  This approach renders the measurements 
independent of possible varying emission regions in different bands.  We 
also decided to skip the most ``suspicious'' egress-times at some stage of 
fitting the orbital model, which is described below.

We note  that the HST observations (three points around $l \sim 14,000$) 
were performed with the FOS instrument in the 120--250~nm range. These 
points were also excluded in our further analysis, falling out of the white 
light and the V~band range.
%
%
\section{Modeling  all recent data}
%
Thanks to the new set of precision OPTIMA mid--egress measurements, as well 
as observations performed at PIRATE, TCS and MONET/N telescopes, we can 
re-fit planetary models to the whole set of data up to November 18th,
2011. We fitted the data with the linear and 
quadratic ephemeris models (Eqs. \ref{eq:linear}, \ref{eq:parabolic}).

%
\subsection{Single-planet models to  all recent data}
%
%
\begin{figure}
\centerline{
   \vbox{
     \hbox{
         \hbox{\includegraphics[width=3.5in]{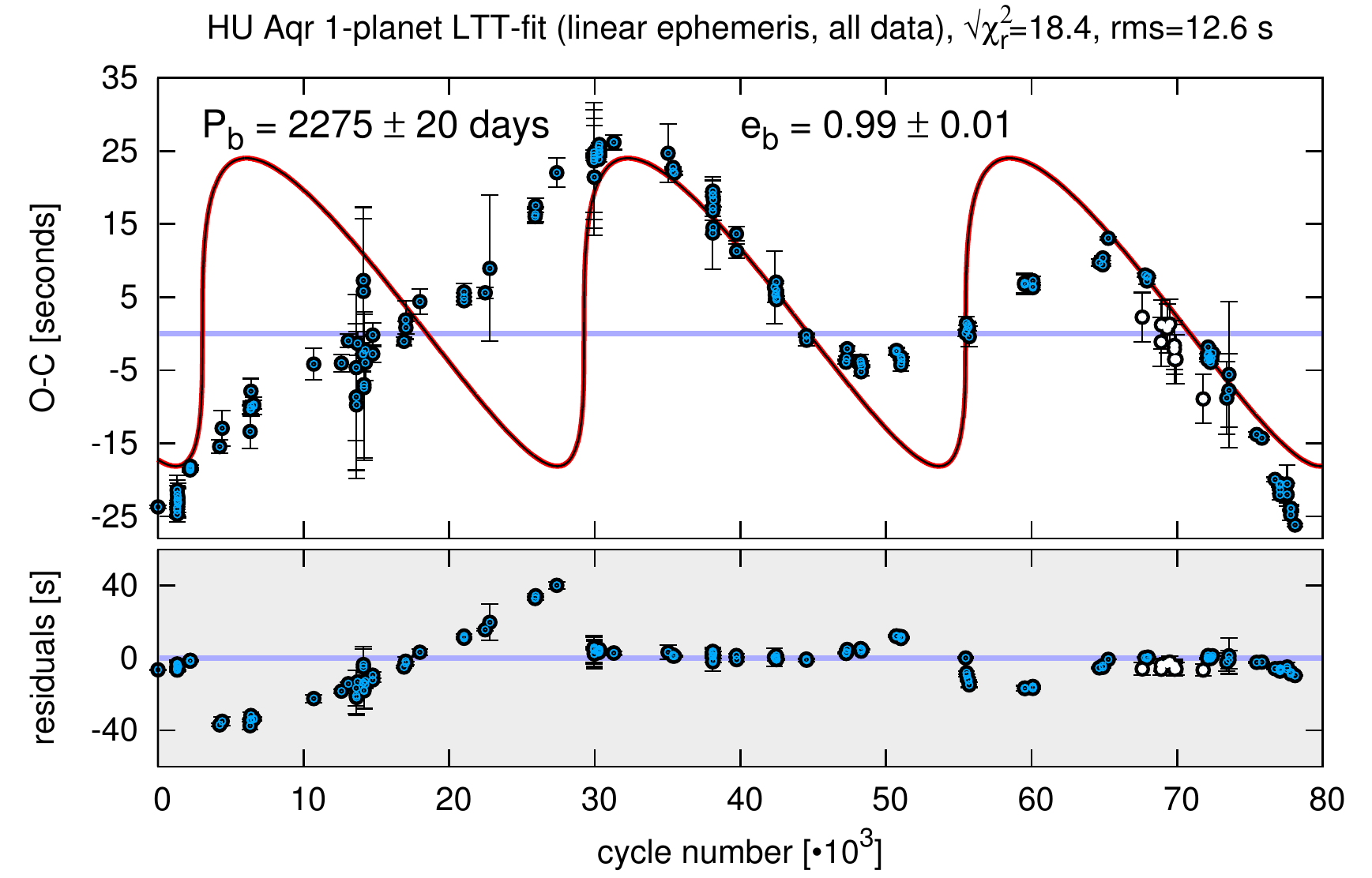}}    
     }
   } 
}
\caption{
Synthetic curve of the 1-planet LTT model with linear ephemeris to all 
available data, including the very recent egress times collected by the 
OPTIMA photometer, as well as PIRATE, TCS and MONET/N telescopes. Open
circles are for measurements in Qian et al. (2011).
}
\label{fig:fig7}
\end{figure}
At the first attempt, we tested the 1-planet hypothesis. For the linear 
ephemeris model, the 1-planet solution {is characterized by extreme 
eccentricity and} displays large residuals and {a} strong trend present in 
the (O-C) diagram (see Fig.~\ref{fig:fig7}). {This suggests a} more general 
quadratic model, on which we focus now. 

\begin{figure*}
\centerline{
   \vbox{
     \hbox{
         \hbox{\includegraphics[width=3.5in]{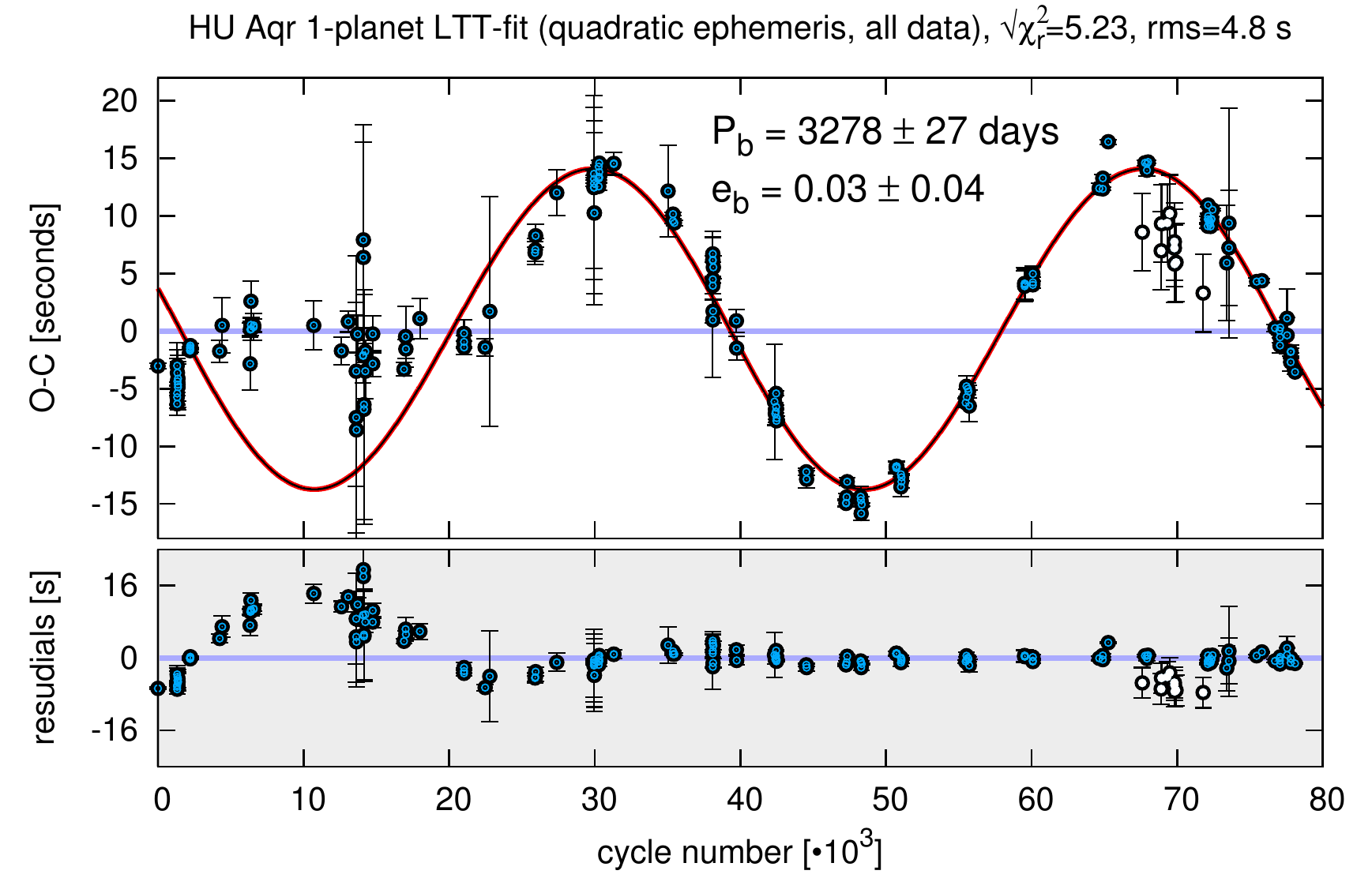}}    
         \hbox{\includegraphics[width=3.5in]{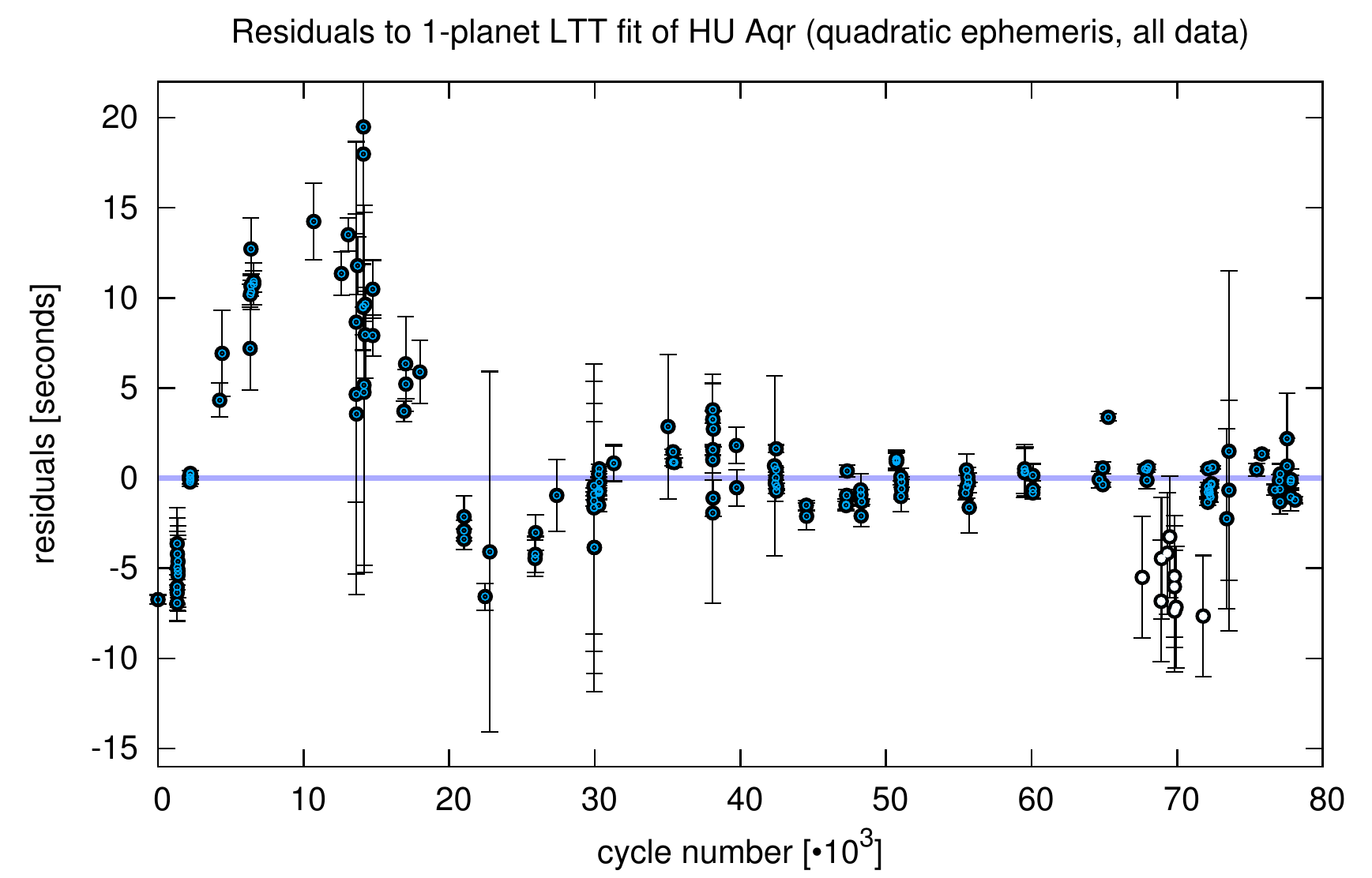}}
     }
     \hbox{
         \hbox{\includegraphics[width=3.5in]{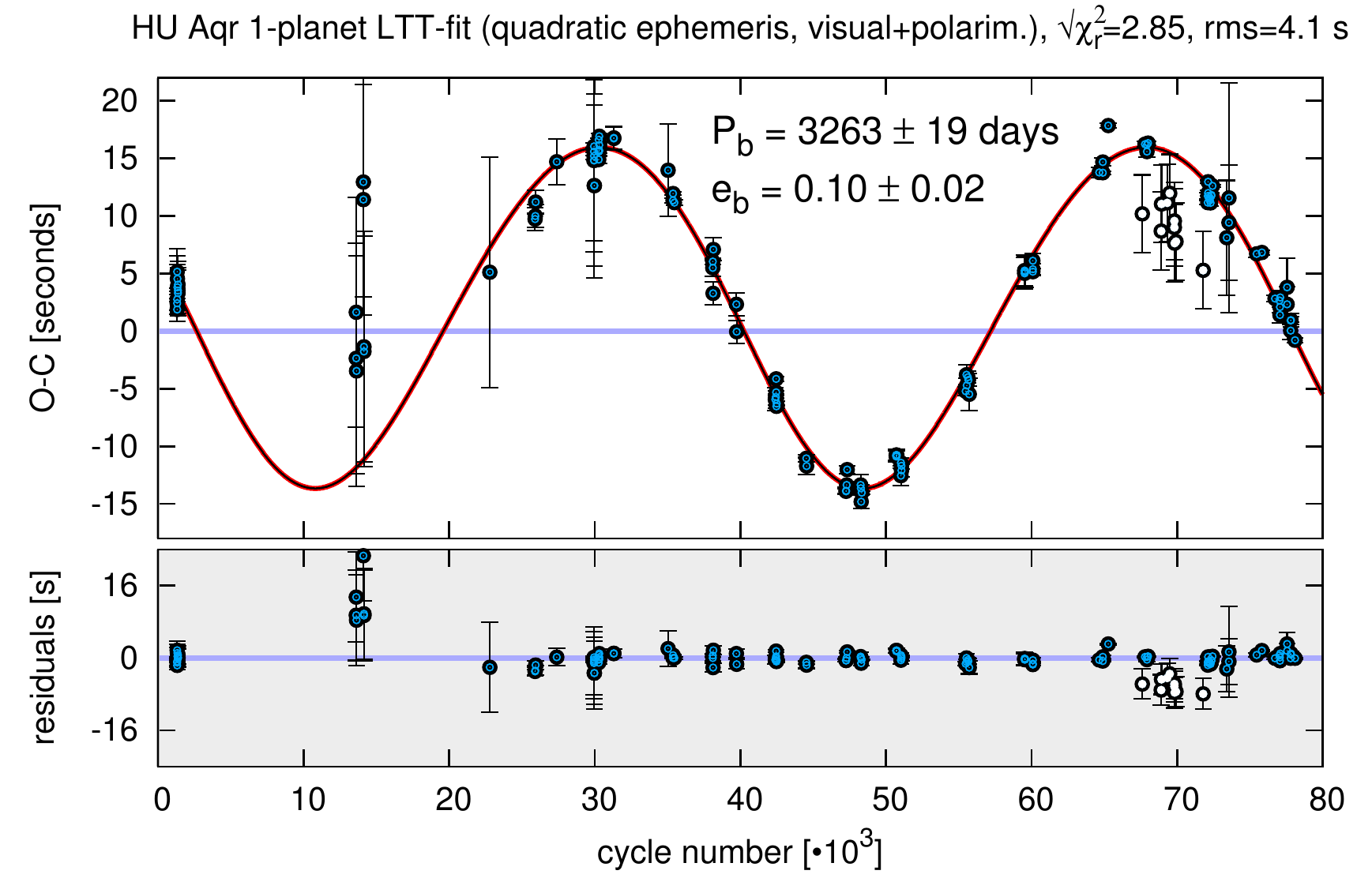}}    
          \hbox{\includegraphics[width=3.5in]{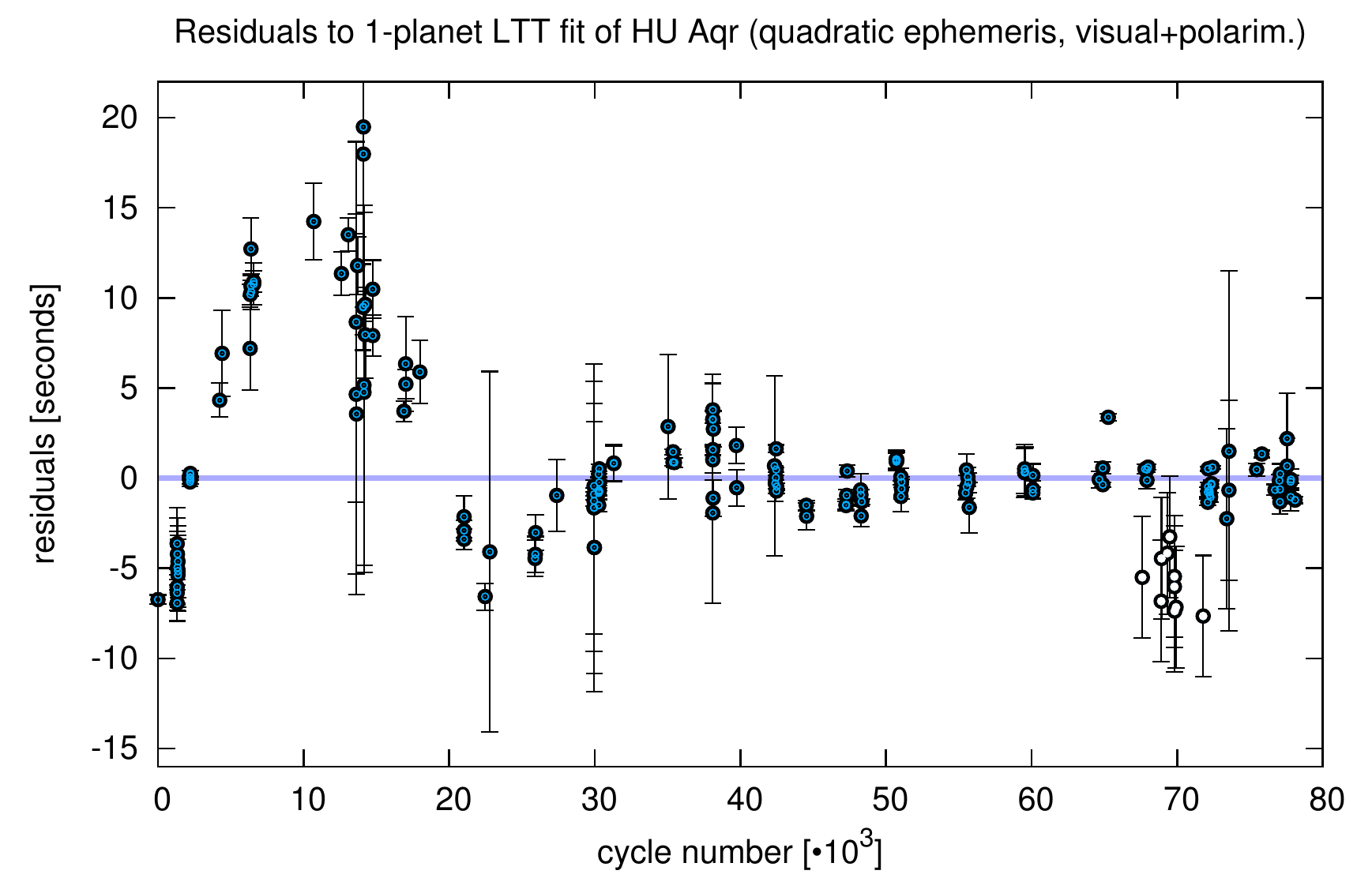}}
     }
   } 
}
\caption{
{\em The top row}: synthetic curve of the 1-planet LTT model with quadratic 
ephemeris to all available data, gathered in this work, including the very 
recent mid--egress times collected by the OPTIMA photometer, as well as 
PIRATE, TCS and MONET/N telescopes ({\em the top left}) with orbital 
parameters given in Table~\ref{tab:tab4} (Fit I), and close-up of residuals 
to that model ({\em the top right panel}). {\em The bottom row:} the same 
for the white light and visual band (V) data, including polarimetric 
observations by OPTIMA (i.e., the UV- and X-band observations are excluded) 
shown at {\em the bottom left panel}, and its residuals ({\em the bottom 
right panel}). The white filled circles mark the Qian et al. (2011) 
measurements. 
}
\label{fig:fig8}
\end{figure*}

The results derived for the whole set of 171 measurements are shown in the 
top panels of Fig.~\ref{fig:fig8}. Interestingly, the 1-planet model fits 
the data very well in a large part of the time-window  between $l=25,000$ 
and $l=80,000$ (see the left-hand panel of Fig.~ \ref{fig:fig8}). However, 
over approximately one fourth of the time-window ({$l=0$ to $l=25,000$}), 
the data fit the synthetic curve {poorly}.  That can be better seen \corr{in} 
the close-up of the residuals shown in the top right panel of Fig.~\ref
{fig:fig8}. It appears that the residuals follow a regular and 
characteristic ``damping'' trend, that could be associated with a 
mass-transfer process ongoing in the binary or solar-like magnetic cycles. 
Results of our experiments show that the recent observations by \cite
{Qian2011} appear to be outliers to our 1-planet solution, as the 
mid--egress times are shifted by about of $3$--$10$~seconds w.r.t. the 
synthetic curve. Because these observations overlap in the time window with 
much more precise OPTIMA data, that discrepancy between these two data sets 
cannot be avoided. Actually, observations by \cite{Qian2011} do not fit any 
model that has been tested with the OPTIMA observations, including 2-planet 
models and both types of the ephemeris (see the Appendix).

\begin{figure*}
\centerline{
   \hbox{
         \hbox{\includegraphics[width=4.8in]{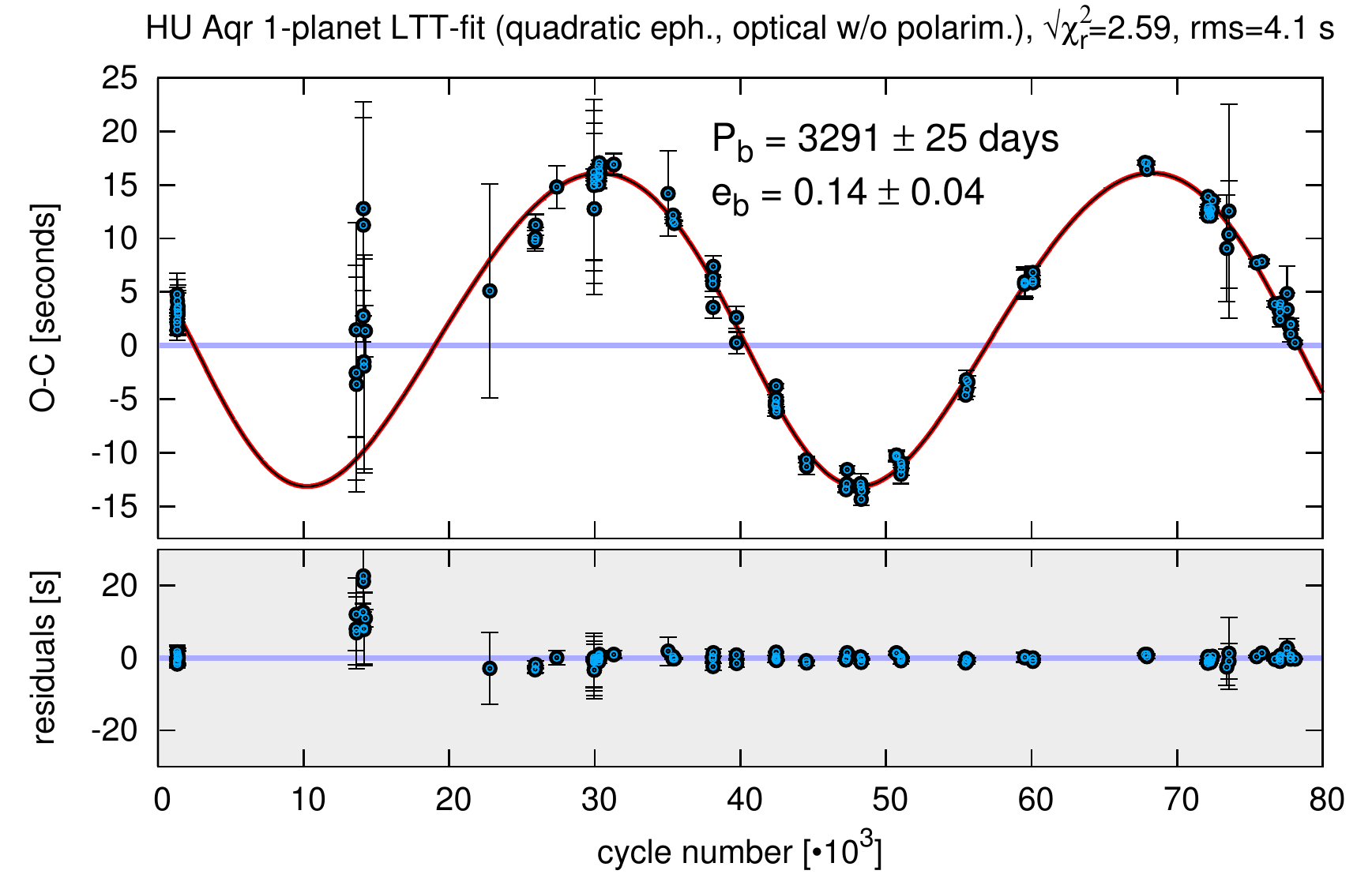}}     
   }  
}
\caption{
Synthetic curve of the 1-planet LTT model with quadratic ephemeris to 
observations in white light + V band (see the text for more details) {\em 
without}  measurements in Qian et. al (2011) {and polarimetric data}. 
Orbital parameters of this solution are given in the Table~\ref{tab:tab4} 
(Fit II).
}
\label{fig:fig9}%
\end{figure*}
\begin{figure*}
\centerline{
\vbox{
   \hbox{
         \hbox{\includegraphics[width=3.2in]{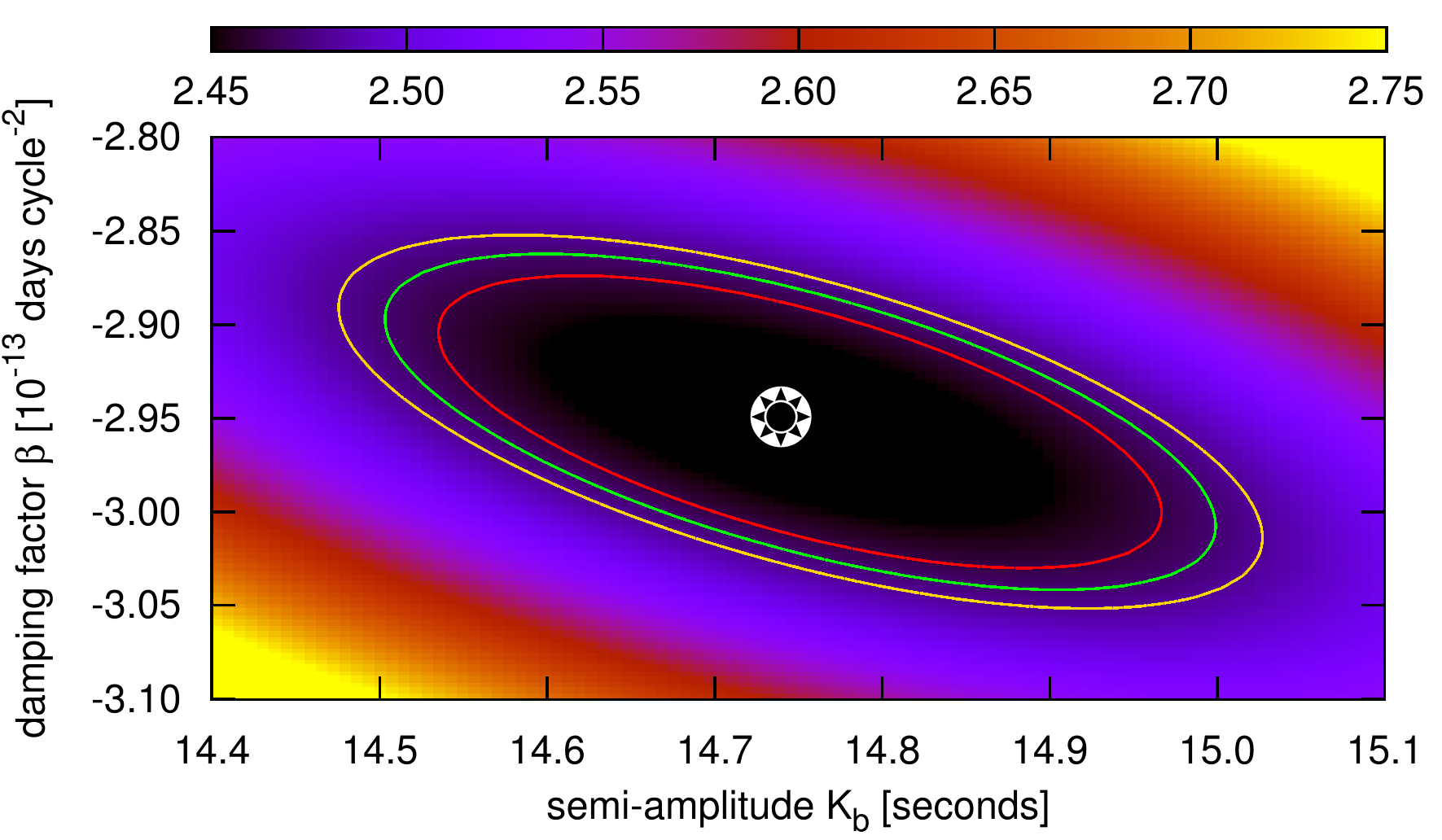}}    
         \hbox{\includegraphics[width=3.2in]{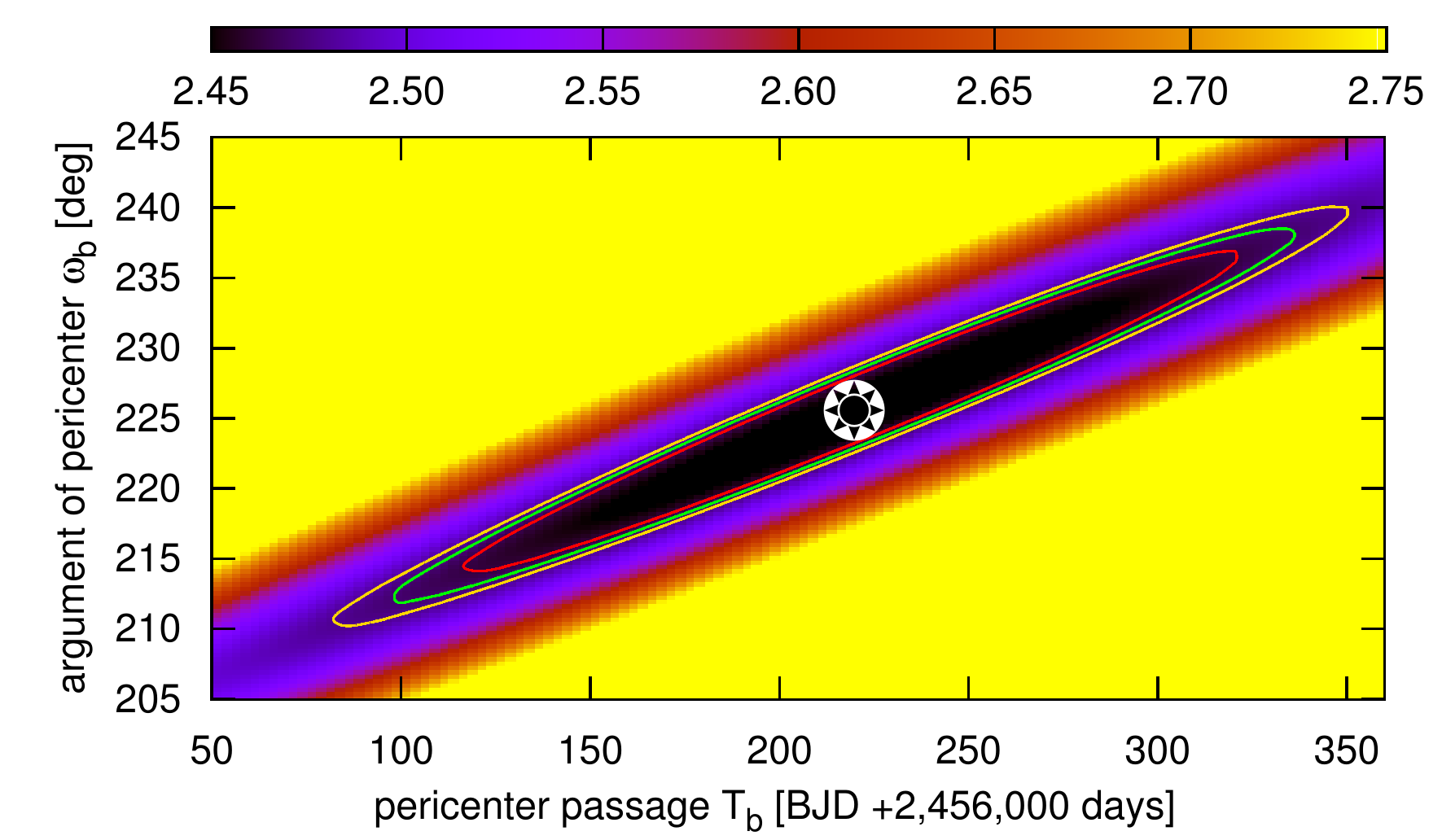}}
   }
  \hbox{
         \hbox{\includegraphics[width=3.2in]{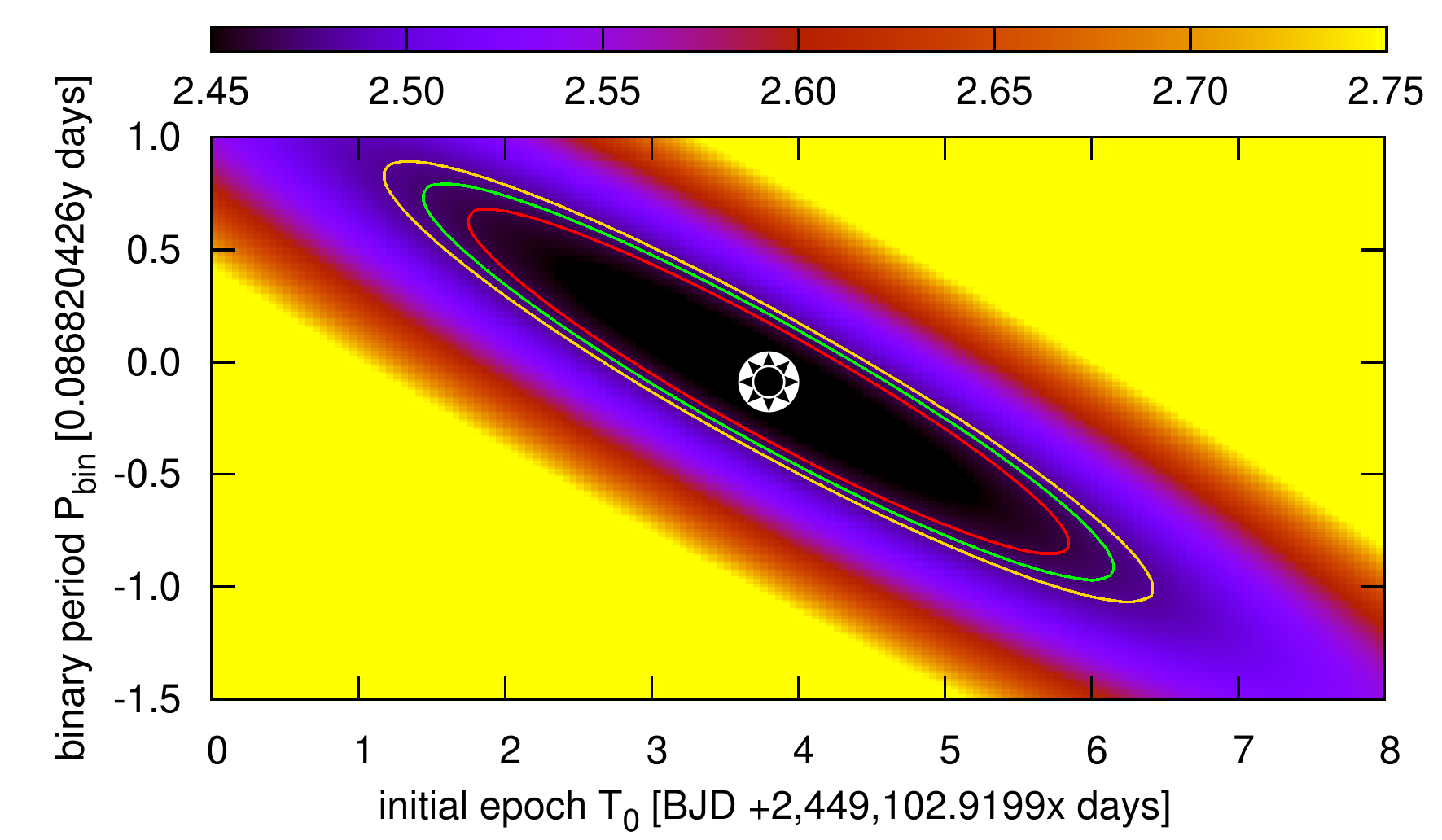}}
         \hbox{\includegraphics[width=3.2in]{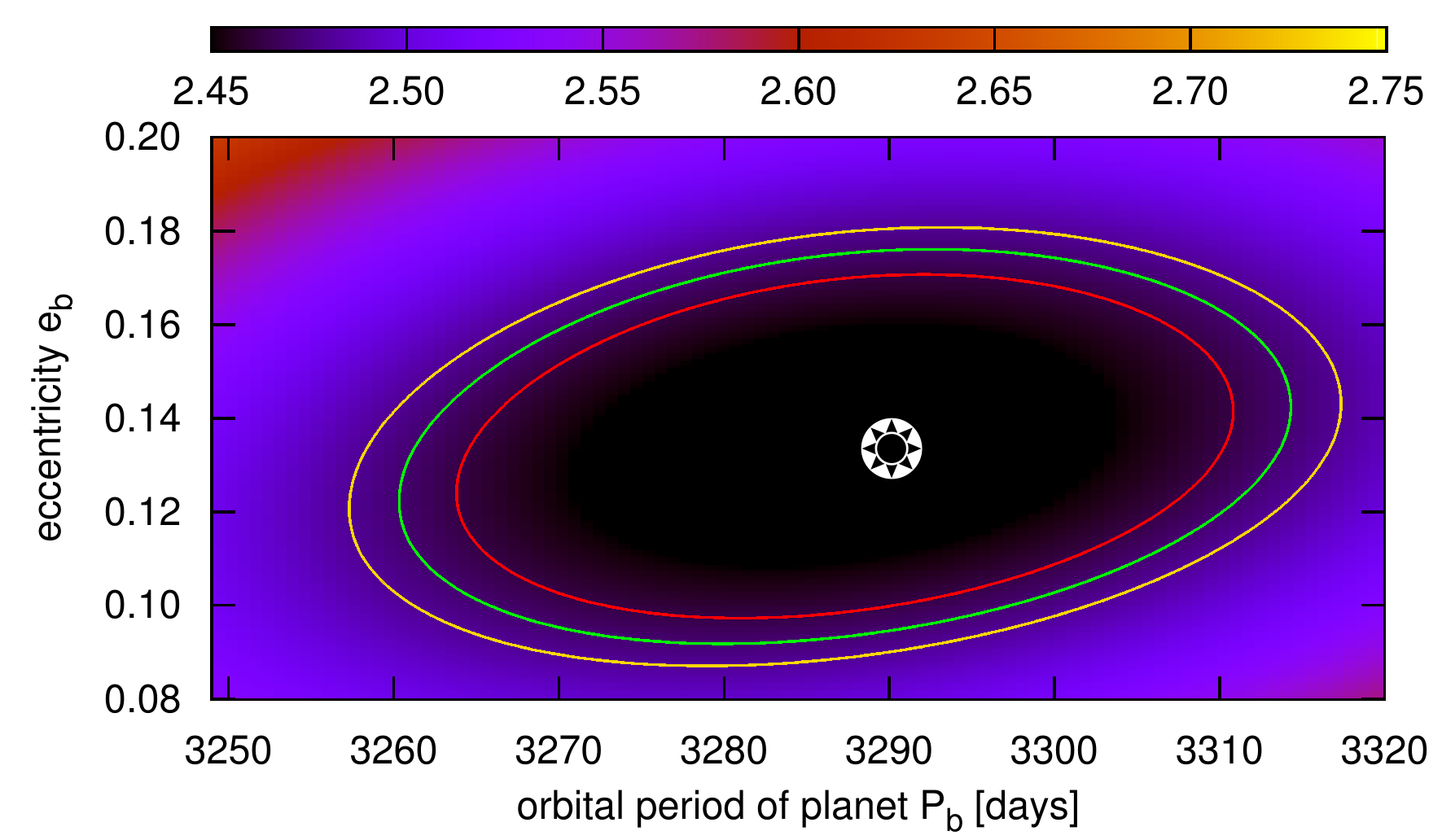}}
   }  
}
}
\caption{
Colour-coded parameter scans of $\Chi$ around the best-fit 1-planet model, 
to quadratic ephemeris and the optical measurements (Fit~II in Table~\ref 
{tab:tab4}). Its synthetic curve with data points overplotted is shown in 
Fig.~\ref{fig:fig9}. The large symbol marks nominal elements of the 
solution. Closed curves are for formal $1,2,3\sigma$ levels of $\Chi$ with 
scale displayed at the colour legend.
}
\label{fig:fig10}%
\end{figure*}
In an effort to explain the strange behaviour of the residuals, we realized,
as it was discussed already, that the available observations come from
different telescopes/instrumentation, and to make the matter worse, the
egress times are measured {on} the basis of light curves in different
spectral windows. In particular, the first part of the data set contains the
egress times derived from X-rays (ROSAT and XMM) and ultraviolet (EUVE, XMM
OM-UVM2 and HST/FOS) light curves, and some eclipses were observed with
OPTIMA in polarimetric mode. To remove the possible inconsistency due to the
different spectral windows and filters, we considered \corr{data sets
consisting of} the egress times measured only in the optical {range (white
light and the V band)}. The results are shown in the bottom panels of Fig.~
\ref {fig:fig8} for the optical data {\em without} X-ray and UV, but {\em
including} polarimetric measurements (\corr{note, that the polarimetry was
done in the white-light band}; compare with the top panels of
Fig.~\ref{fig:fig8} for all data gathered). %
\begin{table}
\caption{
Keplerian parameters {for the} 1-planet LTT fit model with \corr{quadratic} 
ephemeris to all data gathered in this work (Fit I) and to measurements 
selected in the optical and V-band domain (Fit II). Synthetic curves with 
mid--egress times {overplotted} are {shown} in Figs.~\ref{fig:fig8} and \ref
{fig:fig9}. Numbers in parentheses {represent} the uncertainty at the last 
significant digit. Total mass of the binary is 0.98~$M_{\sun}$~\citep 
{Schwope2011}. See the text for {more details}. 
}
\centering
\begin{tabular}{|c|c|c|}
\hline
Model       &  Fit~I  & Fit~II \\
parameter   & all measurements   & optical measurements \\
\hline
$K_{\idm{b}}$~[seconds]   &  13.9 $\pm$ 0.3  &  14.7 $\pm$ 0.2             \\
$P_{\idm{b}}$~[days]      &  3278 $\pm$ 28 &    3287 $\pm$ 19           \\
$e_{\idm{b}}$             &  0.03 $\pm$ 0.04 &  0.13 $\pm$ 0.04           \\
$\omega_{\idm{b}}$~[degrees]  &  211 $\pm$  40 &   226 $\pm$ 10         \\
$T_{\idm{b}}$~[BJD 2,440,000+]   
                &  6233 $\pm$ 360    &  6361 $\pm$ 102             \\
\hline
$\Pb$~[days]    &  0.0868204226(5)   &  0.0868204259(4)            \\
$T_0$~[BJD 2,440,000+]    
                &  9102.92004(2)    &  9102.91994(1)           \\ 
$\beta$ [$\times 10^{-13}$~\dcl] 
                &  -2.61 (5)          & -2.95(4)              \\
\hline
$a_{\idm{b}}$~[au]        &     4.29               &  4.30             \\
$m_{\idm{b}}\sin i$~[M$_{\idm{Jup}}$]        
                &     6.71               &  7.10              \\
\hline
$N$ data             &     171           &       115         \\
$\Chi$          &     5.23          &         2.48     \\ 
rms [seconds]       &     4.8            &       3.7       \\
\hline
\end{tabular}
\label{tab:tab4}
\end{table}
\corr{As can be} seen {from} the bottom panels in Fig.~\ref{fig:fig8}, the 
``damping'' effect has almost \corr{vanished, suggesting} that it could have 
appeared due to the presence of X-ray and UV-derived eclipses. 
Still, there is {a group of data points} with large errors, around $l\sim 
14,000$, which  do not fit well to the clear quasi-sinusoidal variation of 
the (O-C). The deviations of these points may be explained by poor 
time-resolution ($\sim 12$ seconds of the AIP07 CCD camera), that has been 
used to observe the HU~Aqr eclipses \citep{Schwope2001}. Let us also note 
that the \cite{Qian2011} data points are {again} systematically outliers 
with respect to the synthetic signal. After removing these data and
\corr{all points (seven measurements)
in the polarimetric mode}, we obtained a homogeneous optical data set 
to which we fitted the quadratic ephemeris 1-planet model again. The 
synthetic curve of \corr{this fit} with data points over-plotted is shown in 
Fig.~ \ref{fig:fig9}.  Parameters of this fit are presented in Table~\ref 
{tab:tab4} as the final solution Fit~II and are well constrained by the 
observations.  To demonstrate the latter, we show projections of $\Chi$ in 
selected two-dimensional parameter--planes (see Fig.~\ref{fig:fig10}) close 
to the best--fit model. As can be seen, there is a strong correlation 
between the time and  argument of pericentre which can be understood noting 
that the orbital phase ($\lambda_{\idm{b}} = \varpi_{\idm{b}} + M_{\idm{b}}$
) must be preserved.

The best-fit model \corr{seem to constrain} the damping factor $\beta \sim -3\times 
10^{-13}$~\dcl{} very well.  Such a value is close to estimates in the 
literature, e.g., $\sim -5~\times10^{-13}$~\dcl{} by \cite{Schwarz2009} and 
$\sim -2.5\times10^{-13}$~\dcl{} by \cite{Qian2011}.  It is still larger 
\corr{by more than} one order of magnitude to be explained by gravitational 
radiation, but remains in the range of magnetic braking \citep
{Schwarz2009}.  A similar large-magnitude period decrease has been found in 
other CVs, like NN~Ser \citep[$\sim -6\times10^{-13}$~\dcl
{},][]{Brinkworth2006}.  Besides the angular momentum loss, the large 
magnitude of the period change is commonly explained as due to the 
Applegate mechanism (basically excluded in the \huaqr{}) and/or 
a the presence of a very distant, long-period companion body.  Likely, 
a~few astrophysical and/or dynamical effects may be involved that could 
determine apparently secular period decrease. Its definite explanation is 
complex, and we consider this as a subject of \corr{a new, forthcoming work}.

We also fitted the quadratic ephemeris model only to the {highest} precision
OPTIMA data. The results for measurements that {\em include} polarimetric
observations are shown in Fig.~\ref{fig:fig11}. For that case, we found a
period similar to the quadratic ephemeris model for the entire data set. The
fit has very small rms $\sim 0.8$~sec. The relatively large $\Chi \sim 3.4$
of the OPTIMA solution in this case may suggest that the adopted
uncertainties, at the $\sim 0.1$--$0.2$~second level {(in a large sub-set of
the measurements)} are in fact underestimated. We also identified the most
deviating points as coming from the polarimetric measurements (see e.g., a
point marked in the residuals plot around $l=65,000$, and the residuals of
both solutions).  To examine, whether these data may change the solution, we
fitted the quadratic ephemeris model to the white light OPTIMA observations
only, skipping all polarimetric data. The best-fit orbital period of $\sim
3400$~days remains close to the full-coverage window fit. A slightly \corr
{smaller rms {of} $\sim 0.7$~seconds suggest a better fit without the
polarimetric data, indeed}. The orbital periods coincidence cannot be fully
proved \corr{due to the relatively} narrow observational window of the
OPTIMA white light measurements. Actually, the parameter scan (not shown
here) reveals that the $1\sigma$ contour around the $\Chi$ minimum in the (
$P_{\idm{b}},e_{\idm{b}}$)--plane is ``opened'' on the right side of the
orbital period--axis, hence it can not be constrained yet by the OPTIMA data
alone.
%
%
\subsection{Alternate models to  all recent data}
%
Finally, using the hybrid optimiser, we performed additional experiments by 
fitting three models to all available data: the 1-planet model with
a heuristic sine \corr{damping} term, and 2-planet models, 
both in terms of the linear and quadratic ephemeris. We also performed $N$ 
-body modeling of the 2-planet configurations. The results, which are 
described in the Appendix, imply that all these models lead to non-unique 
solutions or configurations with similar orbital periods, for which the 
kinematic model is inadequate, as we discussed above. Some of these 
kinematic best--fit 2-planet solutions are qualitatively similar to 
configurations found for the SSQ data set, with \corr{orbital periods ratios}
close to 1c:1b and 4c:3b, respectively. The extended data set still does not 
constrain the Keplerian 2-planet models. 

The same can be concluded for the $N$-body models (see Appendix, Sect.~A3).
Although we found {\em stable} configurations in terms of the quadratic
ephemeris, the semi-major axis of the outer companion is unconstrained
(between 4~AU and at least 20~AU). \corr{These stable fits exhibit varying
sign of $\beta$ (it means that the binary period might decrease or
increase). Moreover, stable solutions with relativelty small $\beta$ may be
found in very narrow stability zones in the ($a_{\idm{c}},e_{\idm{c}})$
--plane, see the Appendix and Fig.~\ref{fig:fig18}. These areas are
associated with low order MMRs, like 3c:2b~MMR. It is very uncertain though,
how massive companions of \huaqr{} could be locked in such tiny stability
areas. Hence, some larger values of $a_{\idm{c}}$ providing extended zones
of stable motions seems more likely (see panels of stable fits labeled by
IV, V, and~VI of Fig.~\ref{fig:fig18}). However, there is also a correlation
between the magnitude of $\beta$ and the semi-major axis of the outer
planet.  For relatively distant planet~c, $|{\beta}|$ may be $\sim2\times
10^{-12}$~\dcl{}, which is difficult to explain by magnetic braking or
mass--loss. However, this may indicate a presence of a third companion in
\corr{an} unconstrained orbit. }

\begin{figure}
\centerline{
    \hbox{
       \hbox{\includegraphics[width=3.5in]{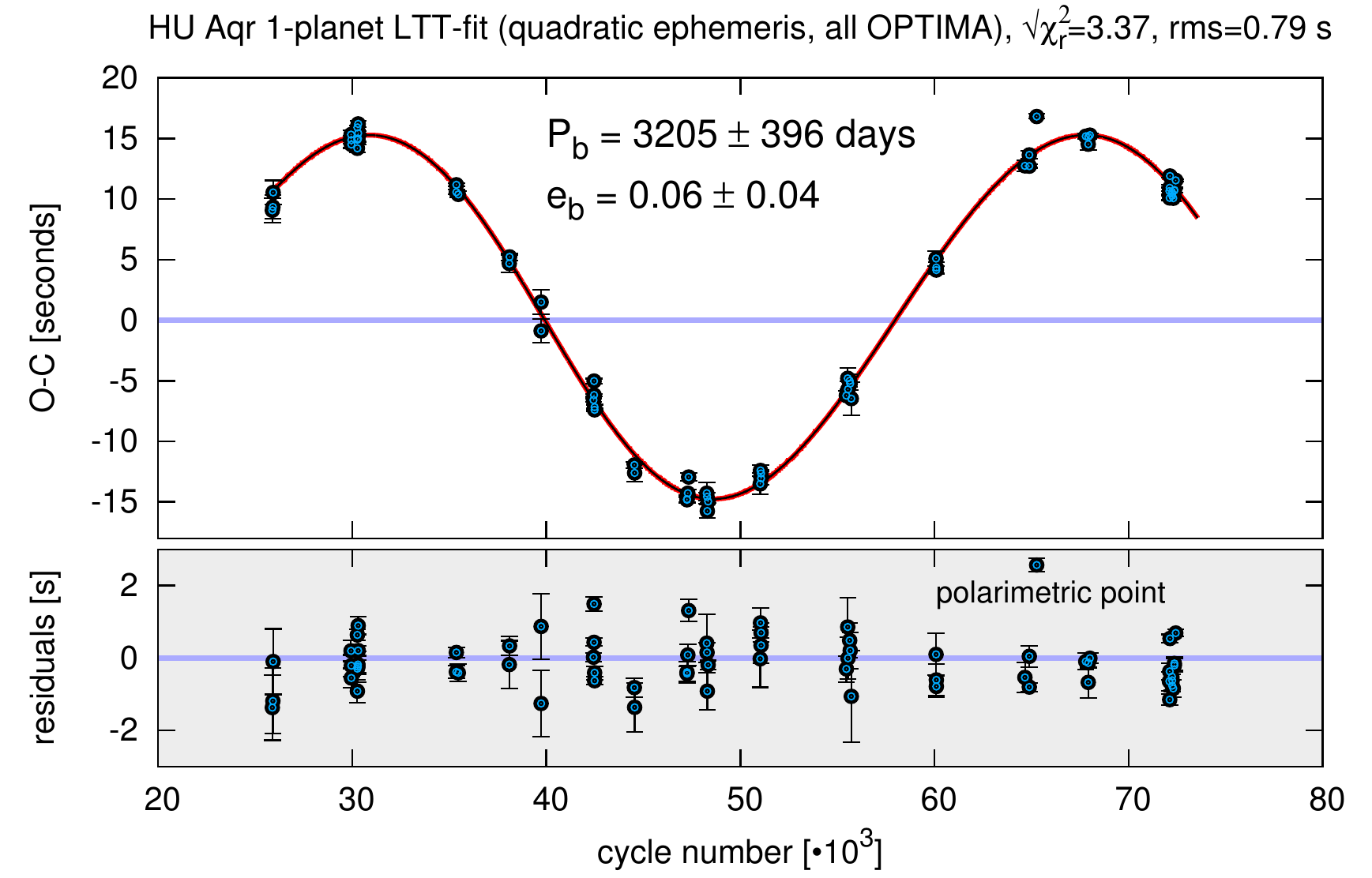}}     
   }  
}
\caption{
Synthetic curves of the 1-planet LTT quadratic ephemeris models to optical 
OPTIMA measurements, including polarimetric data 
One of the most deviating
polarimetric points is labeled in the residuals panel.
}
\label{fig:fig11}
\end{figure}

\corr{We conclude that these results seem to favour the 1-planet hypothesis as
the simplest model explaining the (O-C) variability, particularly in the light
of very small rms of the homogeneous OPTIMA set and apparently perfect
quasi-sinusoidal fit illustrated in Fig.~\ref{fig:fig11}.}
%
\section{Red noise and/or systematic errors?}
%
Analysis of the LTT observations has much in common with pulsar timing,
planetary transits, and precision radial velocity observations, which are
modelled with least-squares under the assumption that the measurement errors
are uncorrelated (white noise). However, as is known particularly by pulsar
observers, the assumption that white noise is the only source of error is
unjustified when aiming at estimating the underlying model parameters and
their uncertainties \citep{Coles2011}. In the past, this effect {had been}
responsible for false detections of planets around pulsars \citep
{Bailes1991}. Similarly,  correlated {(red)} noise {or systematic errors
have} been found in the planetary transit data \citep {Pont2006} and very
recently, in the \corr{radial velocity} measurements \citep {Baluev2011}.
The same type of non-Gaussian, low-frequency correlation of residuals to the
orbital period of the binary may be present in the LTT data collected over
long time intervals.

The danger of such systematic effects in the LTT-analysed binaries is
reinforced due to their activity and complex astrophysical phenomena
responsible for the observed emission.  One of the already well recognized
mechanisms able to produce cyclic variation of the orbital period of the
binary has been proposed by \cite{Applegate1992}.  As shown by this author,
a magnetic star (here, the secondary) changes its internal structure due to
magnetic cycles.  The latter implies a variable zonal harmonic coefficient
$J_2$ and subsequently, a variable gravitational tidal field for the orbital
companion which results in a varying orbital period \citep {Hilditch2001}.
The Applegate mechanism as \corr{a} possible origin of large (O-C)
variations in the \huaqr{} data was studied in detail by \cite{Vogel2008b},
as well as by \cite{Schwarz2009}.  They discarded this possibility since the
\huaqr{} stellar setup does not provide enough energy to drive changes of
the orbital period.  Similar results were obtained for the NN~Ser system,
that likewise has a low-mass, low-luminosity secondary star \citep
{Brinkworth2006} with a conclusion that it is incapable of driving
significant period changes in terms of the Applegate model.

Another mechanism explaining observed long-term periodicities could be a 
slow  precession of the rapidly spinning magnetic WD star, {which} has been 
proposed as a source of long periods detected in a few CVs, for instance 
FS~Aur and V455~And \citep{Tovmassian2007}. However, \huaqr{} is unlikely 
to host such a WD, as this AM Her-like system is known to be synchronously 
locked.
\begin{figure*}
\centerline{
   \vbox{
     \hbox{
         \hbox{\includegraphics[width=2.0in,angle=270]{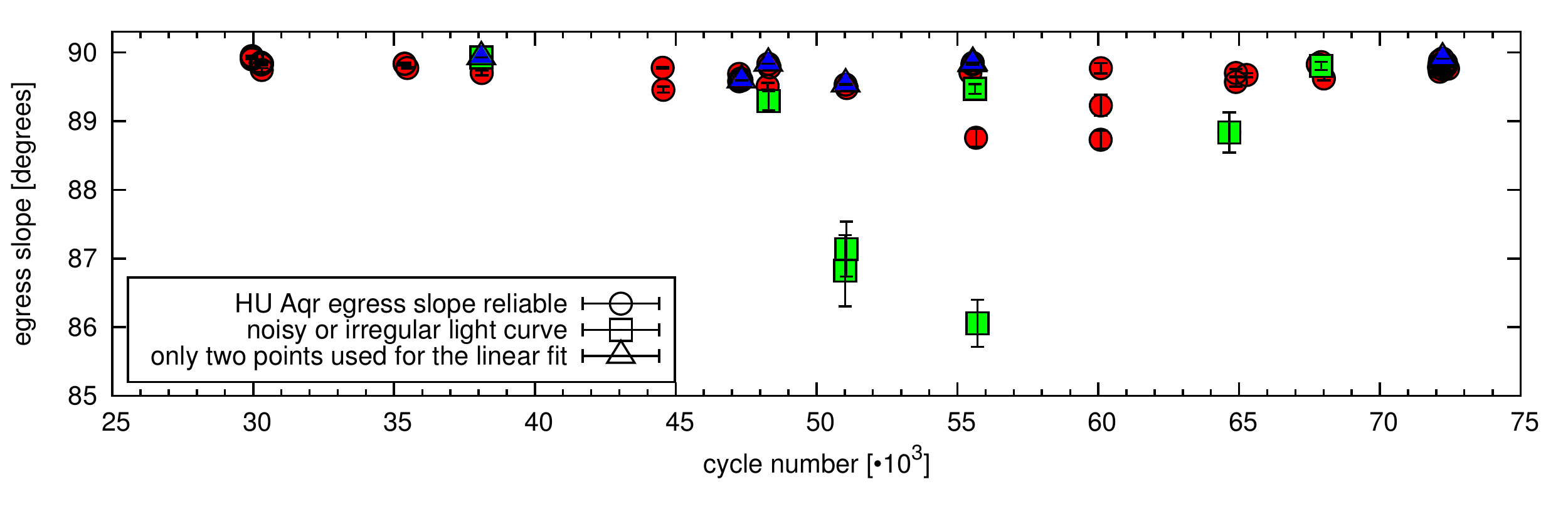}}    
     }
   } 
}
\caption{
Linear slopes of 59 \huaqr{} egresses derived on the basis of OPTIMA light 
curves by fitting {a} linear function only to the egress phase, usually 
spanning not more than a few seconds. In seven cases, the light curves were 
not precise enough (as indicated by green filled squares) because of a bad 
weather. In other seven cases, only two points were taken for the fit, 
therefore no error estimation was possible (blue triangles). See the text 
for more details.
}
\label{fig:fig12}%
\end{figure*}
As a first, yet preliminary attempt, we tried to determine the
characteristic that can be used to quantify the shapes of the \huaqr{} light
curves and  might help to detect their variability and hence astrophysical
sources of the LTT residuals. This approach mimics the bisector \corr
{velocity} span (BVS) technique used to detect distortion of spectral lines
due to stellar spots and chromospheric activity. It is well known that \corr
{stellar spots may produce} apparent radial velocity changes up to $200$~ms
$^{-1}$ \citep {Berdyugina2005}. As a similar characteristic to the BVS, we
choose the slope of the linear function fitted to the egress phase  of the
light curve, usually spanning no more than a few seconds interval. We
analysed 59 available light curves in the precision OPTIMA set. The results
are {shown} in Fig.~\ref{fig:fig12}.  In seven cases, we decided the data
were not precise enough to derive the slope reliably (as indicated by green
filled squares) because of, for instance, bad weather or strong wind that
could {introduce} telescope guidance errors. In a few other cases (seven
again, marked with blue triangles), only two points were taken for the fit,
{and} therefore no error estimation was possible. Nevertheless, the obtained
slopes are uniform and span less than $2$ ~degrees range close to $90$
~degrees. That furthermore indicates a similarly rapid egress phase. The
results of this test are encouraging, and support the planetary hypothesis.

However, the slopes should be best re-computed for all available light
curves that were used to determine the egress-times. The problem  of the
non-homogeneity of the collected light curves still exists. Due to varying
eclipse profiles (e.g. during different accretion states), the determination
of mid--egress dates is often very difficult. For instance, it could be
prone to rather subjective \corr{choices} of the photometric data range to
fit the parameters of the sigmoid function, Eq.~\ref{eq:sigmoid}. That may
introduce significant systematic errors, particularly if the reduction is
performed by different researchers. This issue may be likely resolved by a
re-analysis of the entire set of all available light curves, under similar
conditions paying particular attention to their origin -- the spectral
window, an instrument, and even technical and observational circumstances.

Another direction still open is a study of the binary interactions, to 
eventually eliminate or discover astrophysical causes of the LTT 
variability. The problem is in fact universal and affects other techniques 
of extrasolar planets detection, such as pulsar timing and \corr{radial velocity}
monitoring of active or evolved stars, as well. It is yet possible that
the observed (O-C) signal has both the planetary and unmodeled astrophysical
component \citep{Potter2011}, making its unique resolution even harder. 

To the best of our knowledge, possible effects of the red-noise regarding 
the LTT observations have not been studied in detail. That problem 
certainly deserves a deep and careful investigation. 
%
%
\section{Conclusions}
%
Using a new formulation of the LTT model of the (O-C) to the available data 
of the  HU~Aqr system, we found that {the} 2-planet hypothesis by \cite 
{Qian2011} is not likely.  Our results reinforce recent negative tests of 
dynamical stability of that system in the literature. The self-consistent 
LTT model {presented in this work exhibits} degenerate solutions, {such as} 
(apparently) Trojan objects of $\sim 10^4$ Jupiter masses, or a companion 
in an collisional/open parabolic or hyperbolic orbit. Ironically, two such 
solutions to the literature SSQ data are the best--fit models found in 
extensive, quasi-global searches adopting a hybrid optimization.  

Moreover, on the basis of a much extended, precision data set, collected by 
the OPTIMA network, that increased the number of data points analysed in 
previous works by $\sim50$ \%, we have shown that the observed (O-C) 
variations may be consistently explained by the presence of only one 
circumbinary planet of the minimal mass of $\sim 7$ Jupiter-masses, in an 
orbit {with a} small eccentricity {of} $\sim 0.1$ and {an orbital period 
of} $\sim 10$~years, similar to Jupiter in the Solar system. {Our 
results support the original 1-planet hypothesis by \cite{Schwarz2009} 
rather than the 2-planet model proposed by  \cite {Qian2011}.} If 
confirmed, that planet would be the next circumbinary object detected from 
the ground, shortly after such companions have been announced around HW Vir~
\citep{Lee2009}, NN Ser \citep {Beuermann2010}, UZ For~\citep{Potter2011}, 
SZ Her~\citep{Lee2011}, DP~Leo \citep {Beuermann2011}, followed by recent 
discoveries of Kepler-16b \citep {Doyle2011}, Kepler-34b and Kepler-35b 
planets \citep{Welsh2012}. According {to} estimates by these authors, the 
observed rate of circumbinary planets around close binaries may be $\sim 
1\%$. 

Also, we found that {the} observations {by \cite {Qian2011}} are not 
confirmed by the OPTIMA measurements due to systematic relative shift of 
$\sim3$--$10$~seconds. The nature of this discrepancy is yet unknown.  If 
the shift is caused by an error, all 2-planet models presented in the 
literature that make use of their data are affected.

Besides the disagreement between our conclusions and the previous works, our
results suggest that the kinematic modeling of 2-planet configurations is
not fully justified on the grounds of the dynamics
because the best-fit models may imply large masses (up to stellar range),
large eccentricities, and similar orbital periods indicating a possibility
of strong mean motion resonances.  Moreover, the (O-C) variability that
suggests 2-planet solutions most likely appears due to mixing observations
done in different spectral windows.  That feature of the data set -- as we
have shown here -- introduces systematic effects that may alter the best-fit
solutions significantly.  This conclusion is supported by extensive
numerical simulations of the 2-planet systems dynamics by \cite{Horner2011},
\cite{Wittenmyer2011} and \cite{Hinse2011}.  Considered within statistical
error ranges, the initial conditions lead to catastrophically disruptive
configurations, unconstrained elements of the outermost body, and/or period
damping factor $\beta$. 

In this work, we found best-fit stable 2-planet models within the quadratic 
ephemeris $N$-body model to {\em all available data}, but the semi-major 
axis of the outer planet cannot be yet constrained.  Stable configurations 
are located within low-order MMRs spanning tiny stable zones in the phase 
space, or are characterised by a large magnitude of the period decrease.  
In the first case, it is difficult to explain, how a few Jupiter mass 
companions could be trapped in such particular, isolated resonances.  In 
the second case, a large $|\beta|$ requires an efficient, internal 
mechanism of the binary period change, or indicates a presence of one more 
companion.  Our findings might be a breakthrough after a few cited works 
reporting basically only unstable 2-planet models of the \huaqr{} system, 
but these discrepancies add even more ambiguity to the 2-planet hypothesis.

\corr{However}, the results of our experiments show that the 1-planet solution 
is relatively well constrained by available {\em optical observations} 
selected as a homogeneous data set. 
Because the early optical data (the white light and V-band 
measurements) are coherent with an impressive, very clear quasi-sinusoidal 
signal exhibited by superior-precision OPTIMA measurements, as well as with 
the recent MONET/N, PIRATE and WFC data, a single-companion hypothesis 
seems well justified. A confirmation of the planetary origin of the LTT 
signal still requires long-term monitoring of the system. Due to its very 
long orbital period, it will take many years to confirm or reject the 
signal coherence. Such new data would be also very useful to constrain the 
orbital period by the recent OPTIMA observations alone.

%
%
{\bf Acknowledgements.}
%
We thank the anonymous referee for a review and comments that improved the
manuscript. We would like to thank {Maciej Konacki (CAMK Toru\'n) for a
discussion and a suggestion of the egress slopes test}, as well as Anna
Zajczyk (CAMK, Toru\' n), Andrzej Szary (UZG, Zielona G\'ora), \corr{Alex
Stefanescu, Martin M\"uhlegger, Helmut Steinle, Natalia Primak, Fritz
Schrey, and Christian Straubmeier (all MPE)} for their help with
observations. Krzysztof Go\'zdziewski is supported by the Polish Ministry of
Science and Higher Education Grant No.~N/N203/402739 and POWIEW project of
the European Regional Development Fund in Innovative Economy Programme
POIG.02.03.00-00-018/08.  Ilham Nasiroglu acknowledges support from the EU
FP6 Transfer of Knowledge Project ``Astrophysics of Neutron Stars''
(MKTD-CT-2006-042722).  Aga S\l{}owikowska would like to thank Bronek Rudak
for his support and discussions. She {also} acknowledges support from the
Foundation for Polish Science grant FNP HOM/2009/11B, as well as from the
Marie Curie European Reintegration Grant within the 7$^{\idm{th}}$ European
Community Framework Programme (PERG05-GA-2009-249168). Gottfried Kanbach
acknowledges support from the EU FP6 Transfer of Knowledge Project
ASTROCENTER (MTKD-CT-2006-039965) and the kind hospitality of the Skinakas
team at UoC. B. Gauza thanks the Wide FastCam team for help in performing
the observations. Research by Tobias C.~Hinse is carried out under the KRCF
Young Researcher Fellowship Program at the Korea Astronomy and Space Science
Institute. {NH acknowledges support from the NASA Astrobiology Institute
under Cooperative Agreement NNA09DA77A at the Institute for Astronomy,
University of Hawaii, and from NASA/EXOB program under grant NNX09AN05G.} We
thank the Skinakas Observatory for their support and allocation of telescope
time.  Skinakas Observatory is a collaborative project of the University of
Crete, the Foundation for Research and Technology -- Hellas, and the
Max-Planck-Institute for Extraterrestrial Physics. This article is based on
observations made with the TCS telescope operated on the island of Tenerife
by the Instituto de Astrofisica de Canarias in the Spanish Teide
Observatory. This work is based in part on data obtained with the MOnitoring
NEtwork of Telescopes (MONET), funded by the Alfried Krupp von Bohlen und
Halbach Foundation, Essen, and operated by the Georg-August-Universit\"at G\"
ottingen, the McDonald Observatory of the University of Texas at Austin, and
the South African Astronomical Observatory.
\appendix
%
\section{Alternate models to all data}
%
\corr{
In this section, we display supplementary results illustrating a few
alternative models to the 1-planet solution of the (O-C)
of the HU Aqr binary that was analysed in the main part of the paper.
Basically, all 171~data points are
modeled, although in some cases, we removed outlying data \corr{from}
\cite{Qian2011} \corr{because they clearly introduce a systematic error}. 
We considered 1-planet model with a heuristic,
sine--damping term (Sect.~A1), 2-planet kinematic models (Sect.~A2) and the
full, 2-planet, self-consistent Newtonian model (Sect.~A3). 
The aim of this Appendix is to demonstrate that  2-planet models
lead to non-unique or unconstrained solutions. Hence, these results 
reinforce a hypothesis of a single,
quasi-sinusoidal signal of possibly planetary origin.
}
%
%
\subsection{Quadratic ephemeris 1-planet model with damping term}
%
\begin{figure*}
\centerline{
    \hbox{
       \hbox{\includegraphics[width=3.5in]{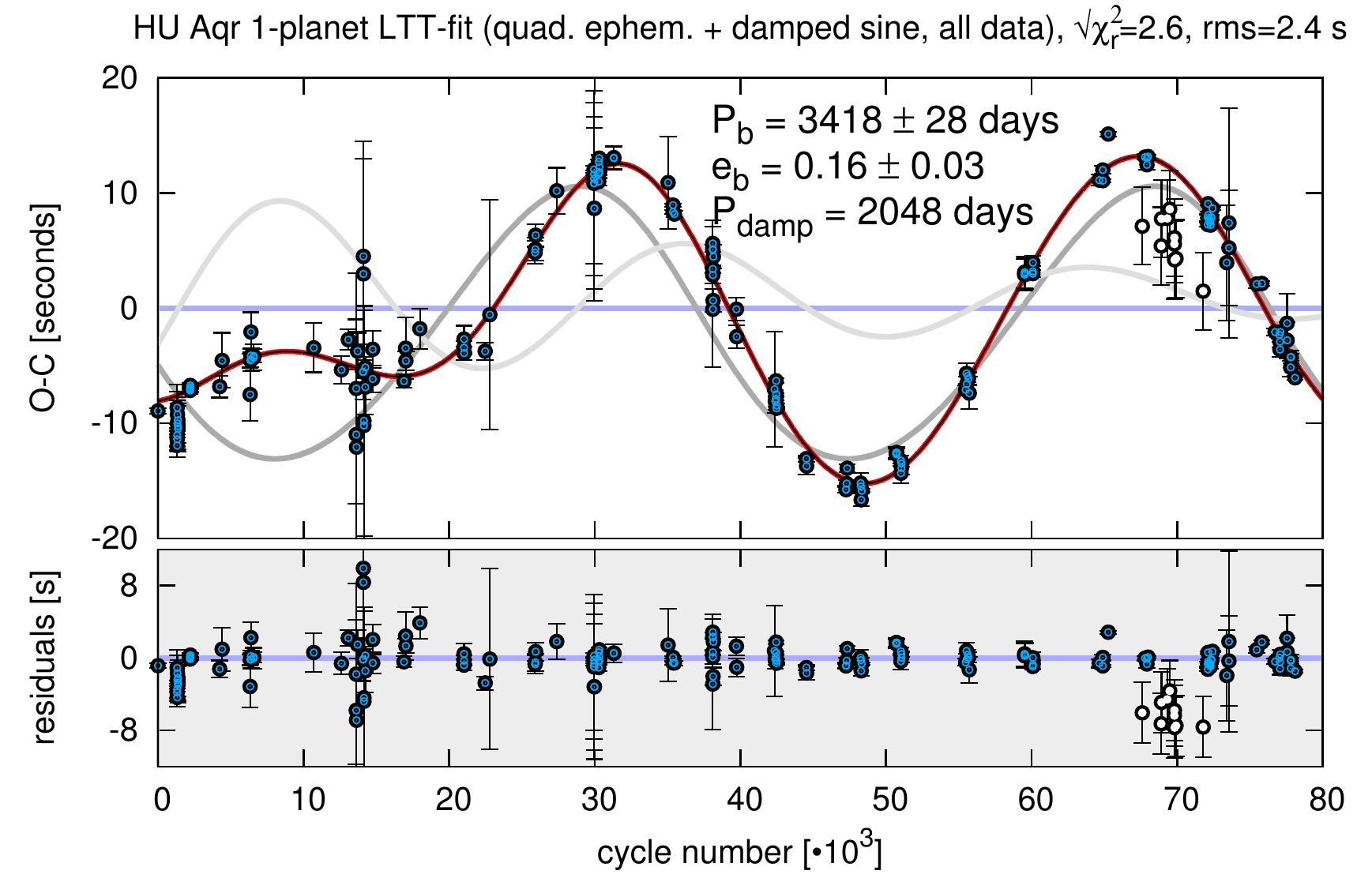}}     
       \hbox{\includegraphics[width=3.5in]{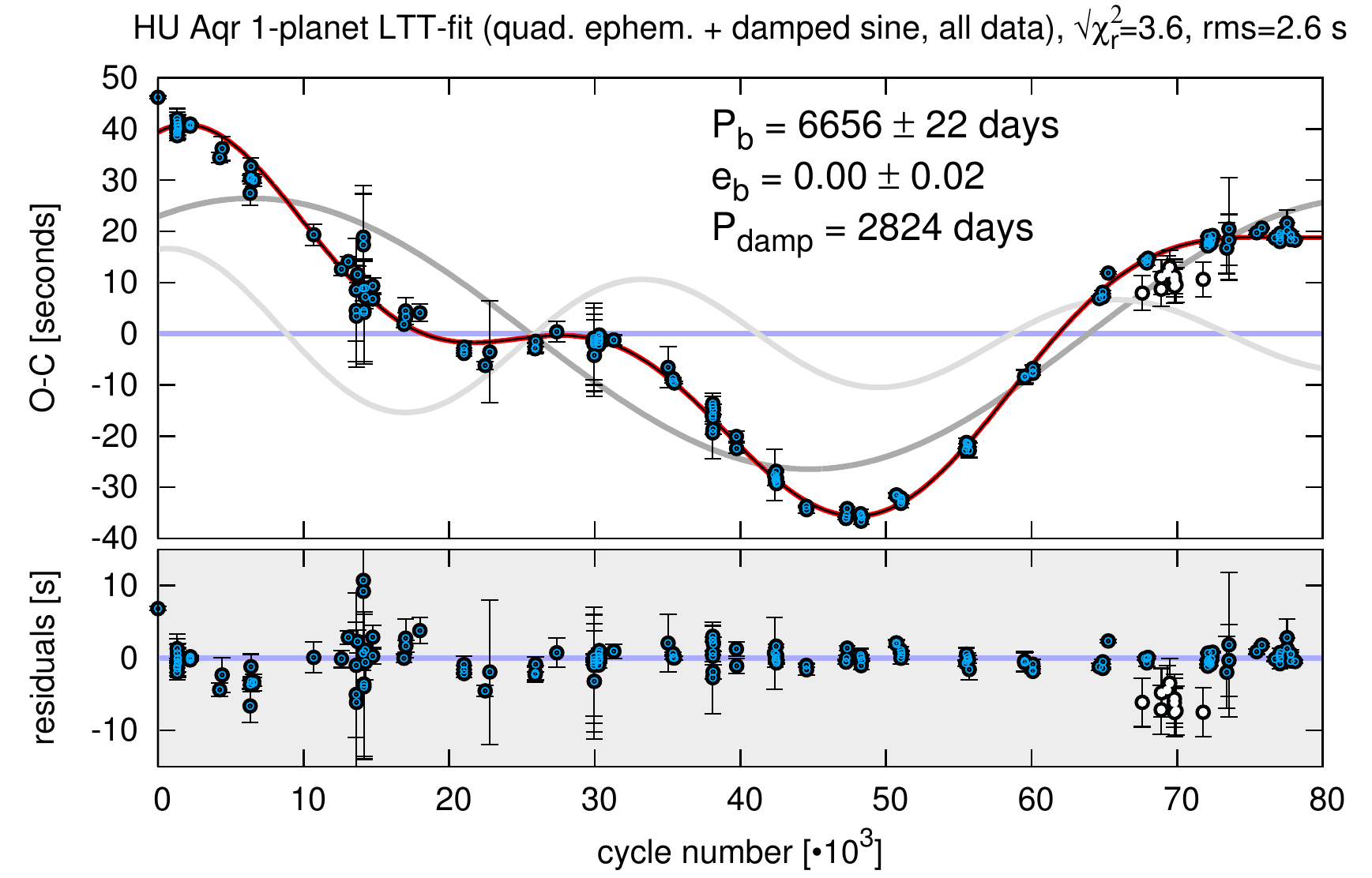}}     
   }
}
\caption{
Examples of synthetic curves of the 1-planet LTT quadratic ephemeris models 
with a sine damping term to all 171 data points gathered in this work, 
including 10 measurements in Qian et al. (2011), marked with white filled 
circles. The shaded curves {represent} the planetary and the damped sine 
signal, respectively.
}
\label{fig:fig13}
\end{figure*}
To describe the \corr{suggested} damping signal visible in Fig.~\ref{fig:fig8} 
(the top right-hand panel), we modified the quadratic ephemeris model by 
{adding a} heuristic term having the following form:
\begin{equation}
\tau_{\idm{damp}}(t) = \tau_0 + A \exp(-t/T_{\idm{damp}}) \sin(
n_{\idm{damp}}\, t +
\phi_0),
\label{eq:damp}
\end{equation}
where $\tau_0$ is an offset, $A$ is the semi-amplitude of the signal, 
$T_{\idm{damp}}$ is the damping time scale, $ n_{\idm{damp}} = 
2\pi/P_{\idm{damp}}$ is the frequency, and $\phi_0$ is the initial phase at 
epoch $l=0$. Two examples of best-fit solutions to all available data (171 
measurements) are shown in Fig.~\ref{fig:fig13}. Let us note that the 
planetary orbital period in the configuration in the right-hand panel of 
this figure is twice the period in the 1-planet model studied earlier. The 
(O-C) of a solution shown in the left-hand panel cannot be distinguished 
from 2-planet models (see the text below).

A physical nature of that damped signal is uncertain. Allowing for some 
speculations, the damping might appear due to a long--term relaxation in 
the binary system which may, for instance, be due to the {binary's} 
magnetic cycles (Applegate mechanism). In such a case, the observed LTT 
signal would be resulted from two distinct phenomena. However, we recall 
here that \cite{Vogel2008b} and \cite{Schwarz2009} estimated that the 
Applegate mechanism cannot be responsible for orbital changes of \huaqr{}.
%
%
\subsection{Kinematic 2-planet models}
\begin{figure*}
\centerline{
\vbox{
    \hbox{
       \hbox{\includegraphics[width=3.5in]{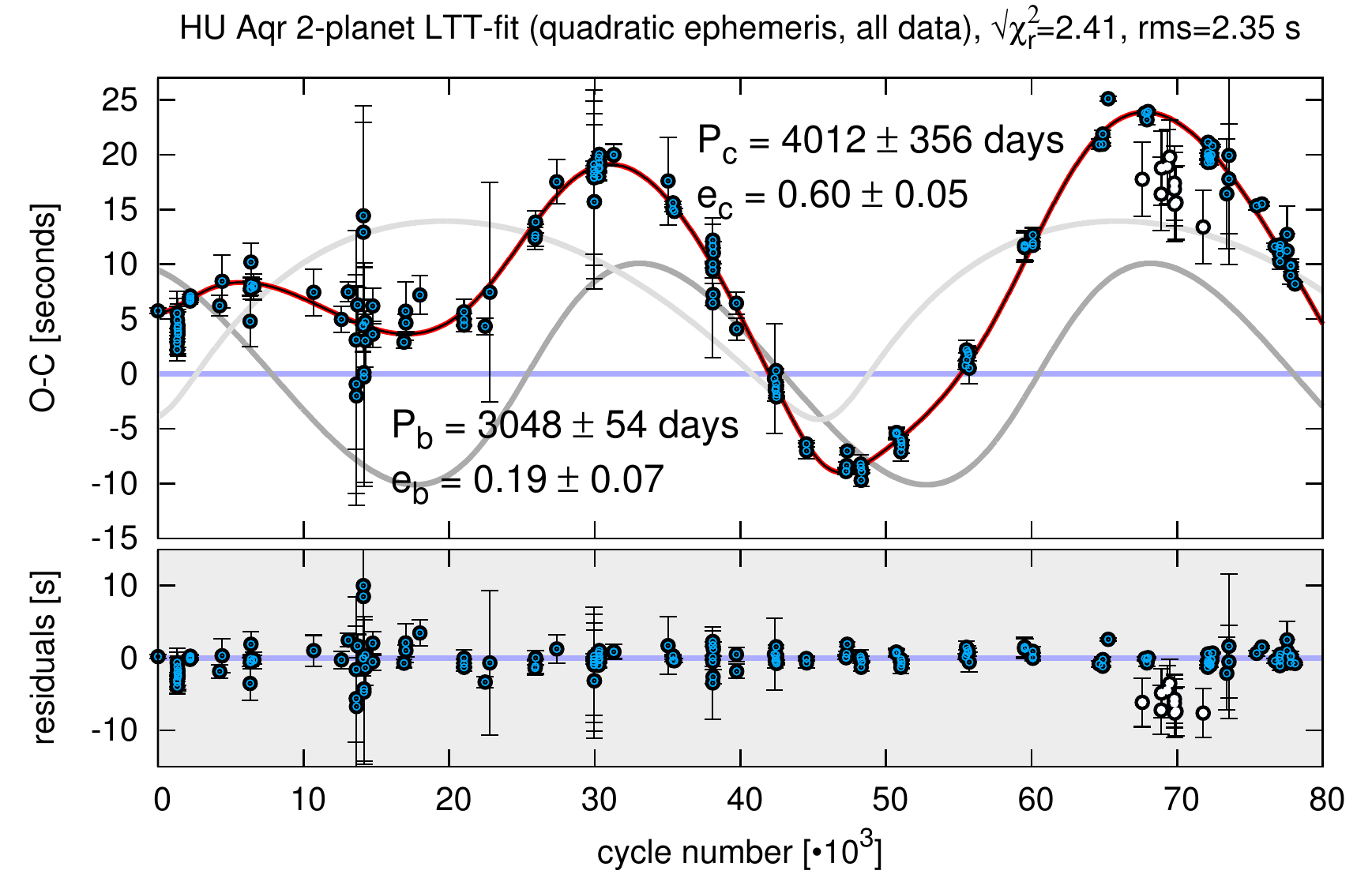}}     
       \hbox{\includegraphics[width=3.5in]{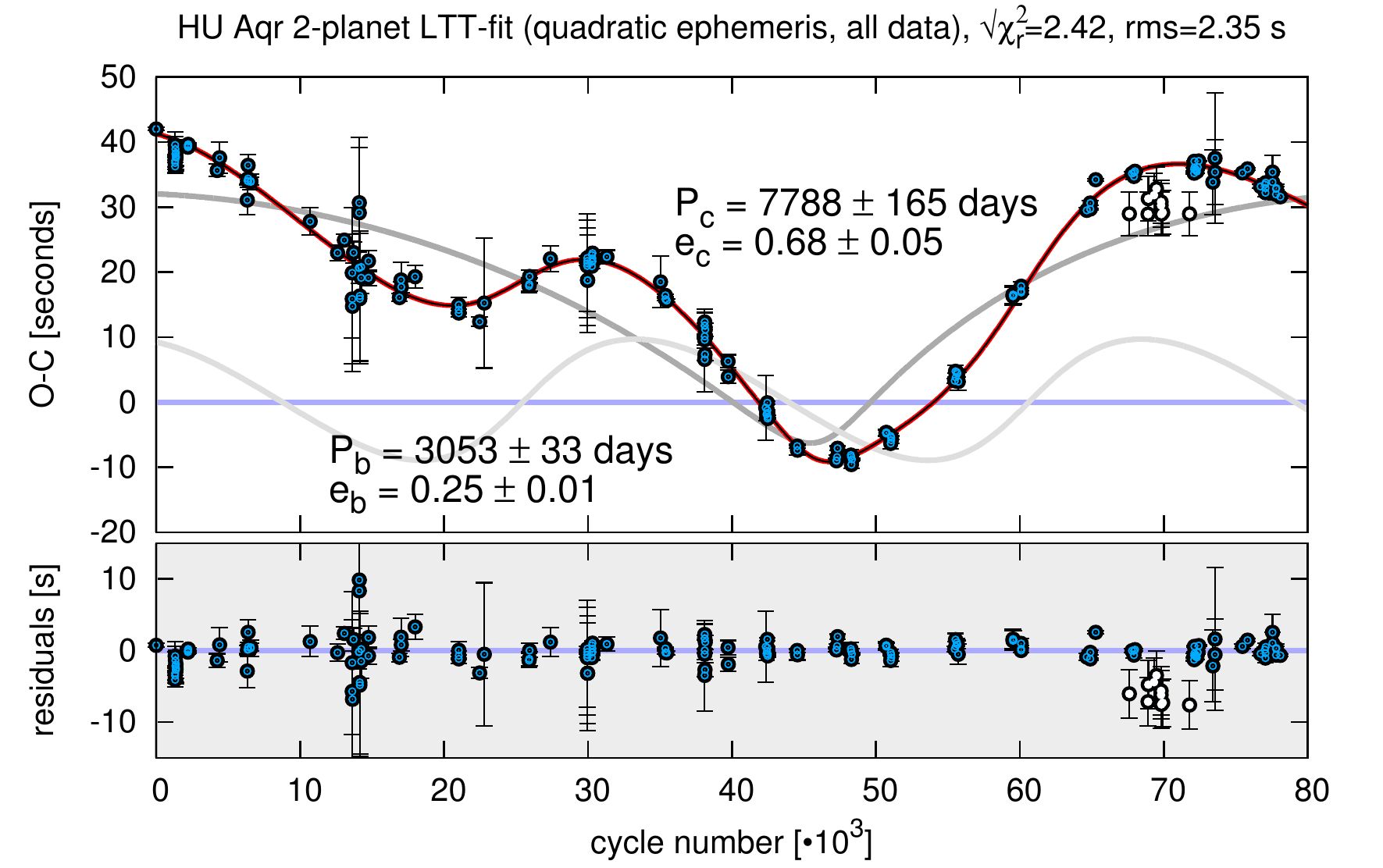}}     
   }
    \hbox{
       \hbox{\includegraphics[width=3.5in]{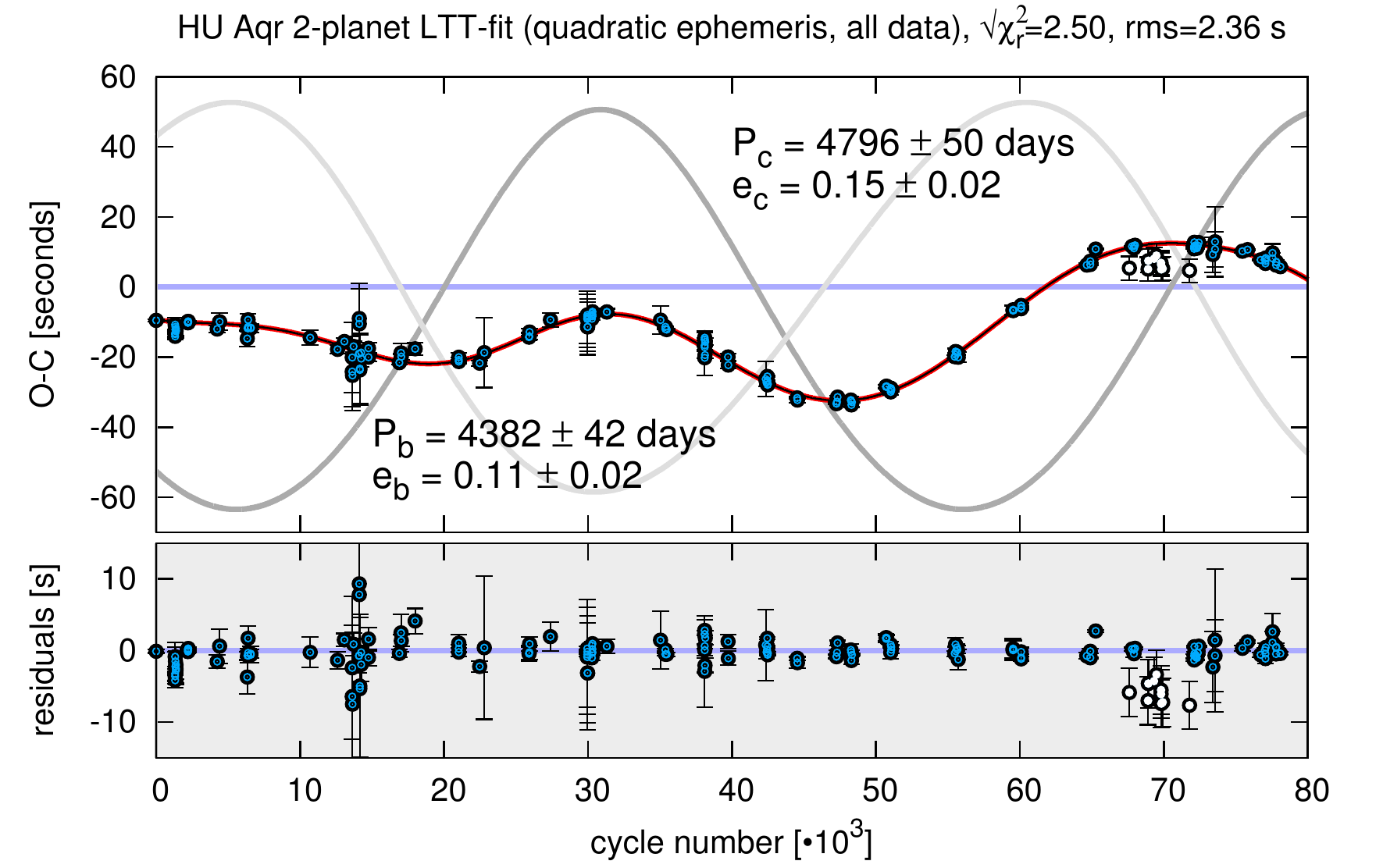}}     
       \hbox{\includegraphics[width=3.5in]{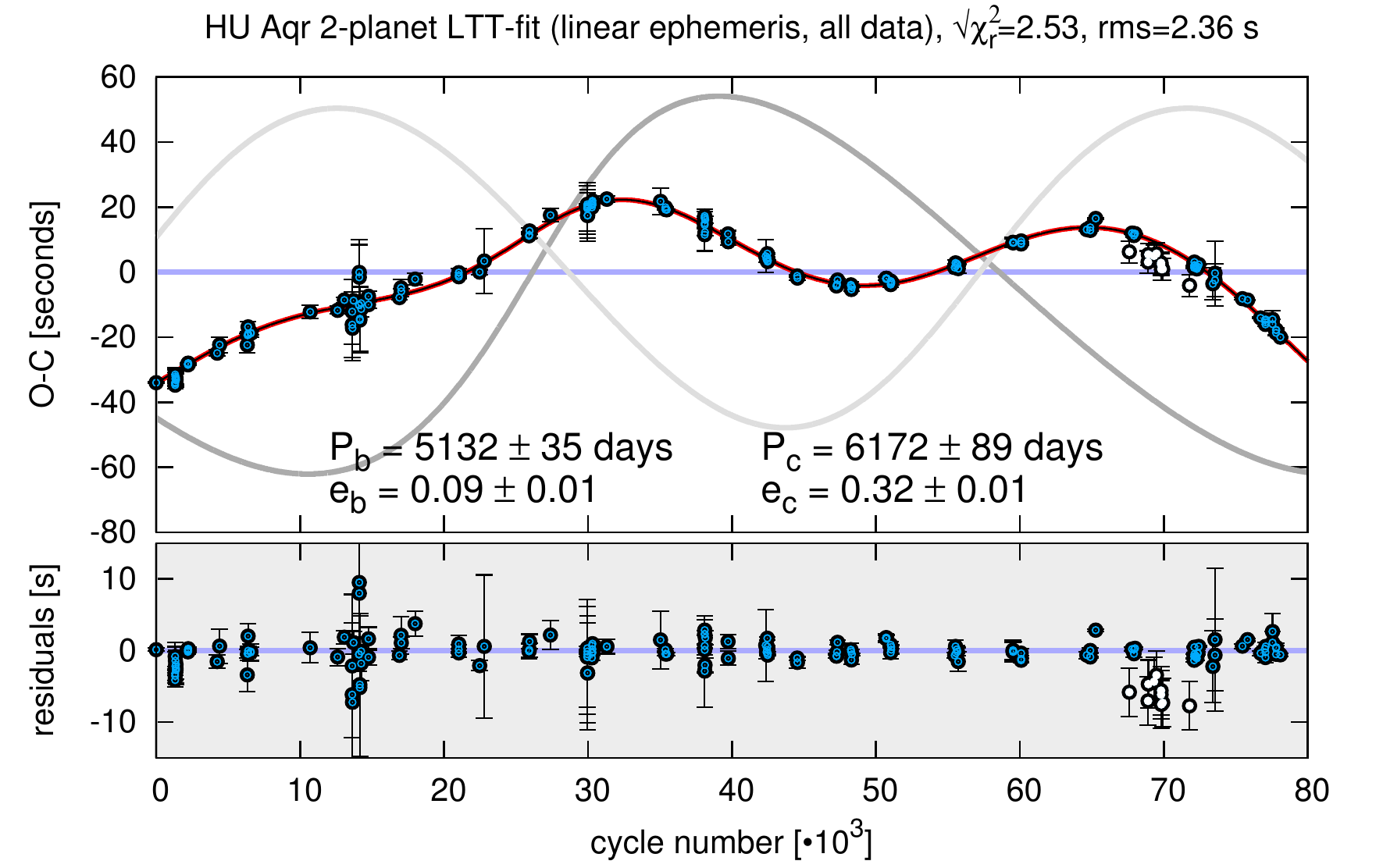}}     
   }
}
}
\caption{
Examples of synthetic curves of the 2-planet LTT quadratic and linear ({\em 
the bottom right panel}) ephemeris models to all 171 data points analysed 
in this work, including 10 data points in Qian et al. (2011) which are 
marked with white filled circles. The shaded curves represent single 
planetary signal terms, respectively. 
}
\label{fig:fig14}%
\end{figure*}

%
We also tested 2-planet models with the linear and quadratic ephemeris, 
(Eqs.~\ref{eq:linear}, \ref{eq:parabolic}). Examples of the best-fit 
configurations with comparable $\Chi$ and an rms are {shown} in Fig.~\ref 
{fig:fig14}. {Similar to} the case of the SSQ data set, no unique solution 
may be {found}. For the parabolic ephemeris, we found many similar-quality 
best-fit solutions. These fits are characterised by the orbital periods 
ratio close to 1c:1b~MMR with inferred planetary masses {of} $\sim 20$~M
$_{\idm{Jup}}$ (the bottom left-hand panel in Fig.~\ref {fig:fig14}), close 
to 4c:3b~MMR with inferred masses {of} $\sim 5.5$~M $_{\idm{Jup}}$ and 
$\sim 4.0$~M$_{\idm{Jup}}$ (the top right-hand panel in Fig.~\ref
{fig:fig14}). A solution close to the 2c:1b ~MMR (the top right-hand panel 
in Fig.~\ref{fig:fig14}), as well as configurations with extreme 
eccentricity $e_{\idm{c}}\sim 0.95$, {\em positive} damping factor $\beta 
\sim 10^{-12}$~cycles day$^{-2}$, and unconstrained $P_{\idm{c}} \sim 
250,000$~days (not {shown} here) {was also found}. In all these cases, the 
rms remains at the level of $2.4$~seconds. Some of these solutions are 
qualitatively similar to the 2-planet fits found for the SSQ data set. 
These results imply that the significantly extended data set still does not 
constrain 2-planet models. 

For the linear ephemeris model, we found one best--fit solution that 
frequently appeared in different runs of the hybrid code. It is shown in 
the bottom right-hand panel of Fig.~\ref{fig:fig14}. This solution is 
characterised by  an orbital periods ratio close to 6:5. Taken literally, 
this fit  corresponds to Trojan brown dwarfs. However, the kinematic model 
is inadequate for such a configuration of massive objects.
%
\subsection{Newtonian, self-consistent $N$-body 2-planet models}
%
%
In the light of the discussion presented above, we performed a preliminary 
modelling of all available data with the help of the hybrid algorithm 
driven by the self-consistent $N$-body model. Moreover, we tested Lagrange 
stability of the best-fit models following their orbital evolution over at 
least $10^6$ orbital periods of the outermost planet. Configurations which 
survived during such time without a collision or remaining on closed orbits 
were regarded stable. In this experiment we use 161~data points, excluding 
data in \citep{Qian2011}, due to the discrepancy with OPTIMA measurements.

\begin{figure*}
\centerline{
\vbox{
    \hbox{
       \hbox{\includegraphics[width=3.5in]{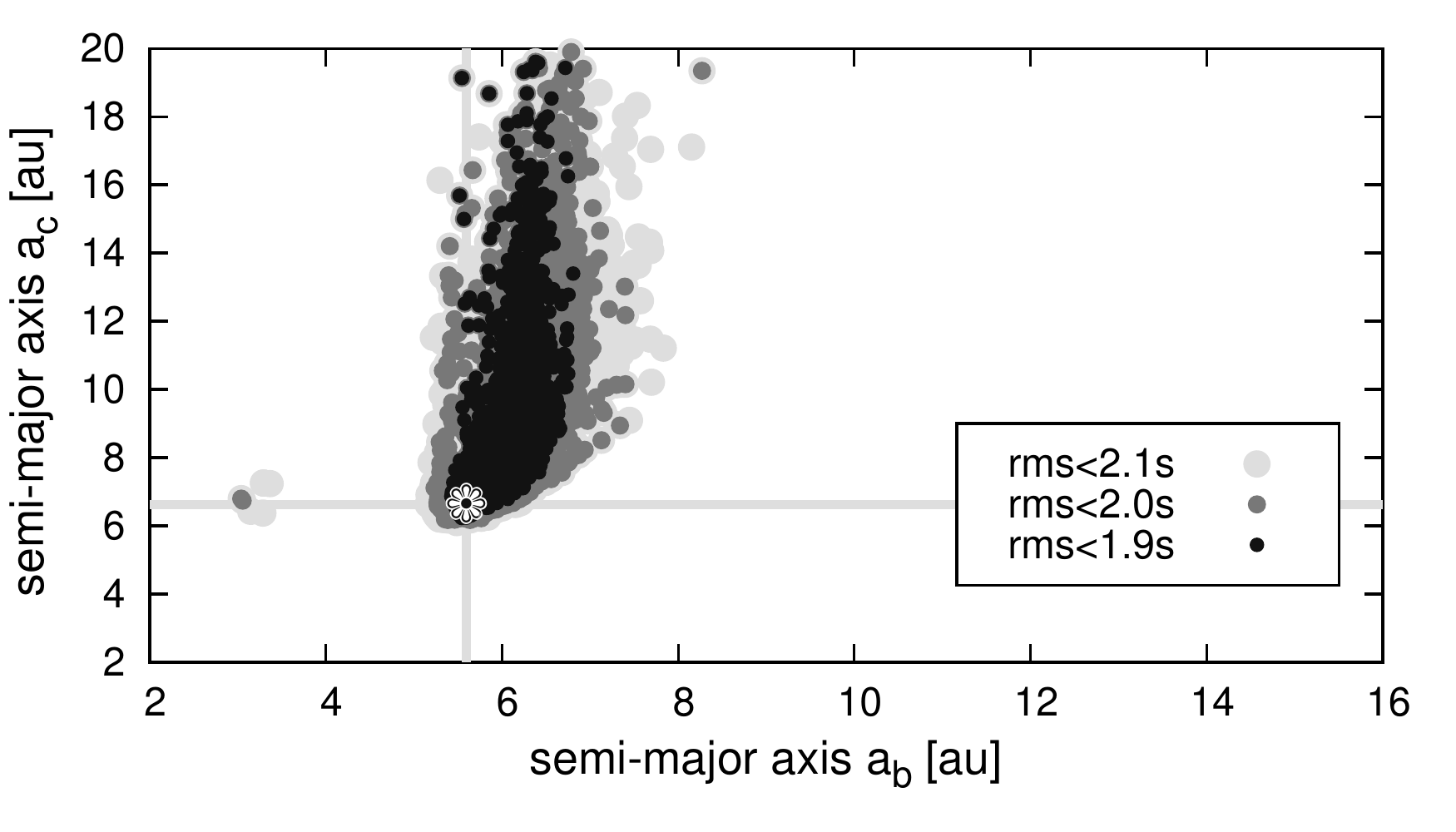}}     
       \hbox{\includegraphics[width=3.5in]{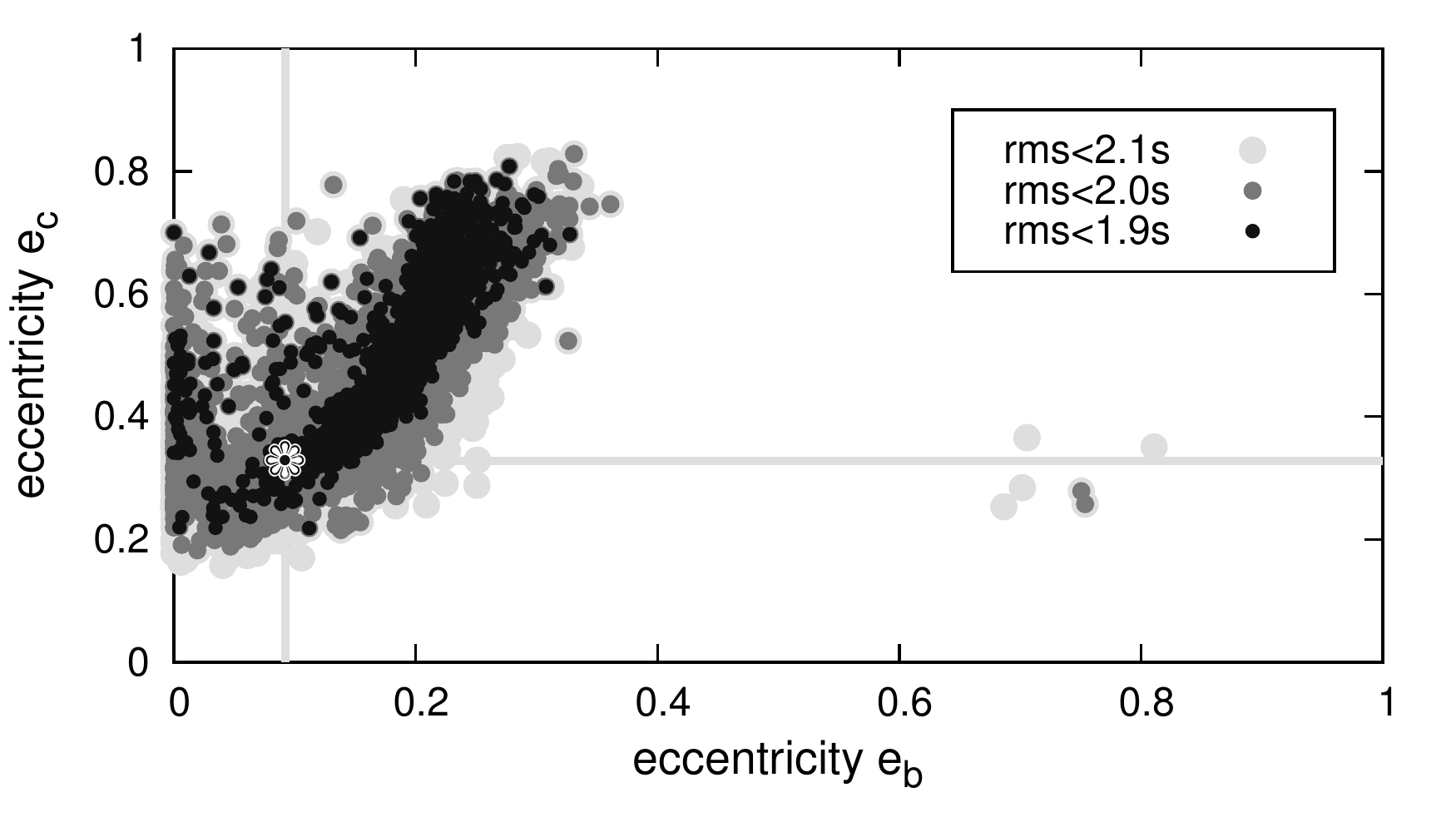}}     
   }
    \hbox{
       \hbox{\includegraphics[width=3.5in]{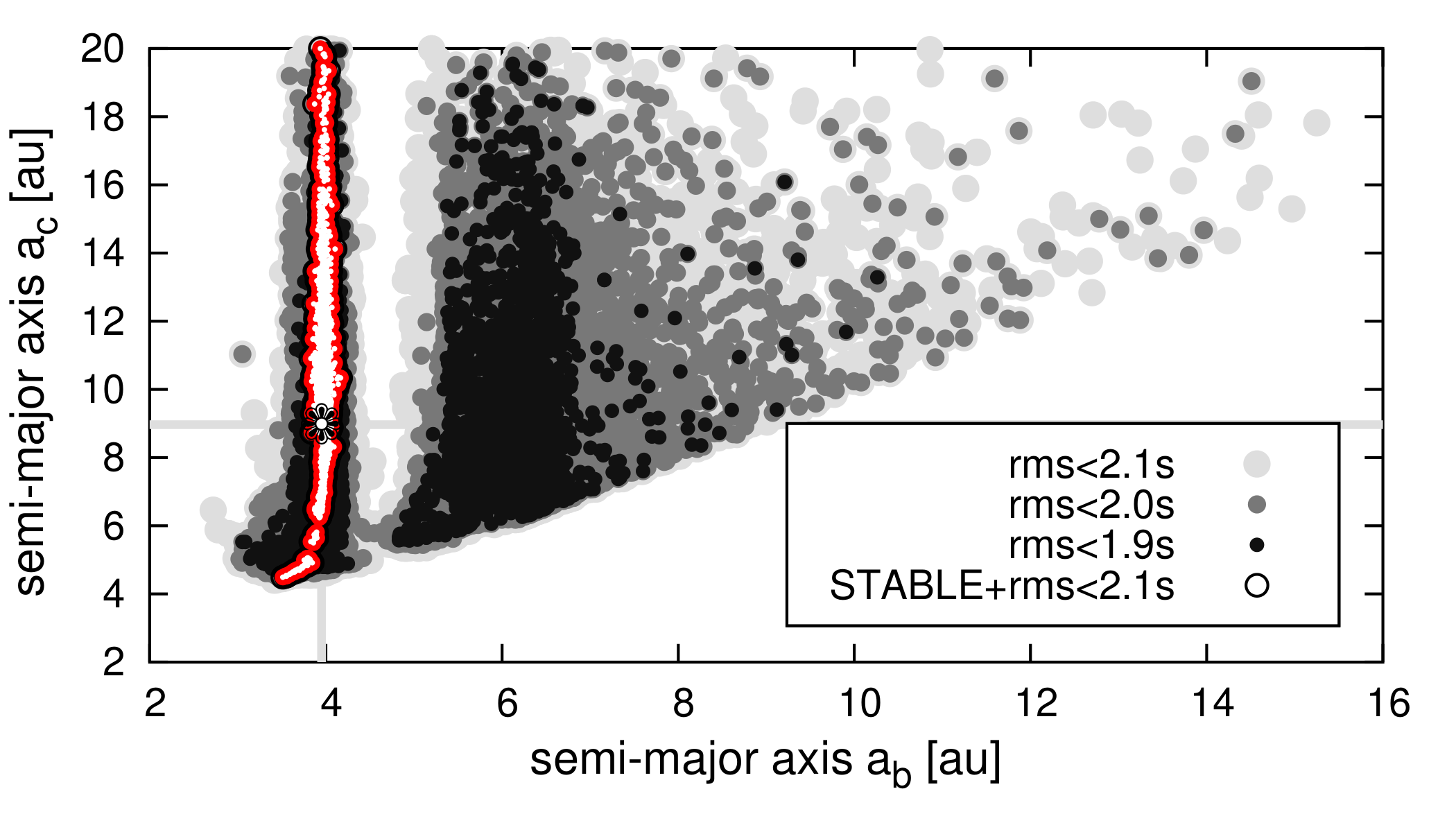}}     
       \hbox{\includegraphics[width=3.5in]{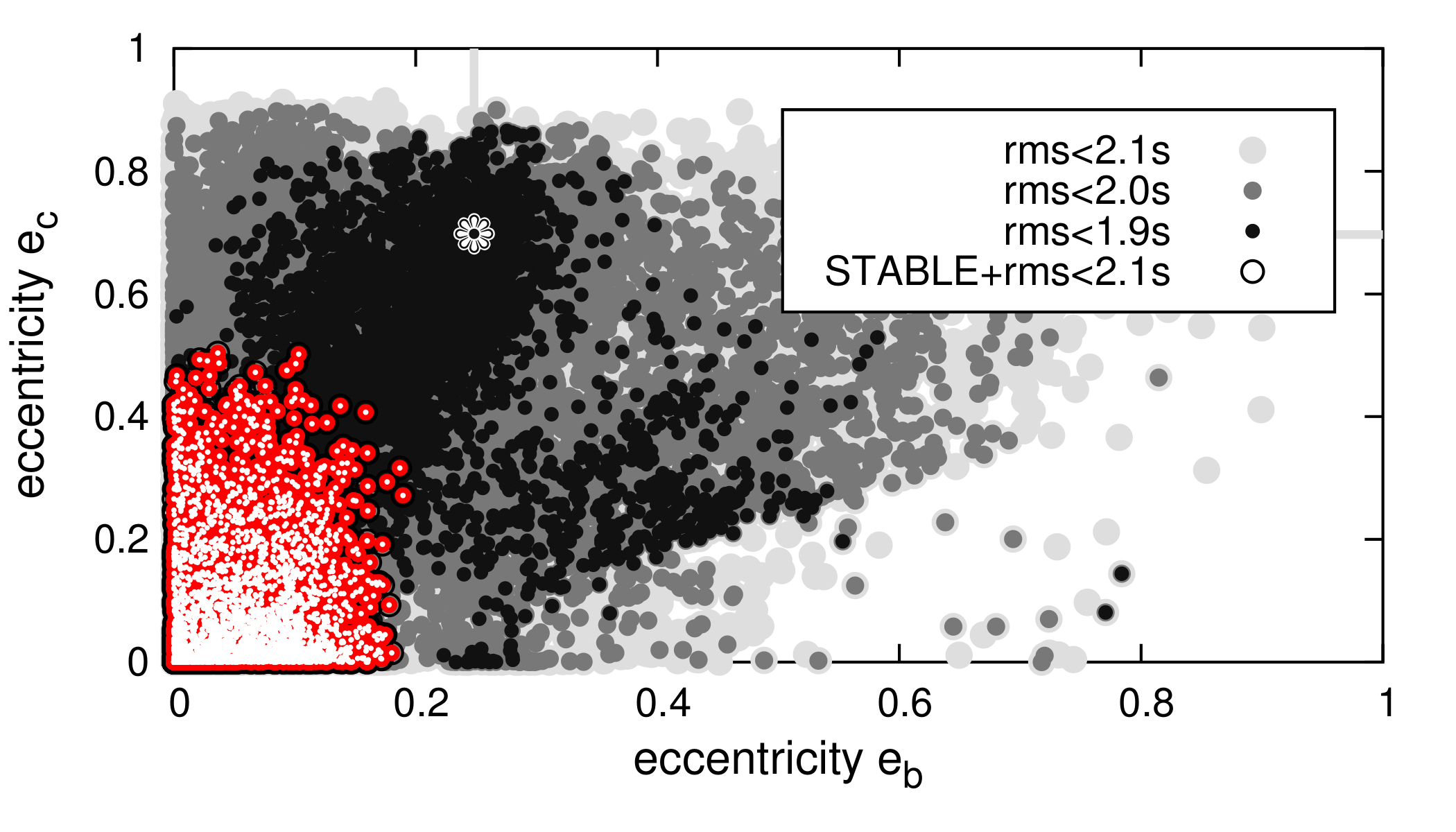}}     
   }
}
}
\caption{
Statistics of 2-planet $N$-body models gathered with the hybrid algorithm, 
projected onto planes of selected parameters. {\em The top row} illustrates 
the results for the linear ephemeris, {\em the bottom row} shows the fits 
for the quadratic ephemeris model. The rms quality of these solutions is 
coded as filled circles: the better fit---the darker colour. Orbital 
parameters of the best-fit solutions are marked with shaded, intersecting 
lines and a flower symbol. Solutions Lagrange--stable over $10^6$ 
revolutions of the outermost planet are marked with red--white circles.
}
\label{fig:fig15}
\end{figure*}

To illustrate the results of the hybrid optimization, we projected the found 
solutions onto particular planes of the Keplerian astrocentric, osculating 
elements of the planets (Fig.~\ref{fig:fig15}) at the epoch of the first 
observation. The general finding is that the $N$-body formulation helps to 
improve the rms, that decreased from $\sim 2.4$~seconds to $\sim1.9$
~seconds as compared to kinematic models.

The top row of Fig.~\ref{fig:fig15} illustrates the results for the linear 
ephemeris. Clearly, the data do not constrain the semi-major axes and 
eccentricities of the companions. The eccentricities tend to be large, up 
to 0.8. Moreover, the best-fit configuration exhibit similar values of 
semi-major axes ($\sim$5.6~AU and $\sim$6.3~AU) and large masses in the 
brown-dwarfs range of $\sim 20$ Jupiter masses. We did not \corr{find}
any stable configurations within this model. It is consistent with the
results for the SSQ data set \citep{Hinse2011}.

\begin{figure*}
\centerline{
\vbox{
    \hbox{
       \hbox{\includegraphics[width=3.5in]{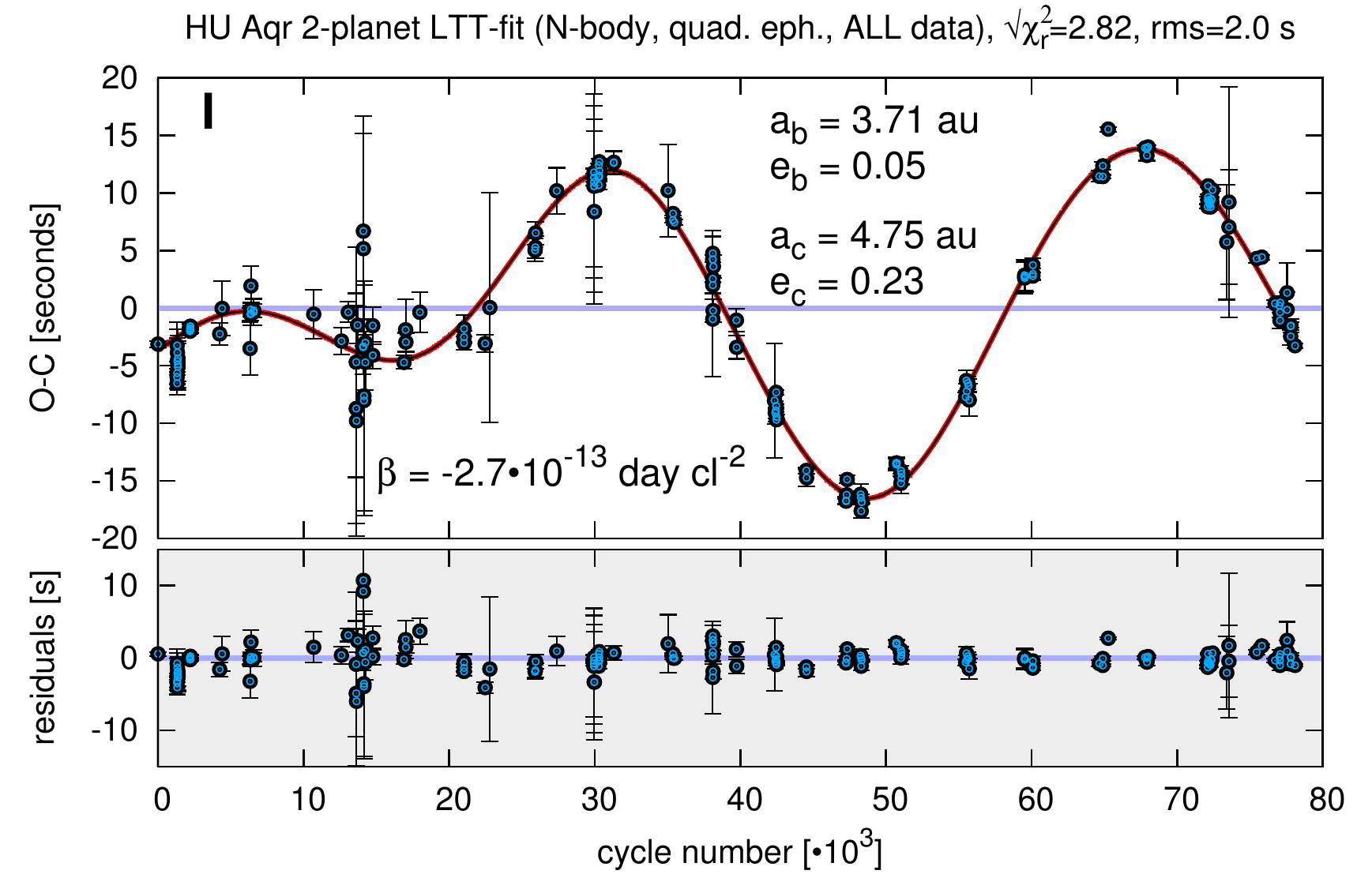}}    
       \hbox{\includegraphics[width=3.5in]{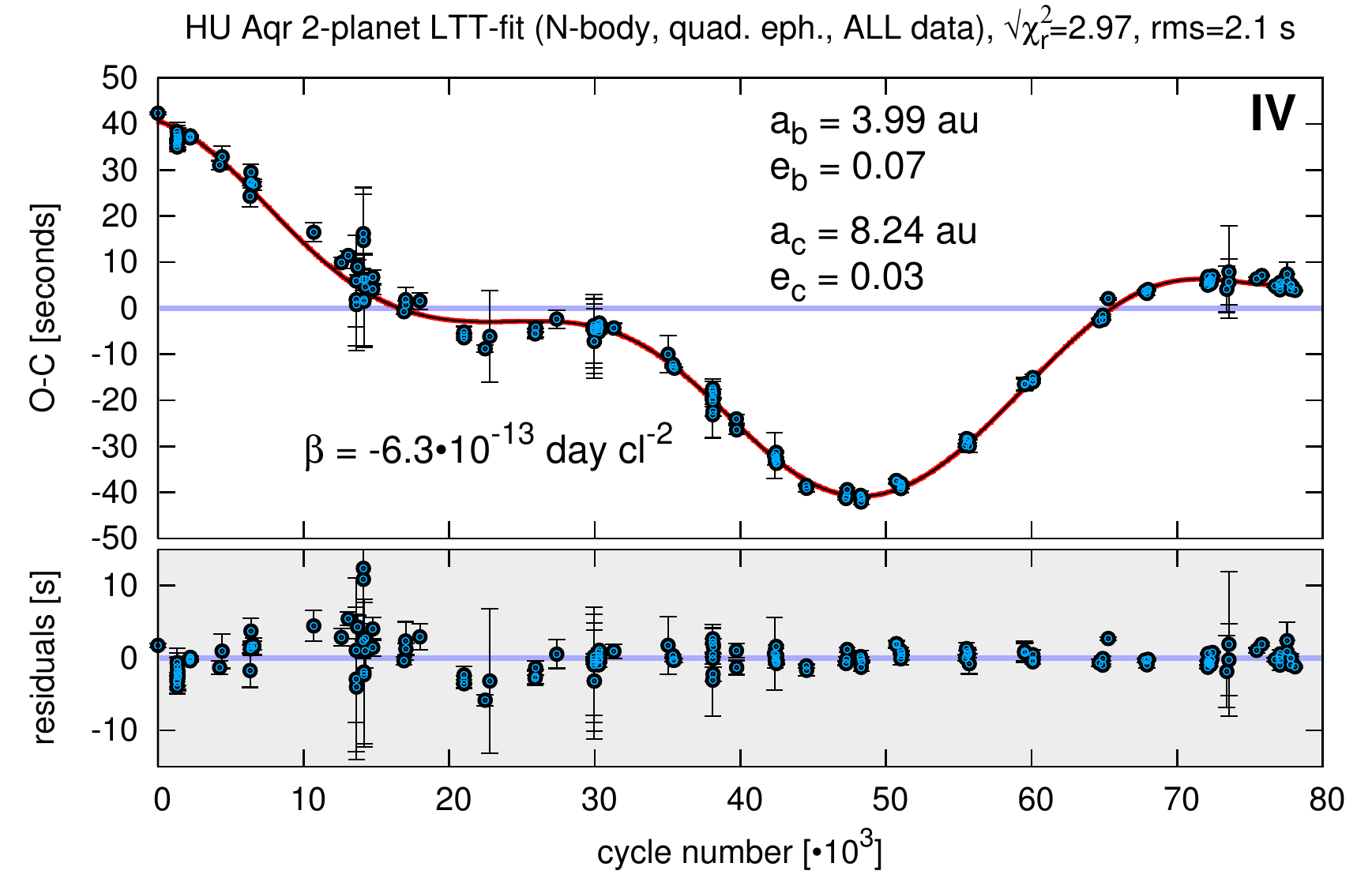}}    
   }
}
}
\caption{
Examples of synthetic LTT curves and (O-C) residuals for {\em stable} 
2-planet quadratic ephemeris Newtonian ($N$--body) models to 161 data 
points analysed in this work, without 10 data points in Qian et al. 
(2011).  These models are selected from a sample illustrated in Fig.~\ref 
{fig:fig15}. Dynamical maps of these solutions shows Fig.~\ref {fig:fig17} 
(they are labelled with the Roman numerals, as I~and IV, respectively). 
}
\label{fig:fig16}%
\end{figure*}

\begin{figure*}
\centerline{
\vbox{
    \hbox{
       \hbox{\includegraphics[width=3.5in]{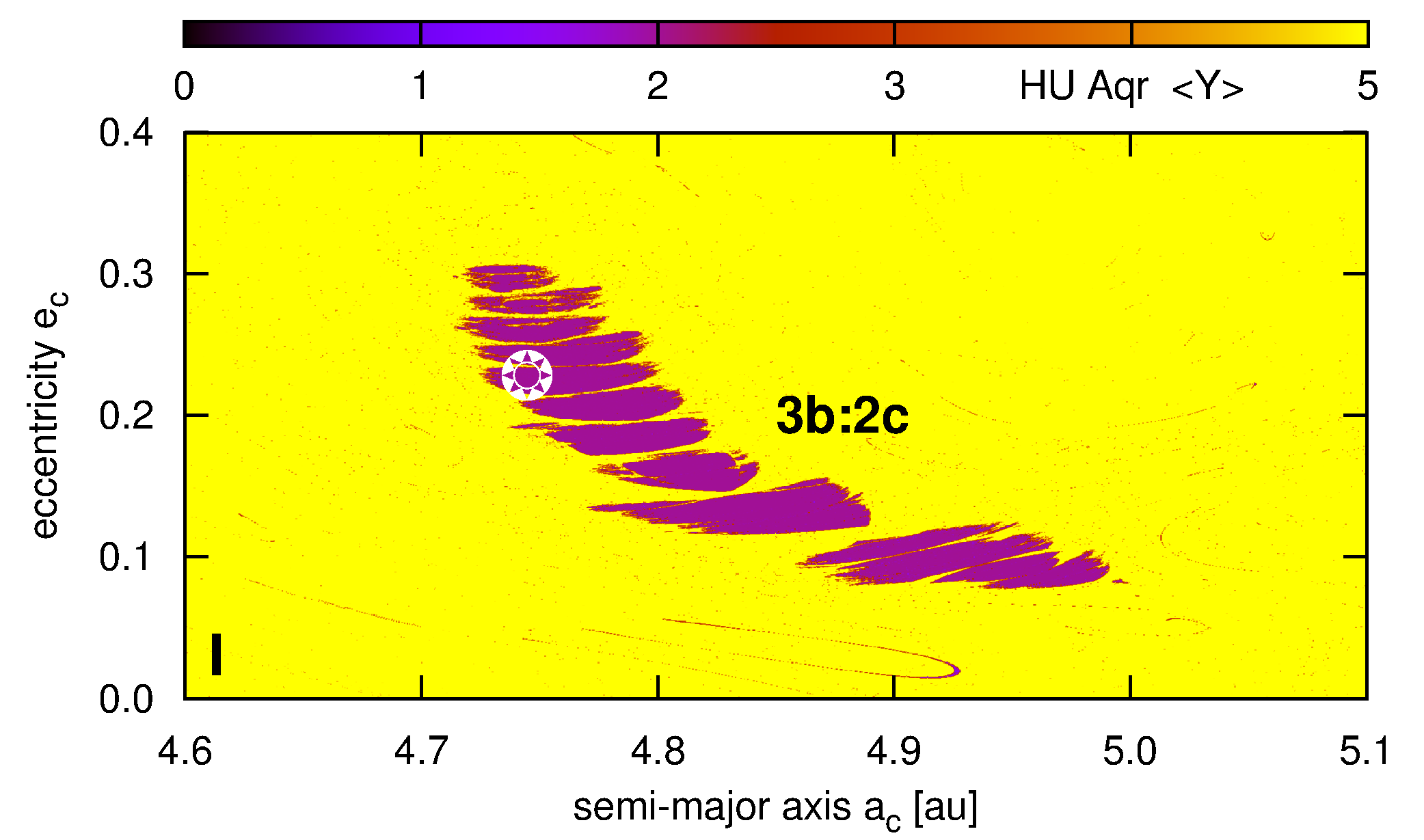}}     
       \hbox{\includegraphics[width=3.5in]{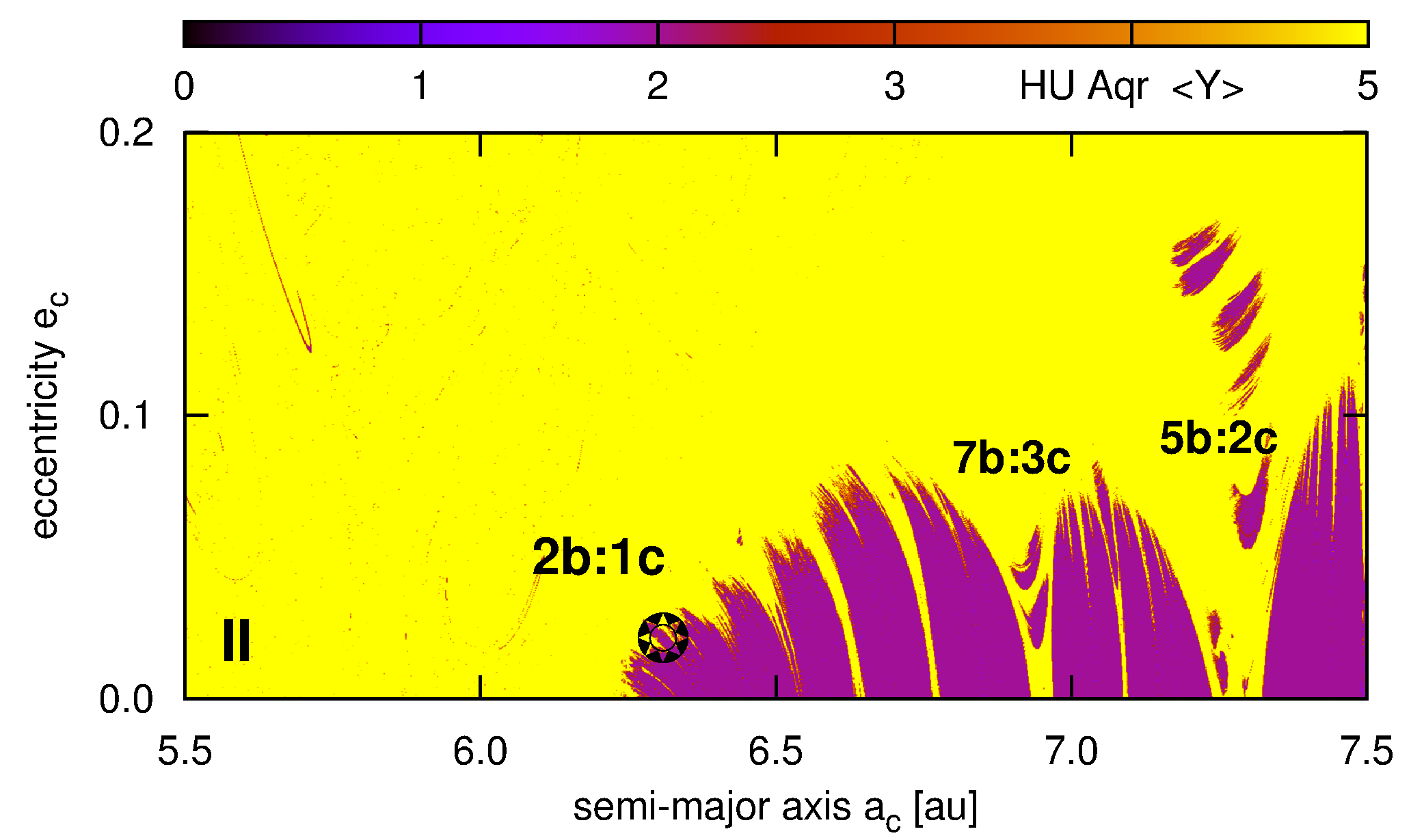}}     
   }
    \hbox{
       \hbox{\includegraphics[width=3.5in]{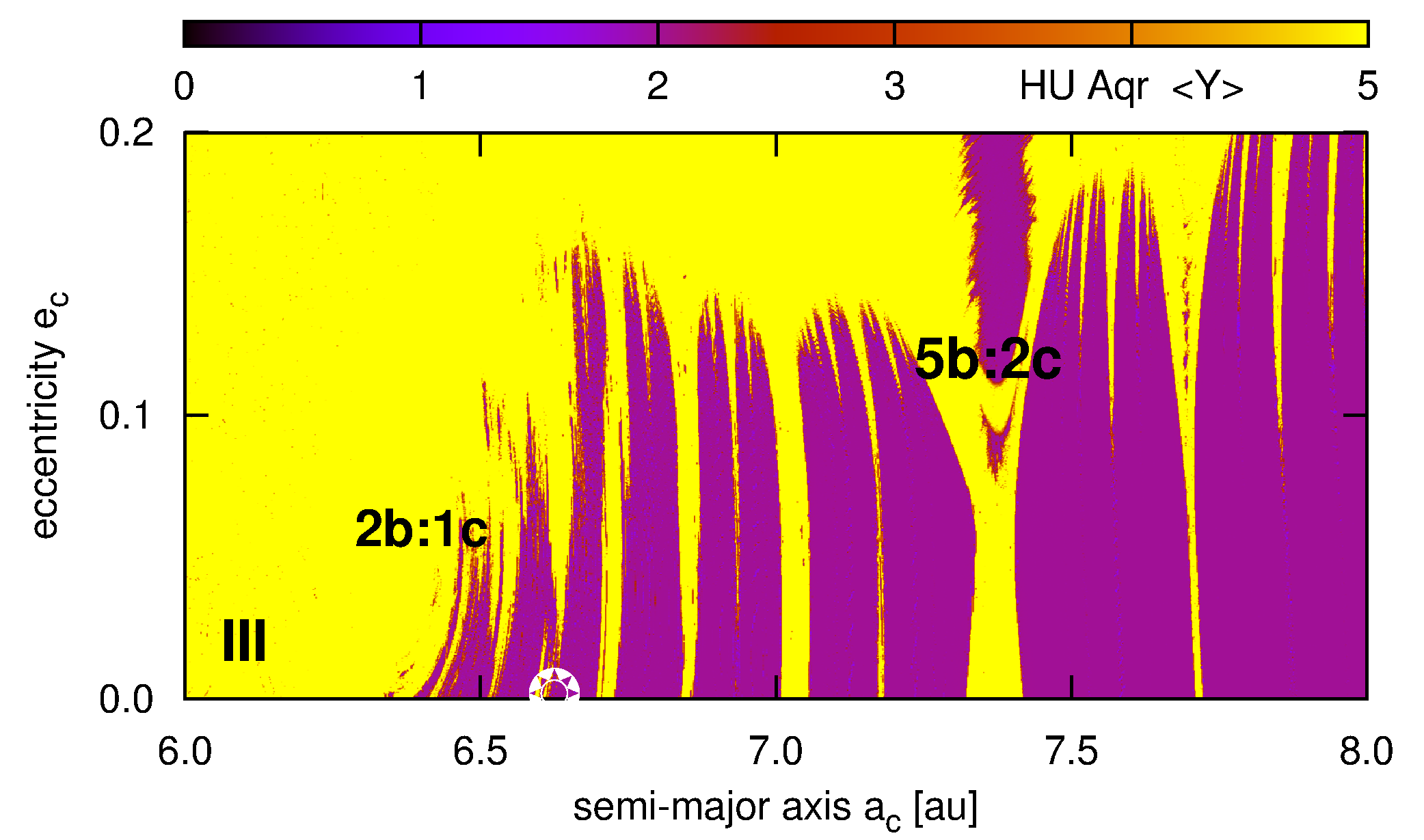}}     
       \hbox{\includegraphics[width=3.5in]{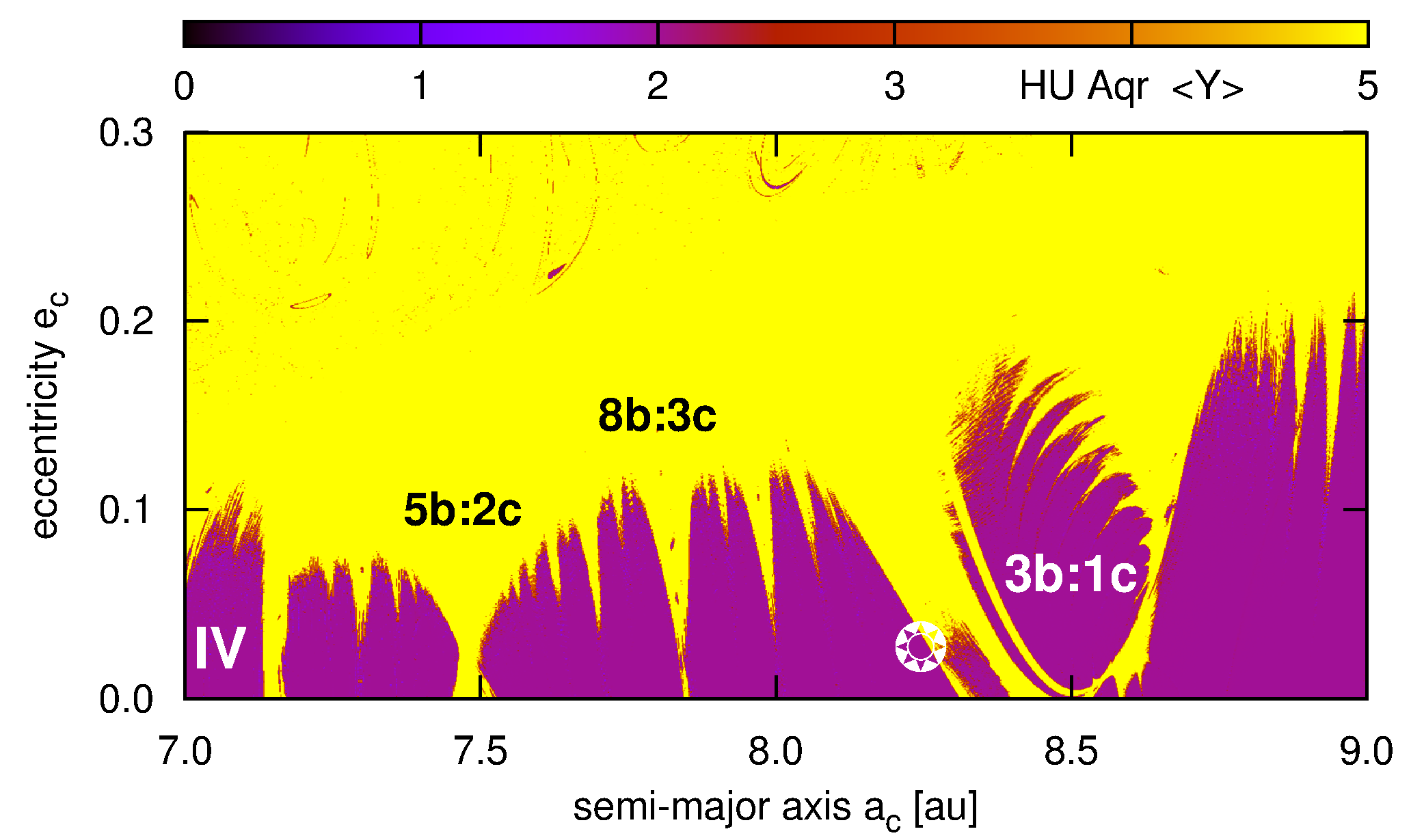}}     
   }
    \hbox{
       \hbox{\includegraphics[width=3.5in]{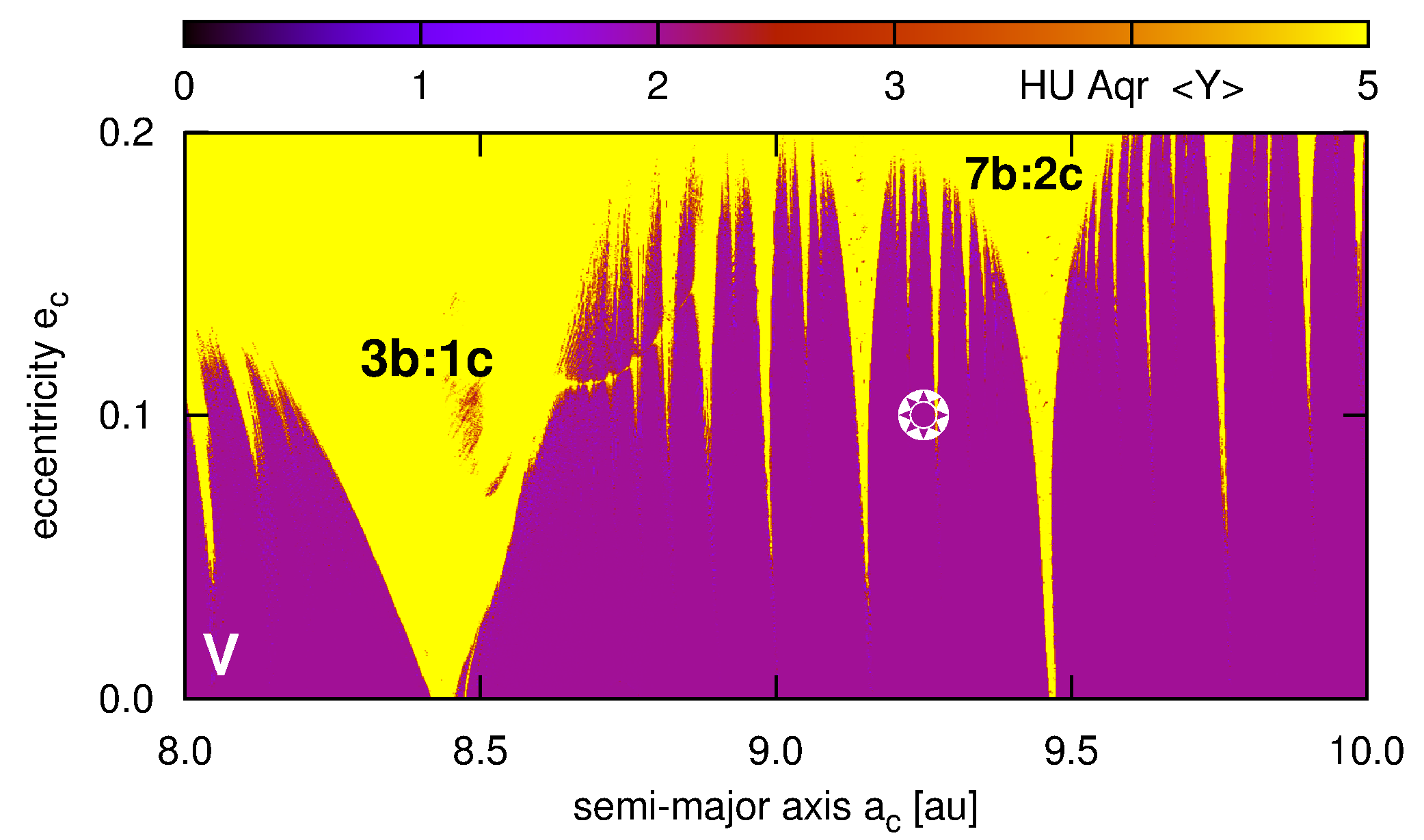}}     
       \hbox{\includegraphics[width=3.5in]{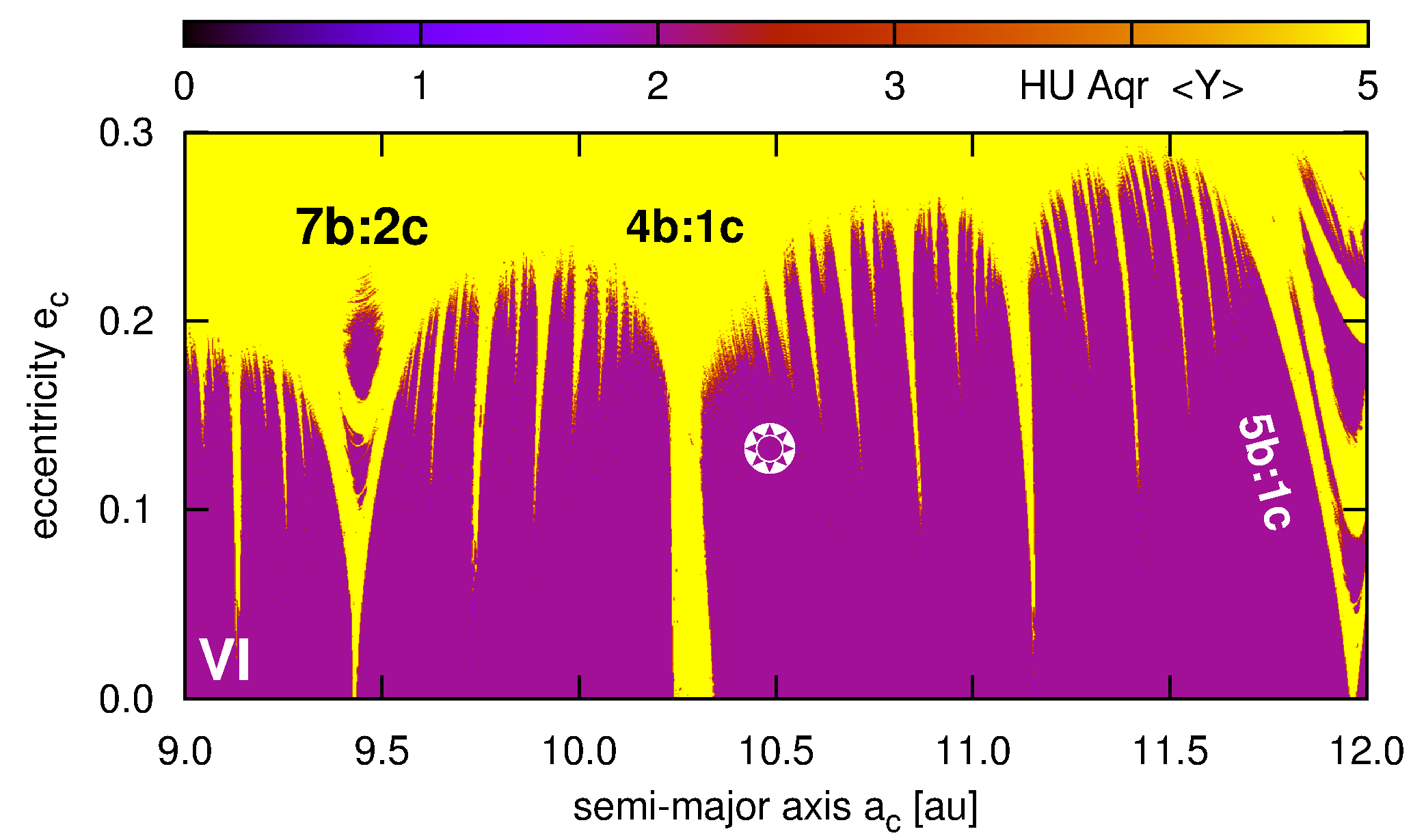}}     
   }
}
}
\caption{
MEGNO dynamical maps  in the $(a_{\idm{c}},e_{\idm{c}})$--plane for a few 
representative $N$-body {\em stable} solutions illustrated in the bottom 
panels of Fig.~\ref{fig:fig16}.  Yellow colour encodes strongly unstable 
(chaotic) configurations, and purple colour (MEGNO $\left<Y\right>\sim 2$) 
is for stable,  quasi-periodic solutions.  Parameters of the 
nominal, tested fits are marked with the star symbol.  The most prominent, 
low-order mean motion resonances are labeled.  The original resolution of 
these dynamical maps is $1440\times900$ data points integrated for $10^4$ 
outermost orbital period each.  The total mass of the binary is 0.98~
$M_{\sun}$~\citep {Schwope2011}.
}
\label{fig:fig17}%
\end{figure*}

Interesting results are obtained for the quadratic ephemeris model (see 
the bottom row in Fig.~\ref{fig:fig14}) although also this model does not 
constrain orbital parameters, due to even larger spread of the semi-major 
axes and eccentricities than in the linear ephemeris model. Two minima of 
$\Chi$ are found, around $a_{\idm{b}}\sim4$~au, and $a_{\idm{b}}\sim6$~au, 
respectively. The best-fit configurations have $\Chi \sim 2.6$ and and an 
rms$~\sim 1.9$~second that is $\sim 20\%$ better than for the best 
kinematic models. In the neighborhood of the first $\Chi$ minimum (
$a_{\idm{b}}\sim4$~AU), we found a few thousands of Lagrange stable models 
characterised by $\Chi<3$ and an rms~$<2.1$ (still better than for the best 
2-planet kinematic models). These fits have well bounded $a_{\idm{b}}\sim4$
~AU and small eccentricities up to 0.4. However, the osculating semi-major 
axis of the outer body is unconstrained and covers many low order mean 
motion resonances, between 3c:2b~MMR and 5c:1b~MMR. Figure~\ref {fig:fig16} 
shows synthetic curves of two example solutions corresponding to the 
3c:2b~MMR (the left panel) and for a model close to 3c:1b~MMR (the right 
panel). To identify these resonances, in the neighborhoods of a few 
selected best fit models, we derived high-resolution dynamical maps (1440
$\times900$ data points)  shown in Fig.~\ref{fig:fig17}. These maps are 
computed in terms of the fast indicator MEGNO \citep {Cincotta2003}, with 
the help of our recently developed CPU cluster software MECHANIC\footnote
{http://git.astri.umk.pl/projects/mechanic} \citep {Slonina2012}. MEGNO 
measures the maximal Lyapunov exponent, which makes it possible to 
distinguish between chaotic and regular solutions. Each point at these maps 
has been integrated over $\sim 10^4$ orbital periods of the outermost 
companion. The dynamical maps confirm that the Lagrange stable models 
examined over a limited time-span are equivalent to \corr{quasi-periodic}, 
stable solutions.

A solution illustrated in the left panel of Fig.~\ref{fig:fig16} is the 
best-fit {\em stable} model found in the hybrid search with $\Chi \sim 2.82$
and an rms$\sim 2$~sec. It is located in a very narrow, isolated stability 
island of the 3c:2b~MMR and characterised by relatively large 
$\beta\sim-2.7 \times 10^{-13}$ day~cycle$^{-2}$, similarly to the 
kinematic model. The right panel of Fig.~ \ref{fig:fig16} shows a 
configuration close to the 3c:1b~MMR, which has even larger $\beta \sim 
-6~\times 10^{-13}$ day cycle$^{-2}$.

\begin{figure}
\centerline{
    \hbox{
       \hbox{\includegraphics[width=3.5in]{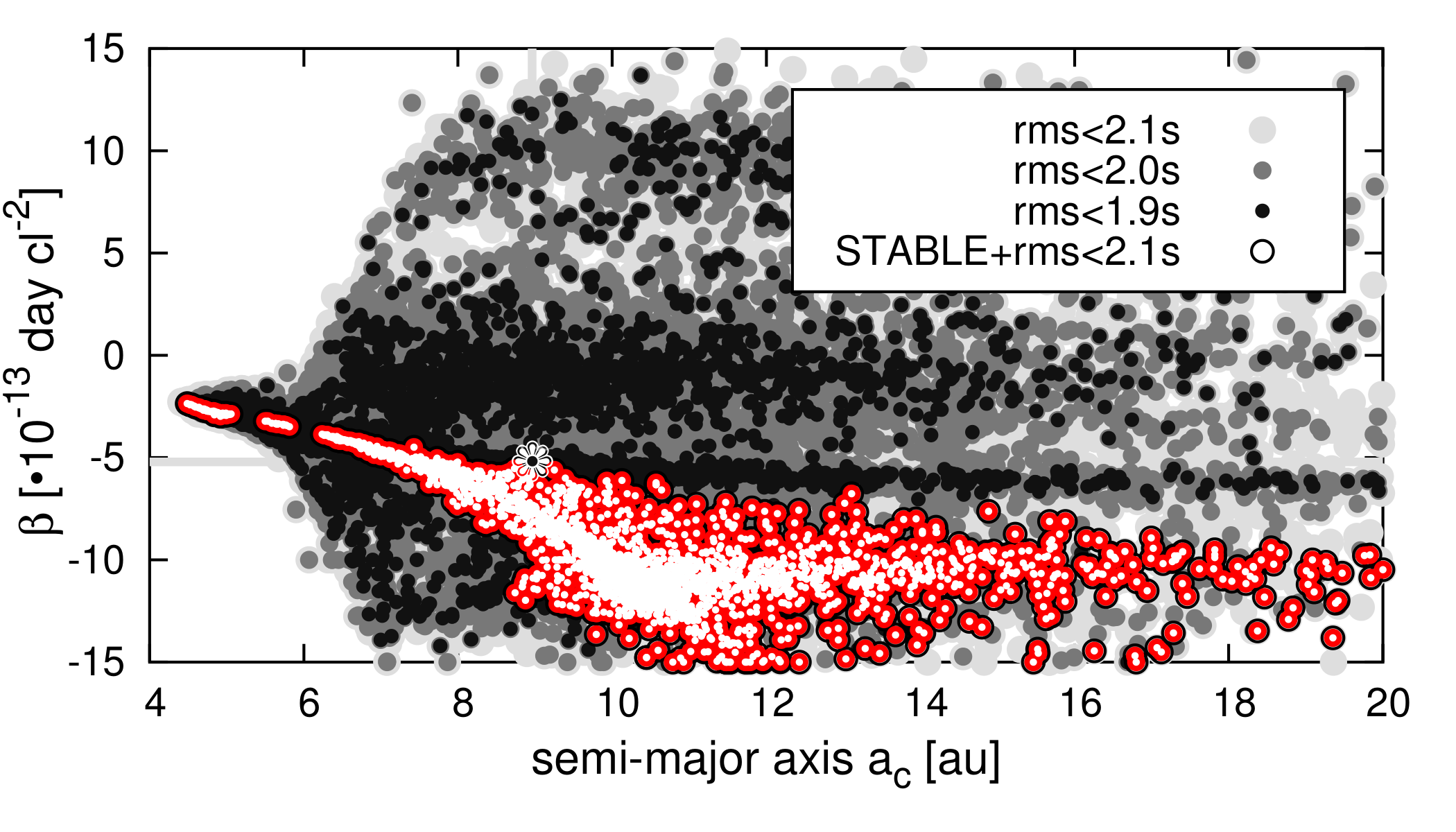}} 
   }
}
\caption{
Statistics of 2-planet $N$-body quadratic ephemeris models gathered with the 
hybrid algorithm, projected onto the ($a_c,\beta$)--plane, for the 
quadratic ephemeris. Meaning of symbols is the same as in Fig.~\ref 
{fig:fig15}.
}
\label{fig:fig18}
\end{figure}

Figure~\ref{fig:fig18} shows a statistics of the best-fit solutions in the 
$(a_{\idm{c}},\beta$)--plane. It reveals that $\beta$ is not constrained, 
regarding even its sign.  Stable models exhibit a strong correlation between 
both these parameters.  A larger value of the semi-major axis of the outer 
planet is related to a larger magnitude of $\beta$.  Due to this 
correlation, an interpretation of stable configurations is complex.  For 
relatively small magnitude of $\beta$, stable configurations are 
characterized by low order MMRs and may be found in tiny areas of stable 
motions (see Fig.~\ref{fig:fig17}).  For more separated planets, when 
stability zones are much more extended, $|\beta|$ increases.  Already 
$|\beta|\sim2\times10^{-13}$ day~cycle$^{-2}$ is difficult to explain by 
physical phenomena in the binary, as we discussed in Sect.~5.  Such large 
values of $|\beta|$ may indicate a third, \corr{long-period companion object
in a very distant orbit}.  However, because already the 2-planet 
model is not constrained by the data, also a 3-planet configuration cannot
be fixed without ambiguity.  We did an attempt to search for such Newtonian 
3-planet models within the linear ephemeris, but we did not find any 
improved, nor stable solutions of this type.

%
%
\bibliographystyle{mn2e}
\bibliography{ms}

\end{document}